\documentclass[11pt,a4paper]{article}
\usepackage{amsmath}
\numberwithin{equation}{section}
\usepackage{amssymb}
\usepackage{float}
\usepackage[section]{placeins}
\usepackage{tensor}
\usepackage{physics}
\usepackage{slashed}
\usepackage{multicol}
\usepackage{graphicx}
\usepackage{subcaption}
\usepackage[normalem]{ulem}
\usepackage{listings}
\usepackage{xcolor}
\usepackage{tikz}
\usepackage{longtable}
\usepackage{tabularx}
\usepackage{booktabs}
\usetikzlibrary{arrows.meta, positioning, fit, shapes.geometric, shapes, calc, arrows}
\usepackage{xurl}
\usepackage[colorlinks=true, allcolors=blue]{hyperref}
\usepackage[backend=biber, style=numeric-comp, sorting=none, sortcites=none]{biblatex}
\AtBeginBibliography{\emergencystretch=4em\sloppy}
\AtEveryBibitem{\clearfield{abstract}\clearfield{note}}
\usepackage{bm}
\usepackage{pifont}
\usepackage{enumitem}

\definecolor{greenbox}{RGB}{46,184,144}
\definecolor{orangediamond}{RGB}{242,148,0}
\definecolor{bluebox}{RGB}{100,160,220}
\definecolor{blueboxlight}{RGB}{200,220,240}
\definecolor{purplebox}{RGB}{160,120,160}
\definecolor{purpleboxlight}{RGB}{220,200,220}

\tikzstyle{process} = [rectangle, rounded corners, minimum width=8cm, minimum height=1.5cm, text centered, draw=bluebox, very thick, fill=blueboxlight, text width=7.5cm, font=\small]
\tikzstyle{preprocessing} = [rectangle, rounded corners, minimum width=8cm, minimum height=1.5cm, text centered, draw=greenbox, very thick, fill=greenbox!20, text width=7.5cm, font=\small]
\tikzstyle{decision} = [diamond, minimum width=2.5cm, minimum height=1.5cm, text centered, draw=orangediamond, very thick, fill=orangediamond!20, aspect=2, font=\small, align=center]
\tikzstyle{terminator} = [rectangle, rounded corners, minimum width=8cm, minimum height=1.2cm, text centered, draw=purplebox, very thick, fill=purpleboxlight, text width=7.5cm, font=\small]
\tikzstyle{arrow} = [->, >=triangle 45, very thick]
\tikzstyle{arroworange} = [->, >=triangle 45, very thick, orangediamond]
\tikzstyle{arrowblue} = [->, >=triangle 45, very thick, bluebox]
\tikzstyle{looparrow} = [->, >=triangle 45, very thick, bluebox]

\IfFileExists{literature.bib}{%
  \addbibresource{literature.bib}%
}{%
  \addbibresource{arxiv/literature.bib}%
}
\graphicspath{{./}{figures/}{arxiv/}{arxiv/figures/}}

\definecolor{codegreen}{rgb}{0,0.6,0}
\definecolor{codegray}{rgb}{0.5,0.5,0.5}
\definecolor{codepurple}{rgb}{0.58,0,0.82}
\definecolor{backcolour}{rgb}{0.95,0.95,0.92}
\definecolor{DESYcyan}{RGB}{0,166,235}
\definecolor{DESYorange}{RGB}{242,142,0}

\lstdefinestyle{mystyle}{
    backgroundcolor=\color{backcolour},
    commentstyle=\color{codegreen},
    keywordstyle=\color{DESYcyan},
    numberstyle=\tiny\color{codegray},
    stringstyle=\color{codepurple},
    basicstyle=\ttfamily\footnotesize,
    breakatwhitespace=false,
    breaklines=true,
    captionpos=b,
    keepspaces=true,
    numbers=left,
    numbersep=5pt,
    showspaces=false,
    showstringspaces=false,
    showtabs=false,
    tabsize=2
}

\lstdefinelanguage{yaml}{
  keywords={true, false, null, yes, no},
  sensitive=false,
  comment=[l]{\#},
  commentstyle=\color{DESYcyan},
  morestring=[b]',
  morestring=[b]",
  morekeywords=[2]{Modelfile, Potential, Scan, Parameters, line_param, other_params,
    grid_params, scan_params, timeout, format, output_path, description,
    plot_description, additional_plots, scale, range, N, name, g, y, x, v_GeV},
  keywordstyle=[2]{\color{DESYorange}},
}

\newcommand{\ba}[1]{\ensuremath{\left( #1 \right)}}
\newcommand{\bb}[1]{\ensuremath{\left[ #1 \right]}}
\newcommand{\bc}[1]{\ensuremath{\left\{ #1 \right\}}}

\newcommand{\pd}[2]{\ensuremath{\frac{\partial #1}{\partial #2}}}

\newcommand{\td}[2]{\ensuremath{\frac{\mathrm d #1}{\mathrm d #2}}}

\newcommand{\nocontentsline}[3]{}
\newcommand{\tocless}[2]{\bgroup\let\addcontentsline=\nocontentsline#1{#2}\egroup}

\newcommand*\diff{\mathop{}\!\mathrm{d}}

\newcommand{\tot}{\ensuremath{\text{tot}}}

\newcommand{\perc}{\ensuremath{\text{perc}}}

\newcommand{\sw}{\ensuremath{\text{sw}}}

\newcommand{\turb}{\ensuremath{\text{turb}}}
\newcommand{\col}{\ensuremath{\text{col}}}

\newcommand{\reh}{\ensuremath{\text{reh}}}

\newcommand{\fancyx}{\textcolor{DESYorange}{\ding{55}}}
\newcommand{\fancyc}{\textcolor{DESYcyan}{\ding{51}}}

\begin{document}

\title{
	\vspace*{-3cm}
	\phantom{h} \hfill\mbox{\small {P3H-26-036, TTP26-016}}
	\vspace*{0.7cm}
\\[-1.1cm]
	\vspace{12mm}
\textbf{\texttt{TransitionListener v2.0} -- Robust gravitational wave predictions for cosmological phase transitions} \\ \vspace{0.5cm} \includegraphics[width=5cm]{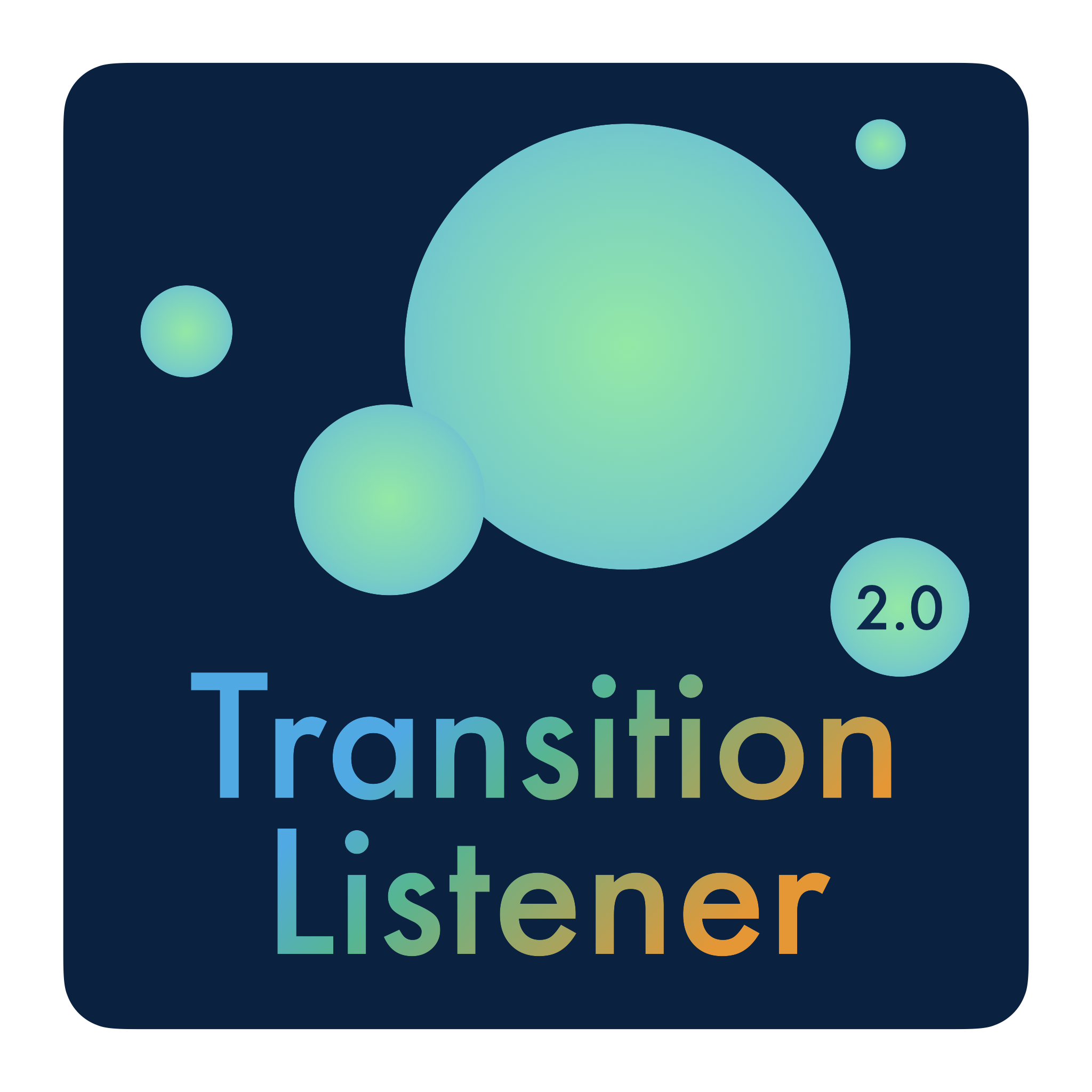}
} 
\date{}

\author{
    Jonas Matuszak$^{1\,}$\footnote{E-mail: \texttt{jonas.matuszak@kit.edu}},
    Carlo Tasillo$^{2,3\,}$\footnote{E-mail: \texttt{carlo.tasillo@ific.uv.es}}
\\[9mm]
{\small\it
$^1$Institute for Theoretical Particle Physics (TTP), Karlsruhe Institute of Technology (KIT),} \\
{\small\it 76128 Karlsruhe, Germany}\\[3mm]
{\small\it
$^2$Department of Physics and Astronomy, Uppsala University,}\\
{\small\it Box 516, SE-751 20 Uppsala, Sweden} \\[3mm]
{\small\it
$^3$Instituto de Física Corpuscular (IFIC), Universitat de Valéncia-CSIC, Parc Científic UV,} \\
{\small\it C/ Catedrático José Beltrán 2, E-46980 Paterna, Spain}\\[3mm]
}

\maketitle

\vspace*{-1.0cm}

\begin{abstract}
  \noindent
  Gravitational wave backgrounds from strong first-order cosmological phase transitions
  are key observational targets predicted by many SM extensions and might be observed by
  current and future observatories like LISA, the Einstein Telescope or pulsar timing
  arrays (PTAs). Still, their precise forecast given a specific model remains a challenge.
  In this article, we present \texttt{TransitionListener v2.0}, a Python framework for
  precision studies of cosmological phase transitions and their associated gravitational
  wave (GW) signals. The code provides an end-to-end pipeline from a user-defined scalar
  potential to GW spectra and signal-to-noise ratios, enabling both benchmark studies and
  large-scale parameter scans. Version 2 introduces a self-consistent treatment of the
  transition dynamics, including the evolution of the true-vacuum fraction and its
  backreaction on the Hubble expansion, as well as a consistent description of reheating
  during percolation. A direct computation of the mean bubble separation allows to
  faithfully map to the GW spectral templates from bubble collisions, sound waves, and
  turbulence stemming from state-of-the-art simulations. \texttt{TransitionListener}
  includes built-in sensitivity curves for space- and ground-based detectors and PTAs,
  interfaces to PTA likelihoods, and wrappers for Bayesian model inference and
  high-dimensional parameter scans. Compared to existing public tools,
  \texttt{TransitionListener v2.0} improves the physical consistency and numerical
  stability of GW predictions across a wide range of models, with particular emphasis on
  the strongly supercooled and ultraslow transition regime where conventional
  approximations break down and the most promising GW signals are expected.
\end{abstract}

\newpage
{\bf PROGRAM SUMMARY/NEW VERSION PROGRAM SUMMARY}
  %Delete as appropriate.
  \vspace{1cm}

  \begin{small}
    \noindent
    {\em Program Title:} \texttt{TransitionListener v2.0} \\
    {\em CPC Library link to program files:} (to be added by Technical Editor) \\
    {\em Developer's repository link:} \href{https://github.com/tasicarl/TransitionListener}{github.com/tasicarl/TransitionListener} \\
    {\em Code Ocean capsule:} (to be added by Technical Editor)\\
    {\em Licensing provisions:} GPLv3  \\
    {\em Programming language:} python 3                            \\
    {\em Journal reference of previous version:} \href{https://doi.org/10.1088/1475-7516/2022/02/014}{10.1088/1475-7516}                  \\
    {\em Does the new version supersede the previous version?:} Yes   \\
    {\em Reasons for the new version:} Improve the physical consistency and reproducibility of GW predictions from first-order phase transitions, especially for strong supercooling and multi-field models, and extend the framework from single-point studies to large scans and inference. \\
    {\em Summary of revisions:}  Iterative percolation with true-vacuum fraction backreaction on $H(T)$; mean bubble separation as default GW length scale; LTE wall velocity; thermodynamics from the full effective potential including field-independent radiation terms; updated GW spectra; built-in detector/PTA sensitivities with optional PTArcade likelihoods; expanded scan and parameter inference interfaces. \\
    {\em Nature of problem:}  Starting from a user-defined scalar potential and mass spectrum of the theory, determine whether and when a first-order transition nucleates, percolates, and succeeds, then compute consistent hydrodynamic quantities, GW spectra and observability with controlled numerical uncertainties. \\
      %Describe the nature of the problem here. \\
    {\em Solution method:} Modular Python pipeline: phase tracing, bounce-action/tunnelling solvers, iterative Picard-sweeps for percolation, formulated as an ODE system, thermodynamic and wall-dynamics modules, GW spectrum prediction for multiple sources, and detector likelihood/SNR evaluation with configurable accuracy controls and standardised outputs. \\ 
      %Describe the method solution here.
    {\em Additional comments including restrictions and unusual features:}
    Public GPLv3 code with CLI and Python API; optional dependencies activate PTA likelihoods and nested-sampling scans; a detailed error-code system reports non-physical or numerically unresolved transitions (e.g.\ non-percolation). \\
    \end{small}

\thispagestyle{empty}
\vfill
\newpage

\tableofcontents

\section{Introduction}

\begin{figure}[t]
  \centering
  \includegraphics[width=\linewidth]{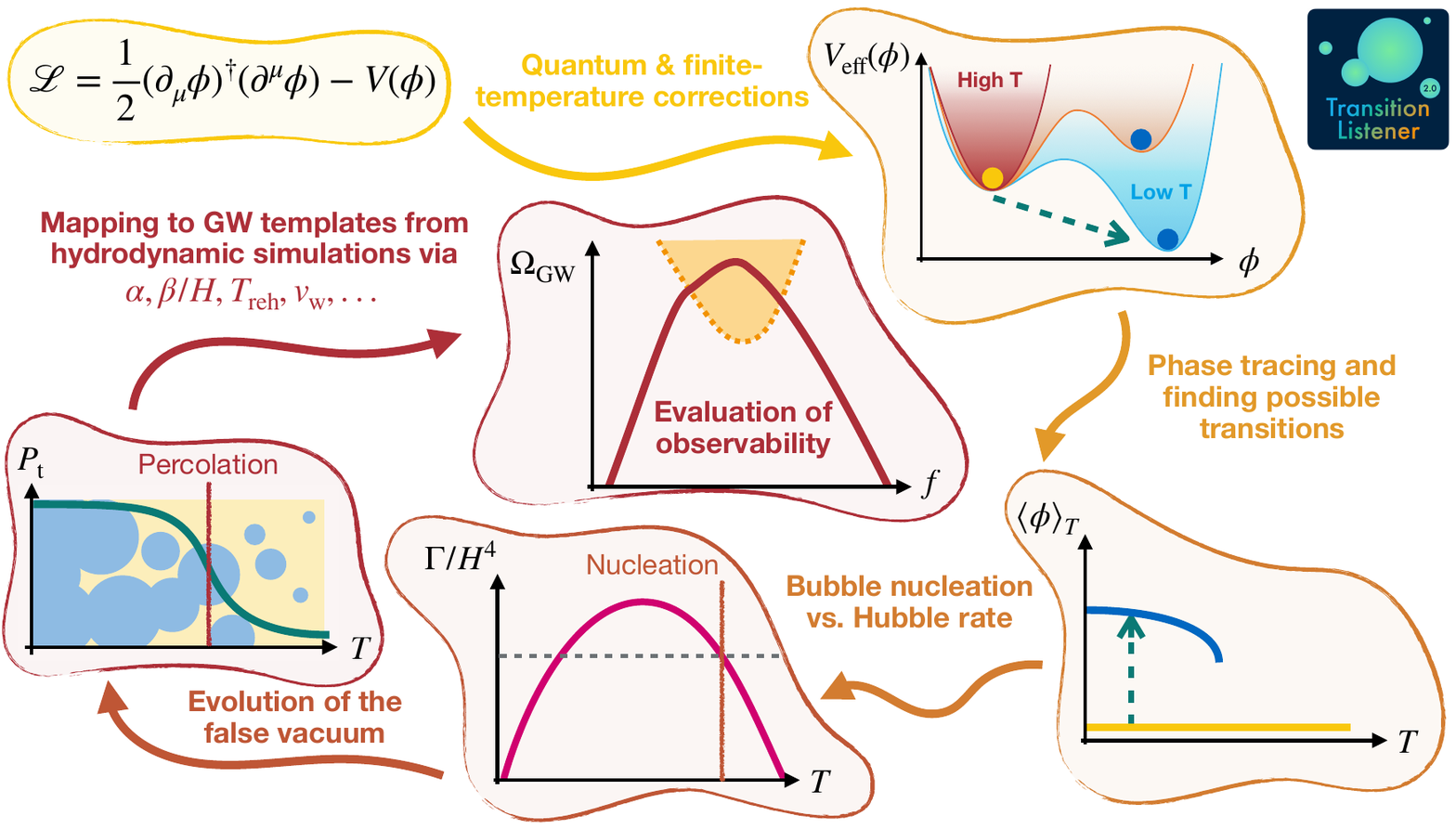}
  \caption{Workflow of the \texttt{TransitionListener} pipeline. Given a user-defined tree-level potential
  and mass spectrum of a theory, the 1-loop effective potential is constructed, the phase
  structure is traced and possible first-order transitions are identified. For each possible transition, the
  nucleation history and the evolution of the true-vacuum fraction is computed. Using this, the percolation
  and reheating temperatures are found and the thermodynamic quantities needed for the GW templates
  are calculated. Lastly the GW spectrum is evaluated and compared to existing and future GW
  observatories' sensitivities.}
  \label{fig:fromLtoObs} 
\end{figure}

Much of the early universe's thermal history remains hidden from direct observation. The
primordial plasma was opaque to electromagnetic radiation until recombination, roughly
$380,000$ years after the end of inflation, when the universe finally became transparent
and photons could free-stream to form the cosmic microwave background we observe
today~\cite{Planck:2018vyg}. Gravitational waves (GWs), by contrast, propagate essentially
unimpeded once produced, offering a unique probe of the universe's earliest moments and
the fundamental physics governing them~\cite{Caprini:2018mtu}. With
LISA~\cite{LISA:2017pwj} and the Einstein Telescope~\cite{Maggiore:2019uih} planned for
the coming decades, and pulsar timing arrays
(PTAs)~\cite{Antoniadis:2022pcn,NANOGrav:2023gor,EPTA:2023fyk,Xu:2023wog,Reardon:2023gzh,Miles:2024seg}
already reporting evidence for a stochastic GW background, the challenge of connecting BSM
model building to quantitative GW predictions has become both timely and urgent.

As the universe cooled after reheating, decreasing temperatures triggered symmetry
breaking and qualitative changes in the vacuum structure~\cite{Kirzhnits:1972iw,
  Dolan:1973qd, Weinberg:1974hy, Kirzhnits:1976ts,}. The Standard Model (SM) predicts that
both the electroweak and QCD phase transitions proceeded as smooth crossovers, producing
no observable GW signals~\cite{Kajantie:1996mn, Aoki:2009sc}. Yet the SM leaves
fundamental questions unanswered: the origin of the baryon
asymmetry~\cite{Sakharov:1967dj}, the identity of dark matter~\cite{Zwicky:1933gu,
  Rubin:1970zza}, the hierarchy problem~\cite{Weinberg:1975gm, Susskind:1978ms}, and the
mechanism behind neutrino masses~\cite{Minkowski:1977sc, Yanagida:1979as} all point toward
physics beyond the Standard Model. Many proposed solutions -- whether through additional
scalar sectors~\cite{Lee:1973iz}, new gauge groups~\cite{Pati:1974yy},
supersymmetry~\cite{Nilles:1983ge}, grand unification~\cite{Georgi:1974sy}, or dark sector
dynamics~\cite{Holdom:1985ag,Schwaller:2015tja} -- predict a richer phase structure in
which some transitions are or become first-order. Such transitions are compelling
for multiple reasons: they can provide the out-of-equilibrium conditions necessary for
baryogenesis~\cite{Kuzmin:1985mm, Baldes:2021vyz, Cataldi:2024pgt, Bhusal:2025lvm},
and produce the observed dark matter abundance~\cite{Baker:2019ndr,Azatov:2021ifm,
Bringmann:2023iuz,Balan:2025uke},
while simultaneously sourcing a stochastic GW background
through the violent dynamics of bubble nucleation, expansion, and
collision~\cite{Witten:1984rs, Hogan:1986dsh}, thus possibly explaining
the reported PTA signal~\cite{NANOGrav:2023hvm}.
Further, GW backgrounds at high frequencies allow tests of physics far above the energy
scales accessible at colliders~\cite{Caprini:2018mtu, Aggarwal:2025noe}, thereby providing a
potentially unique window into early-universe physics.

Translating these theoretical possibilities into testable predictions requires solving a
multi-scale problem that spans many orders of magnitude in both energy and
length~\cite{Athron:2023xlk}. Starting from the microscopic physics encoded in an
effective potential, one must trace the temperature-dependent vacuum structure, determine
when and how bubbles of the new phase nucleate~\cite{Coleman:1977py,Linde:1981zj}, and track
their expansion and the true-vacuum fraction throughout percolation~\cite{Guth:1981uk,
  Turner:1992tz}. One has to compute the hydrodynamic response of the surrounding
plasma~\cite{Kamionkowski:1993fg,Espinosa:2010hh,Ekstedt:2024fyq}, relate the bubble size
distributions with GW templates obtained from numerical
simulations~\cite{Caprini:2009yp,Caprini:2009fx,Hindmarsh:2013xza,Hindmarsh:2015qta,
  Jinno:2016vai,Hindmarsh:2017gnf,Cutting:2019zws,RoperPol:2019wvy,Jinno:2022mie,Caprini:2024gyk,
  Caprini:2024hue}, and finally propagate the resulting GW spectrum forward through cosmic
history to predict what observatories might detect today. The computational workflow,
sketched in Fig.~\ref{fig:fromLtoObs}, must maintain physical consistency at each step
while remaining efficient enough to explore the high-dimensional parameter spaces of
realistic BSM theories in order to connect with observations. This challenge is particularly
acute for strong first-order transitions with significant supercooling, precisely the regime where GW signals are
strongest and most likely to be
observable~\cite{Ellis:2018mja,Ellis:2019oqb,Bringmann:2026xcx}.

Existing numerical tools address different parts of this pipeline. \texttt{BSMPT}~\cite{Basler:2018cwe,
Basler:2020nrq, Basler:2024aaf}
specialises in extended Higgs sectors and the electroweak phase transition,
\texttt{PhaseTracer}~\cite{Athron:2020sbe,Athron:2024xrh} provides robust multi-field phase tracing,
\texttt{ELENA}~\cite{Costa:2025pew} offers efficient
single-field evolution with PTA likelihood interfaces, and \texttt{PT2GWFinder}~\cite{Brdar:2025hxw} delivers
a Mathematica-based workflow. The original
\texttt{TransitionListener} code integrated these steps for specific dark sector models, 
established the complementarity between dark matter searches and GW observations, and motivated
the search for specific dark matter candidates~\cite{Ertas:2021xeh,Bringmann:2023opz,Bringmann:2026xcx}.
Numerous specialised codes handle individual aspects such as the construction of effective
potentials~\cite{Ekstedt:2022bff}, bounce solutions~\cite{Guada:2020xnz,Sato:2019wpo}, bubble
nucleation rates~\cite{Ekstedt:2023sqc}, wall velocities~\cite{Ekstedt:2024fyq}, or spectral
templates~\cite{Caprini:2019egz}.
Despite the progress on all fronts of this challenge, significant gaps remain, particularly in
the consistent treatment of thermodynamics during strong
supercooling, the handling of multi-field potentials, and the
seamless integration of physical calculations with observational likelihood
machinery~\cite{GAMBITModelsWorkgroup:2017ilg,GAMBITCosmologyWorkgroup:2020htv}.

\texttt{TransitionListener v2.0} addresses these limitations through a comprehensive
\texttt{python} framework that advances the state of the art in several key areas. The
code implements a self-consistent treatment of the true-vacuum fraction evolution,
including backreaction on the Hubble rate and accounting of reheating in the broken phase
during percolation by considering local energy conservation at the bubble wall. It handles multi-field
effective potentials with automated
one-loop corrections, computes bubble wall velocities in the local thermal equilibrium
approximation~\cite{Ai:2024uyw,Ai:2023see}, and evaluates GW spectra using the latest
simulation-calibrated templates. The framework incorporates built-in sensitivity curves
for current and planned detectors across twelve orders of magnitude in frequency, provides
direct interfaces to \texttt{PTArcade}~\cite{Mitridate:2023oar} for pulsar timing
likelihoods, and enables Bayesian parameter inference through integration with
\texttt{UltraNest}~\cite{Buchner:2021cql}. A comparison
with \texttt{BSMPTv3} through a benchmark scan in sec.~\ref{sec:6} shows excellent agreement
for intermediately strong phase transitions, whereas for strong transitions
$\mathcal{O}(1)$ deviations in the expected LISA signal-to-noise ratios are obtained.
We show that these differences can be understood from a more
precise treatment of thermodynamics and cosmological consistency during the transition,
as well as different conventions for the computation of GW spectra and signal-to-noise
ratios.

This paper is organised as follows. Section~\ref{sec:2} describes the underlying physics
of FOPTs that went into the development of the algorithms described in section
\ref{sec:3}. Section~\ref{sec:scan-and-plotting} then describes the various scan modes for
parameter space exploration and documents the many different visualisation capabilities. In
section~\ref{sec:6} we compare our code with existing implementations and validate our
results through detailed comparison with \texttt{BSMPTv3}. We conclude in
section~\ref{sec:conclusion}. A set of appendices provides technical details and discusses
the models implemented in \texttt{TransitionListener v2.0}, including three dark sector
models with distinct phenomenological properties, as well as the (real) Two-Higgs Doublet Model
(2HDM).

\section{Phase transitions and gravitational waves}
\label{sec:2}

In this section we review the physics behind strong first-order phase 
transitions, which went into the algorithms implemented in \texttt{TransitionListener v2.0}.
Most of the physics presented here has already been reviewed in detail, see for instance
ref.~\cite{Athron:2023xlk} for a neat compendium of references and more detailed derivations
of many of the computations presented here. As opposed to previous works and going
beyond these in certain aspects, we put our focus on the self-consistent treatment of the
time-dependent thermodynamics of the two phases during the transition, the time-temperature
relation, and the relevant length scales at which the dynamics of bubbles takes place. In
particular, we will show that previously made assumptions on the size of bubbles and the
temperature of the broken-symmetric phase are violated strongly in the most interesting
cases of strong and slow FOPTs, having a large influence on the forecasted
GW backgrounds.

This section is organised as follows: In sec.~\ref{subsec:Effective potential} we
describe the effective potential at finite temperature and its role in determining
the phase structure of a given theory; examples are given in Appendix~\ref{sec:models},
where we expand on dark sector models and the 2HDM, which are released as model files
together with the code. Sec.~\ref{subsec:thermodynamic-quantities} then defines
thermodynamic quantities and their evaluation during the phase transition.
The following sec.~\ref{subsec:truevacuumfraction} delineates the computation
of the true-vacuum fraction during the phase transition and defines different
milestone temperatures. Sections~\ref{subsec:strengthspeed} and \ref{subsec:hydro}
discuss the specific set of strength and speed parameters we recommend for
performing phase transition computations as well as the energy budgets available
for GW emission. The GW spectral templates used in this work as well as the
general description of how we evaluate a given GW signal's observability
are described in sections \ref{subsec:grav-wave-spectr} and \ref{subsec:observability}.

\subsection{The perturbative description of vacuum decay}
\label{subsec:Effective potential}

The dynamics of a phase transition is encoded in the free energy of a thermodynamical
system in dependence on the temperature and the order parameter. In the case of weakly
coupled quantum field theories with
scalar fields, like the electroweak theory, the transition dynamics are governed by the
evolution of the effective potential $V_\text{eff}$. This
potential is a function of the scalar fields $\bm{\phi} = (\phi_1, \phi_2, \dots)$ and
temperature, and it determines the stability of the vacuum states. Given a general
Lagrangian of scalar fields
\begin{align}\label{eq:general-lagrangian}
  \mathcal{L} \supset \frac{1}{2}\ba{ \partial_{\mu} \bm{\phi}} \ba{\partial^{\mu} \bm{\phi}} - V_\text{tree}(\bm{\phi})\,,
\end{align}
with a tree-level potential $V_\text{tree}(\bm{\phi})$, the classical vacuum expectation
value $\langle \bm{\phi} \rangle_\text{tree}$ is given by the global minimum of the potential. Quantum
and thermal corrections can shift or completely alter the vacuum structure of the theory.
The aim of the construction of an effective potential is to include these corrections in a
systematic way, allowing us to trace the different potential minima throughout the thermal
evolution of the cosmos. In the following we will make use of the background field method,
meaning that the field $\bm{\phi}$ is split into a classical background field and quantum
fluctuations\footnote{In the case of strongly coupled quantum field theories, the
  perturbative approach of constructing an effective potential breaks down and different
  techniques have to be employed to obtain a thermodynamic description of the
  plasma~\cite{Polyakov:1975rs,Fukushima:2003fw,
    Fukushima:2017csk,Sagunski:2023ynd,Aarts:2023vsf}. Here, we focus only on weakly
  coupled cases in which a perturbative expansion is justified.}.
In order to not obscure our notation and as we are only interested in the dynamics of the
classical, spatially homogeneous background field, we do not distinguish between these
components.

At one-loop order in the $\overline{\text{MS}}$ scheme and at zero-temperature the
corrections to the tree-level potential are given by the Coleman-Weinberg potential~\cite{Coleman:1973jx}
\begin{align}\label{eq:coleman-weinberg}
  V_{\text{CW}}(\bm{\phi})=\sum_i \pm \frac{n_i}{64\pi^2} m_i^4(\bm{\phi})
\left[\ln\left(\frac{m_i^2(\bm{\phi})}{\bar{\mu}^2}\right)-C_i\right],
\end{align}
which is understood to be added to $V_\text{tree}(\bm{\phi})$. Here, the sum runs
over all fields coupled to the scalar $\bm{\phi}$; the $+$ ($-$) sign applies to
bosons (fermions), $n_{i}$ are the degrees of freedom, $\bar{\mu}$ is the
renormalisation scale and the renormalisation constants read $C_i=3/2$ for
scalars and fermions and $5/6$ for gauge bosons. At finite temperature,
the one-loop corrections receive an additional term, given here in terms of the thermal
functions $J_{\mathrm{b/f}}$ for bosons ($\text{b}$) and fermions ($\text{f}$),
\begin{align}\label{eq:thermal-potential}
  V_\text{T}(\bm{\phi},T)=\sum_i \pm  \frac{n_i T^4}{2\pi^2}
  J_{\text{b}/\text{f}}\,\left(\frac{m_i^2(\bm{\phi})}{T^2}\right)\,, \qquad 
  J_{\mathrm{b/f}}(z^2) = \int_{0}^{\infty} \dd x \,x^2 \ln
  \left( 1 \mp \mathrm{e}^{-\sqrt{x^2 + z^2}}  \right)\,.
\end{align}
The introduction of counterterms to cancel the UV divergence of the Coleman-Weinberg
potential adds another contribution $V_{\mathrm{ct}}(\bm{\phi})$ to the potential.
Additionally, corrections arising from the resummation of the longitudinal, bosonic
Matsubara zero-modes cure the IR divergences in $V_\text{T}$ in the Arnold-Espinosa scheme
and give rise to the daisy correction term\footnote{Special care needs to be taken when
  dealing with states coupled to $\bm{\phi}$ which mix, like the electroweak gauge bosons in
  the SM. In that case, the correction has to be understood as a sum over the eigenvalues
  of the temperature-corrected mass matrix of the longitudinal, bosonic
  modes.}~\cite{Arnold:1992rz}
\begin{align}\label{eq:daisy-potential}
  V_{\text{daisy}}(\bm{\phi},T) = -\frac{T}{12\pi}\sum_{i\in B_\text{L}} n_i
  \left[\left(m_i^2(\bm{\phi})+\Pi_i(T)\right)^{3/2}-m_i^3(\bm{\phi})\right]\,.
\end{align}
Alternatively the Parwani resummation scheme~\cite{Parwani:1991gq} replaces
$m_i^2\to m_i^2+\Pi_i$ in $V_\text{T}$. In summary, the one-loop,
daisy-corrected finite-temperature effective potential reads
\begin{equation}
  V(\bm{\phi},T) = V_\text{tree}(\bm{\phi}) + V_{\text{CW}}(\bm{\phi}) +
  V_{\text{ct}}(\bm{\phi}) + V_\text{T}(\bm{\phi},T) + V_{\text{daisy}}(\bm{\phi},T)
+ V_{\text{rad}}(T) - V_{T=0}(\bm{v}) \, .
\label{eq:Vtot}
\end{equation}
The field-independent terms $V_{\mathrm{rad}}(T) - V_{T=0}(\bm{v})$ have been
added for the sake of consistency of the following sections; they do not play a
role in the dynamics of the field theory, but rather become relevant when computing
thermodynamic quantities and relating them to the Hubble rate through the Friedmann
equations: The first term accounts for additional degrees of freedom which are not
coupled to $\bm{\phi}$, but still contribute to the energy density and pressure
of the plasma. Further, we make sure that a constant term $- V_{T=0}(\bm{v})$ with
$\bm{v} \equiv \langle \bm{\phi} \rangle_{T=0}$ is included, such that at $T=0$,
the potential, and hence also the pressure, vanishes\footnote{It is possible to
include a parametrically small term here to further account for the relevance of
a cosmological constant in our universe, which however does not play any role in
the dynamics of finite-temperature phase transitions in the early universe.}.

In case of two competing vacua, i.e.~solutions of $\nabla_{\bm{\phi}} V(\bm{\phi},T)=0$
with positive potential curvature, we speak about the coexistence of
two phases.
The moment that a previously global vacuum turns unstable by the other vacuum state
moving to smaller potential values (cf.~fig.~\ref{fig:fromLtoObs}), the false vacuum
$\bm{\phi}_{\text{f}}$ can decay to the true one if thermal fluctuations are sufficiently
large and supercritical bubbles form. In the absence of seeds which could trigger the decay~\cite{Blasi:2023rqi},
bubbles are spherical and expand satisfying Klein-Gordon equations of motion,
\begin{equation}\label{eq:bounce-equation}
  \frac{\mathrm{d}^2 \phi_i}{\mathrm{d}r^2}+\frac{2}{r}\frac{\mathrm{d}\phi_i}{\mathrm{d}r}
  =\frac{\partial V(\bm{\phi},T)}{\partial \phi_i}
\end{equation}
with boundary conditions $\mathrm{d}\phi_i/\mathrm{d}r|_{r=0}=0$ and
$\phi_i(r\to\infty)=\phi_{i,\text{f}}$. The Euclidean action of these
$O(3)$-symmetric bounce solutions can be computed from the instanton solution
of eq.~\eqref{eq:bounce-equation} (the bubble profile) $\bm{\phi}(r)$ following
\begin{align}\label{eq:bounce-action-O3}
  S_3(T)=4\pi\int_0^\infty \mathrm{d}r\,r^2\left[
  \frac{1}{2}\sum_i\left(\frac{\mathrm{d}\phi_i}{\mathrm{d}r}\right)^2
  +V(\bm{\phi},T)-V(\bm{\phi}_{\text{f}},T)\right].
\end{align}
The decay rate of the false vacuum to the true vacuum (i.e., the bubble nucleation rate)
can then be computed as~\cite{Coleman:1973jx,Linde:1981zj}
\begin{equation} \label{eq:nucrate}
\Gamma(T)=T^4\left(\frac{S_3(T)}{2\pi T}\right)^{3/2}\exp\!\left[-\frac{S_3(T)}{T}\right]\,.
\end{equation}

We note that while the construction of the effective potential and the computation of the
bubble nucleation rate as demonstrated here are a standard approach found across the
literature, higher-order perturbative improvements and dimensionally reduced effective
field theory approaches can significantly improve the validity and convergence of the
perturbative expansion~\cite{Lewicki:2024xan}. In our code, we treat the renormalisation
scale as an open parameter and employ the Landau gauge. Further work will be required
to also increase the accuracy of phase transition
computations by the automated use of dimensionally reduced effective field theories,
impacting both the effective potential as well as the precise description of how to
compute the nucleation rate, cf.~\cite{Croon:2020cgk,Ekstedt:2022bff,Kierkla:2023von}.

\subsection{The thermodynamics of two phases in an expanding universe}
\label{subsec:thermodynamic-quantities}

Each of the phases defined by the effective potential has distinct thermodynamic properties.
For the study of cosmological phase transitions, the most important ones are the pressure,
entropy, and energy density, which can all be obtained from the effective potential,
within a given phase $\bm{\phi}_\text{min}(T)$,
\begin{align}
  p = -V(\bm{\phi}_{\mathrm{min}}, T) \, , && s = \pd{p}{T}
  = -\pd{V(\bm{\phi}_{\mathrm{min}}, T)}{T} \, , && \rho = Ts - p
  = V(\bm{\phi}_{\mathrm{min}}, T) - T \pd{V(\bm{\phi}_{\mathrm{min}}, T)}{T} \, .
\end{align}
Out of these, the energy density of a given phase will be of the biggest
importance for the computation of GW signals and the self-consistency
when solving the Friedmann equations, so it makes sense to study its
relationship with the effective potential in more detail: To do so,
let us split the effective potential in a temperature-independent part
($V_\text{CW}$ and $V_\text{ct}$), a temperature-dependent part
($V_{\mathrm{T}}$, $V_{\mathrm{daisy}}$ and $V_{\text{rad}}$),
as well as a normalisation constant ($-V_{T=0}(\bm{v})$),
\begin{align}
  V(\bm{\phi},T) = V_{T=0}(\bm{\phi}) + V_{T \neq 0}(\bm{\phi}, T) - V_{T=0}(\bm{v}) \, .
\end{align}
Using this, the energy density within a portion of space in a
given phase splits into the components
\begin{align}
  \label{eq:rho}
  \rho(\bm{\phi}, T)
  &= \underbrace{V_{T=0}(\bm{\phi}) - V_{T=0}(\bm{v})}_{\rho_{\mathrm{vac}}} +
    \underbrace{V_{\mathrm{T}} - T \partial_{T} V_{\mathrm{T}}}_{\rho_{\mathrm{rad}}} +
    \underbrace{V_{\mathrm{rad}} - T \partial_{T} V_{\mathrm{rad}}}_{\rho_{\mathrm{rad, ext}}} +
    \underbrace{V_{\mathrm{daisy}} - T \partial_{T} V_{\mathrm{daisy}}}_{\rho_\text{int}} \\
    &\simeq \rho_{\mathrm{vac}} + \frac{\pi^{2}}{30} g_{\mathrm{eff}}(T) T^{4}
    + \frac{\pi^{2}}{30} g_{\mathrm{eff}}^\text{ext}(T) T^{4}
      % + \rho_\text{int}
      %+ \frac{T^{2}}{8\pi} \sum_{i \in B_\text{L}} \left( m^{2}_{i}(\phi) + \Pi(T) \right)^{1/2} \frac{\partial \Pi(T)}{\partial T}
\end{align}
In the second step, we rewrote the thermal one-loop contribution in terms of the effective
degrees of freedom $g_{\mathrm{eff}}(T)$ and neglected the contribution from the daisy
term. In Apprendix~\ref{app:effective_dof} we provide more details on the relationship
between the thermal functions $J_\text{b/f}$ and the effective degrees of freedom
$g_\text{eff}$ and $g_\text{eff}^\text{ext}$ (relevant for the description of
non-interacting species in a plasma), and discuss the relevance of the daisy contributions
for modelling the energy density in a plasma of interacting fields.

During a first-order phase transition, two distinct phases coexist and fill
up regions of space with different thermodynamic properties: Assuming the instant thermalisation
of the fluid within a given bubble (see ref.~\cite{Bringmann:2023iuz} for a
discussion of the validity of this assumption), the temperature inside the
bubble will be higher due to the release of vacuum energy $\rho_\text{vac}$
(hence the name \textit{reheating}); the pressure will be higher, as bubbles
otherwise would not be able to expand; and the energy density will be only
slightly lower than outside the bubbles. The latter is due to the energy
density stored in the bubble walls and emitted in the form of gravitational
waves being negligible compared to the energy densities themselves~\cite{Espinosa:2010hh}.
The volume-averaged total energy density and the volume-averaged pressure are
then given by the weighted sum of the energy densities and pressures of the
two phases, each multiplied with the respective true and false-vacuum
fractions, $P_\text{t}$ and $P_\text{f}=1-P_\text{t}$,
\begin{align} \label{eq:rhobar}
  \bar{\rho} = P_\text{t} \rho_{\text{t}} + (1 - P_\text{t} ) \rho_{\text{f}}
  \qquad \text{and} \qquad \bar{p} = P_\text{t}  p_{\text{t}} + (1 - P_\text{t} ) p_{\text{f}} \,.
\end{align}
Assuming statistical homogeneity, i.e.~taking local overdensities to be sufficiently
small, the Friedmann equations can be written as
\begin{align}
  \label{eq:Friedmann}
  H^2 = \frac{ \bar{\rho}}{3 m_\text{Pl}^2}  \qquad \text{and}
  \qquad \dot{\bar{\rho}} = -3 H (\bar{\rho} + \bar{p})\,.
\end{align}
Assuming further that the expansion of bubbles succeeds sufficiently quickly, 
which is the case for strong transitions in the detonation regime,
there is no net heat transfer from the true vacuum to the false vacuum in front of the
bubble walls, and hence
\begin{align} \label{eq:adiabaticFalsePhase}
  \dot{\rho}_\text{f} = -3 H (\rho_\text{f} + p_\text{f})
\end{align}
holds\footnote{Note that the same relation does \textit{not} hold for the true vacuum. A
  FOPT generates entropy and heat, such that
  $\dot{\rho}_\text{t} + 3 H (\rho_\text{t} + p_\text{t}) > 0$. A derivation of the heating rate
  is provided in eq.~\eqref{eq:truevacenergy}.} also for the energy density of the false
vacuum. Let us point out that $\bar{\rho}$ will in general not follow the time-evolution
known from radiation domination: For strongly supercooled phase transitions, $P_\text{t}$
can remain small for a significant period of time, while the equation-of-state parameter
goes from $1/3$ (in radiation) to $-1$ (for vacuum domination). Note further that even in
the instantaneous reheating assumption, it is no longer possible to unambiguously define a
temperature playing the role of a time variable, as the temperature in the true vacuum in
general does not coincide with the temperature in the false vacuum. It is hence necessary
to specify the temperature of each phase separately at a given time. For a given
transition, we choose the false vacuum temperature $T_{\mathrm{f}}(t)$ as a viable time
variable and express the true-vacuum temperature $T_{\mathrm{t}}$ as a function of it.

By using the chain rule on eq.~\eqref{eq:adiabaticFalsePhase}, the time-temperature
relation in false-vacuum phase during the phase transition can then
be derived to read
\begin{align}\label{eq:time-temperature}
  \td{T_\text{f}}{t} &= - 3 H \frac{{\rho}_\text{f} + {p}_\text{f}}{\partial {\rho}_\text{f}/ \partial T_\text{f}}
  = - 3 H T_\text{f} \,  c_{\text{s,f}}^2 \, , \quad \text{where} \quad 
  c_{\text{s,f}}^2 = \frac{\partial_T V_\text{f}}{T_\text{f}\partial_T^2 V_\text{f}} \, .
\end{align}
It is easy to see that in the limit of a bag equation of state, the time-temperature
relation simplifies to the usual $\dot{T}_\text{f} = - H T_\text{f}$ relation known from
radiation domination, corresponding to $c_{\text{s,f}}^2 = 1/3$, where $H$, however,
still includes an implicit dependence on the true-vacuum fraction through $\bar{\rho}$.
Without simplifying assumptions, the above time-temperature relation does not have a
simple analytic form and needs to be solved self-consistently. In practice, we often
find that $\mathcal{O}(10\%)$ deviations from $c_\text{s}^2 = 1/3$ occur in either
phase and in either direction, see also ref.~\cite{Giese:2020znk}, thus affecting
all other quantities relevant for the computation of the GW spectrum. This further
motivates detailed numerical computations in the case of strongly supercooled phase
transitions.

To determine the temperature $T_\text{t}$ within the true vacuum as a function of
time (or equivalently as a function of the false-vacuum temperature $T_\text{f}$),
we can expand
$\dot{\bar{\rho}} = \td{}{t} (P_\text{t}  \rho_\text{t} + (1-P_\text{t}) \rho_\text{f})$
using the chain rule and plug in the expression for $\dot{\rho}_\text{f}$ from
eq.~\eqref{eq:adiabaticFalsePhase}
in order to obtain
\begin{align} \label{eq:truevacenergy}
  \dot{\rho}_\text{t} = -3 H (\rho_\text{t} + p_\text{t}) +
  \frac{\dot{P}_\text{t}}{P_\text{t}} (\rho_\text{f} - \rho_\text{t}) \, .
\end{align}
The first term just describes the usual energy density redshift of a
perfect fluid; the second term instead describes the contribution from the
released energy due to the changing true-vacuum fraction in a given volume.
Using energy conservation and starting at some initial time when the bubble nucleation rate
$\Gamma \ll H^4$ is still negligible, i.e.
$\rho_\text{t}(t_{\text{init}}) = \rho_\text{f}(t_{\text{init}})$, we can compute
the energy density in the true vacuum at an arbitrary time by integrating
eq.~\eqref{eq:truevacenergy}. Using eq.~\eqref{eq:rho}, we can then also infer
the temperature within the
true vacuum at a given time, by solving $\rho_\text{t}(T_\text{t})$ for $T_\text{t}$.

\begin{figure}[t]
  \centering
  \includegraphics[width=0.85\textwidth]{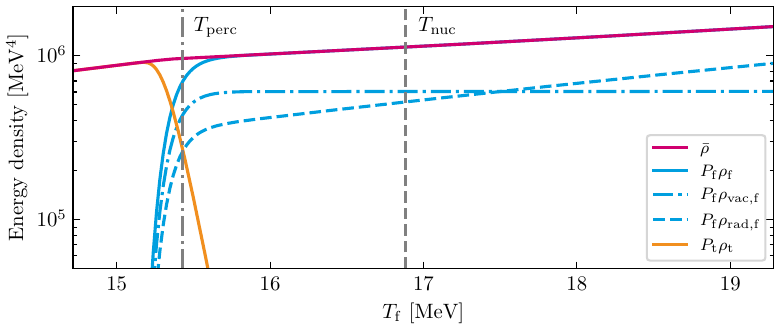}
  \caption{Evolution of the energy densities in the false and true vacuum
  around the phase transition in the conformal $U(1)^\prime$ model
  ($g = 0.925$, $v = 140$\,MeV, $y = 0.01$). The percolation (nucleation)
  temperature is indicated by a dash-dotted (dashed) vertical line.}
  \label{fig:energy-density-dof}
\end{figure}

In fig.~\ref{fig:energy-density-dof} we show the evolution of the energy densities
$\bar{\rho}$, $\rho_\text{t}$ and $\rho_\text{f}$ for a moderately supercooled phase transition in
the conformal $U(1)^\prime$ model, defined in appendix~\ref{sec:models}. During the transition,
the energy fraction in the true vacuum
$P_{t}\rho_\text{t}$ continuously increases, until virtually all of the vacuum energy has
been released and the redshift of $\rho_\text{t}$ starts to dominate the evolution of the
true vacuum. Before and after the transition, the universe is radiation dominated; during
the transition, vacuum energy dominates, indicating that conventional approximations
assuming a specific and constant time-temperature-relation become invalid.

\subsection{The fate of the false vacuum}
\label{subsec:truevacuumfraction}

Given two distinct phases coexisting in the universe, the evolution of the
true-vacuum fraction can be tracked as a function of time. Assuming
spherical bubbles, nucleated homogeneously throughout the universe,
and expanding with a steady-state velocity $v_\text{w}$, the true-vacuum
fraction $P_\text{t}$ can be shown to follow~\cite{Athron:2023xlk}
\begin{equation} \label{eq:false-vacuum-time}
  1 - P_\text{t}(t) = \mathrm{e}^{-I(t)} =  \exp \left( -\frac{4\pi v_{\text{w}}^3}{3}
    \int_{t_{\mathrm{init}}}^{t} \dd t^{\prime} \, \Gamma(t^{\prime}) \frac{a^3(t^{\prime})}{a^3(t)}
    \left[\int_{t^{\prime}}^t \dd t^{\prime \prime} \, \frac{a(t)}{a(t^{\prime \prime})}
    \right]^3 \right)\,.
\end{equation}
Here, $t_{\mathrm{init}}$ indicates the earliest time at which
tunnelling is possible, which usually coincides with the
critical temperature, $T_\text{c}$, at which the potential
minima degenerate and at which bubble nucleation becomes possible.
Long before the transition, at $t \to t_{\mathrm{init}}$, the
argument $I(t) \to 1$, such that the true-vacuum fraction
vanishes. Long after a successful transition, at $t \to \infty$,
the argument $I(t) \to \infty$, such that the true-vacuum
fraction approaches its maximum value\footnote{In fine-tuned
cases of ultraslow transitions, the bubble
nucleation rate can decrease during the phase transition, leading
to the saturation of the true-vacuum fraction at values below 1.
In these cases, the bubble nucleation rate cannot compete with the
Hubble expansion and it is not guaranteed that we would
find ourselves in the true vacuum state today. Special care needs
to be taken in these cases, sometimes referred to as vaccum
trapping or used phenomenologically within the eternal inflation
framework, see e.g.~ref.~\cite{Blau:1986cw,Franciolini:2026ujb}.
We are not interested in such pathological scenarios, as the GW
signal from bubble collisions and bulk fluid motion will be absent
or strongly suppressed in these cases.}, which is typically 1.

For our purposes, it is more convenient to compute the percolation
integral $I(t)$ in terms of the false-vacuum temperature $T_\text{f}$
instead of time: Using the time-temperature relation found above,
we can substitute
$\diff t = - 3 H(T_\text{f}) T_\text{f} \, c_{\text{s,f}}^2 \, \diff T_\text{f}$
and obtain (neglecting the subscript ``false'' in the following equations for brevity)
\begin{align}\label{eq:percol-integral}
  I(T)&= \frac{4\pi}{3}v_{\text{w}}^3 \int_{T}^{T_{\text{init}}} \frac{\diff T' }{H(T') T' }
  \frac{\Gamma(T')}{3 c_{\text{s}}^2(T')} \frac{a^3(T^{\prime})}{a^3(T)}
  \left[\int_{T}^{T'}\frac{\diff T''}{H(T'') T^{\prime \prime}} \frac{1}{3 c_{\text{s}}^2(T^{\prime \prime})}
  \frac{a(T)}{a(T^{\prime \prime})} \right]^3 \, .
\end{align}
Percolation is defined as the point in time
when a connected cluster of bubbles spans through a Hubble patch. The
threshold relevant for cosmological phase
transitions is found to be
$P_\text{t} (T_{\perc}) \equiv f_\perc \simeq 0.28957$~\cite{Ziff:2001aa,
Athron:2023xlk}. Since $P_\text{t} (T) \to 1$ only asymptotically, the
definition of completion is somewhat arbitrary and chosen to be
$P_\text{t} (T_{\mathrm{f}}) \equiv f_\text{final} = 1 - \varepsilon$
with $\varepsilon \ll 1$.
In practice we set $f_\text{final} = 0.99$.
We further define the reheating temperature as the true vacuum temperature
at the moment of percolation
$T_{\mathrm{reh}} \equiv T_{\mathrm{t}}(t_\perc)$.

Another often described milestone temperature is the nucleation
temperature, $T_{\mathrm{nuc}}$, which is defined as the temperature
at which the average number of bubbles per Hubble patch reaches one,
\begin{align} \label{eq:bubblenumberNuc}
  N(T_{\mathrm{nuc}}) = \frac{4 \pi}{3} \int_{t_\text{init}}^{t_{\mathrm{nuc}}}
  \diff t \, \frac{\Gamma(t) (1 - P_\text{t}(t))}{H^3(t)} = 1 \, .
\end{align}
Typically, nucleation happens long before percolation, such that the above
expression can be simplified using $P_\text{t}(t < t_\text{nuc}) \approx 0$.
For the case of an exponentially growing bubble nucleation rate,
the integrand is dominated by the temperatures near the nucleation
temperature and $N(T_{\mathrm{nuc}}) \approx \Gamma(T_{\mathrm{nuc}})
/ H(T_{\mathrm{nuc}})^4 \overset{!}{=} 1$ becomes a good approximation.
Further approximations of this nucleation criterion exist for the case of
radiation domination, when $H \propto T^2 / M_\text{pl}$, such that
$S_3/T \simeq \mathcal{O}(10^2)$ at nucleation. Note that these
approximations break down for the case of the strongest supercooled
phase transitions, expected to emit the largest possible GW signals.
Note further that the nucleation temperature is \textit{not} directly
related to the GW signal, as no GWs are emitted during the nucleation
process itself. In ref.~\cite{Athron:2022mmm} it has further been argued
that there can even be pathological cases in which
$T_{\mathrm{nuc}} < T_{\perc}$ or in which $N<1$ for all times, even though
the transition succeeds.

\subsection{On the strength and speed of the transition}
\label{subsec:strengthspeed}
The most important characteristics of a phase transition, in order to predict
its gravitational wave signature, are its strength and the size of bubbles
during the transition. The first is typically quantified by the energy
release during the transition, while the second can be characterised in
two ways: either through the speed with which the nucleation rate is
growing, or through the mean bubble separation at percolation.

For weak FOPTs, the bubble nucleation rate is typically a quickly increasing
function of time, such that around percolation one can define the speed of
the transition as
\begin{align}\label{eq:beta}
  \left.\frac{\dot{\Gamma}}{\Gamma}\right|_\perc = \td{}{t} \!
  \left.\bb{\ln \Gamma_\perc  + \left. \td{\ln \Gamma}{t}  \right|_\perc 
  \left( t - t_{\perc} \right) + \mathcal{O}(t^2) }\right|_\perc\simeq - \td{}{t}
  \! \left.\ba{\frac{S_3}{T}} \right|_\perc \, .
\end{align}
To arrive at this expression, we have Taylor-expanded $\ln \Gamma$ around percolation,
used that the bubble nucleation rate is exponentially growing in time
(such that $\mathcal{O}(t^2)$ terms could be ignored),  and that the exponent
of $\Gamma$ reads $-S_3/T$, see eq.~\eqref{eq:nucrate}.
For a bag equation of state, i.e.~$c_{\text{s,f}}^2 = 1/3$ in eq.~\eqref{eq:time-temperature},
the time derivative can be replaced by a temperature
derivative and $(\dot{\Gamma}/\Gamma)|_\text{perc} \simeq TH  \td{}{T} (S_3/T)|_\perc$,
motivating the definition~\cite{Caprini:2015zlo}
\begin{equation}\label{eq:transition-speed}
  \ba{\frac{\beta}{H}}_{S_3} \equiv T\,\frac{\mathrm{d}}{\mathrm{d}T}\!
  \left.\left(\frac{S_3}{T}\right) \right|_\perc \, ,
\end{equation}
which is a good measure of the transition speed if the above assumptions are met.
This is typically the case when $\beta/H \ge \mathcal{O}(10^3)$, i.e.~when the
transition will \textit{not} be observable by any planned GW observatories.
Stronger transitions, however, typically feature a slower growth of the bubble
nucleation rate, which can even decrease after having reached a peak at a
given time. This leads to a lesser amount of bubbles, percolating at a
later stage where they could already grow larger, thus emitting stronger
GW signals. In particular, for cases in which $\Gamma(T)$ peaks just before
percolation, $\ba{\beta/H}_{S_3}$ can even obtain negative values. We therefore refrain
from treating $\ba{\beta/H}_{S_3}$ as a good indicator of the speed of an arbitrarily
strong FOPT.

We instead compute the characteristic length scale directly from the
nucleation history using the average bubble number density in a Hubble patch
\begin{align}\label{eq:bubble-density-temp}
  n(T) &= \int_{t_\text{init}}^{t} \diff t^{\prime} \, \Gamma(t^{\prime})
  (1 - P_\text{t}(t^{\prime})) \left(\frac{a(t^{\prime})}{a(t)}\right)^{3}\,.
\end{align}
From this the mean separation of bubbles can be obtained by evaluating
\begin{equation}\label{eq:mean-bubble-separation}
  R_{\text{sep}}(T)=  \frac{1}{n(T)^{{1/3}}} = \left[\int_{T}^{T_{\max}}\!
    \mathrm{d}T'\, \frac{\Gamma(T')}{H(T') T' }
    \frac{1 - P_\text{t}(T')}{3 c_\text{s}^2(T^{\prime})}
    \left(\frac{a(T^{\prime})}{a(T)} \right)^3\right]^{-1/3}\,,
\end{equation}
where in the last step we again plugged in the time-temperature relationship in the false vacuum,
see eq.~\eqref{eq:time-temperature}. At the percolation temperature $T_{\mathrm{p}}$,
one can show that for exponentially growing bubble nucleation rates and
$c_\text{s,f}^2 = 1/3$~\cite{Megevand:2016lpr}
\begin{align}\label{eq:RH-betaH}
  RH \equiv R_\text{sep}(T_\perc) H(T_\perc) \simeq \left(\frac{8\pi}{f_\perc}\right)^{1/3}
  \frac{\max \left( v_\text{w}, c_\text{s} \right)}{(\beta/H)_{S_3}}
\end{align}
holds, relating the two measures. In appendix~\ref{app:betaH-RH}
we present a derivation of this relation.
The factor $\max(v_\text{w}, c_\text{s})$ instead of $v_\text{w}$ is included to account
for the possibility that the sound speed in the false vacuum is larger than the wall
velocity, meaning that there can be shock waves expanding as subsonic deflagrations
in front of the bubble wall, see ref.~\cite{Caprini:2019egz} for details.
Note however, that for sufficiently slow transitions, the above relation will break
down because of the exponential-in-time approximation no longer being valid.
Note further that the prefactor $1/f_\perc^{1/3}$ is
often missing in the literature, even in standard references and
reviews~\cite{Caprini:2019egz,Athron:2023xlk,Caprini:2024hue}. Based on the above
relation, we define
\begin{align} \label{eq:betaH_RH}
  \ba{\frac{\beta}{H}}_{RH} \equiv \left(\frac{8\pi}{f_\perc}\right)^{1/3}
  \frac{\max \left( v_\text{w}, c_\text{s} \right)}{RH}
\end{align}
in order to compare with the existing literature and conventions in the field.

\begin{figure}[t]
  \centering
  \includegraphics[width=0.98\linewidth]{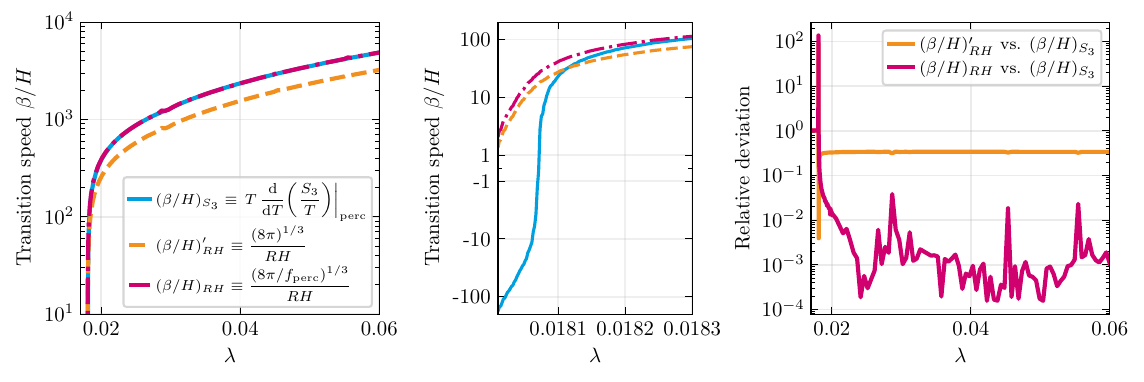}
  \caption{Numerical validation of the $f_\perc^{1/3}$ correction relating
  $RH$ with $\beta/H$, based on a line scan over the quartic coupling $\lambda$
  in the dark Abelian Higgs model ($g = 1$, $v = 100 \, \text{MeV}$),
  defined in appendix~\ref{sec:models}.
  \textit{Left:} Three different definitions of $\beta/H$.
  The factor $f_\perc^{1/3}$ is needed to achieve consistency, see also
  appendix~\ref{app:betaH-RH} for a derivation. 
  \textit{Middle:} For ultraslow transitions, the conventional definition of
  $(\beta/H)_{S_3}$, based on the logarithmic temperature derivative of $S_3/T$, becomes
  physically meaningless, while $(\beta/H)_{RH}$ remains reliable.
  \textit{Right:} Relative
  deviations between the different quantities. The residual sub-percent-level inaccuracies
  between $(\beta/H)_{S_3}$ and $(\beta/H)_{RH}$ at $\lambda>0.02$ are due to numerical
  limitations. In this plot, we explicitly set $v_\text{w} = 1$ to
  isolate the effect of different definitions of the transition speed on the GW signal.}
  \label{fig:bubble-separation}
\end{figure}

In the left panel of fig.~\ref{fig:bubble-separation} we compare 
the transition speed obtained from
eq.~\eqref{eq:transition-speed}
(blue) with the transition speed obtained from eq.~\eqref{eq:betaH_RH}, once with
(magenta) and once without (orange) the factor $f_\perc$. For the comparison
we show a slice of the
Abelian dark Higgs model parameter space ($g = 1$, $v = 100 \, \text{MeV}$),
cf.~appendix~\ref{sec:models} for the model definition.
The smaller the self-coupling $\lambda$,
the more supercooled and slower the FOPT becomes. For very small
$\lambda < 0.0181$, corresponding to ultraslow and
supercooled transitions, $(\beta/H)_{S_3}$ becomes negative because the bubble nucleation
rate decreases at percolation, whereas 
$(\beta/H)_{RH}$ remains positive,
since the mean bubble separation is always positive. For relatively fast transitions with
$\beta/H \gg 100$ (corresponding to $\lambda \gtrsim 0.02$ in this model),
$(\beta/H)_{S_3} = (\beta/H)_{RH}$ and eq.~\eqref{eq:RH-betaH}
correctly translates between the mean bubble separation and the
transition speed parameter $(\beta/H)_{S_3}$. The remaining sub-percent-level
discrepancies between $(\beta/H)_{S_3}$ and $(\beta/H)_{RH}$ for $\lambda > 0.02$,
depicted in the right panel, 
are due to numerical limitations. If one omits the factor
$1/f_\perc^{1/3}$, the inferred transition speed
$(8 \pi)^{1/3} / RH$ is too small by about $30\%$, leading to an over-estimate
of the GW signal by a factor $\mathcal{O}(10)$.
For other cases in which $(\beta/H)_{S_3} \to 0$ is used to infer the strength
of a GW signal, an artificial, arbitrarily large increase of
the GW signal can be achieved. We therefore use and recommend the
use of $RH$ as defined above.

For the transition strength we follow the prescription developed in
refs.~\cite{Giese:2020rtr,Giese:2020znk} using the pseudo-trace of
the energy-momentum tensor of the fluid described by the effective potential
\begin{align}\label{eq:pseudo-trace-diff}
  \bar{\theta}_{i}(T) = \left( \rho_{i}(T) - \frac{p_{i}(T)}{c^2_{\text{s},i}(T)} \right)\,,
\end{align}
where $i = (\text{t, f})$ labels either vacuum and $c_{\text{s},i}(T)$ is the speed
of sound in the respective phase, see eq.~\eqref{eq:time-temperature}.
With this, the strength of the phase transition can be defined as the
difference in the pseudo-trace in both phases at the percolation temperature
\begin{align}\label{eq:pt-strength}
  \alpha =
\frac{\bar{\theta}_{\text{f}}(T_\perc) - \bar{\theta}_{\text{t}}(T_\perc) }{3 (\rho_{\mathrm{f}}(T_\perc) + p_{\mathrm{f}}(T_\perc)) }\,.
\end{align}
Note that previously also other definitions of $\alpha$ have been used in the literature, based
on $\Delta \rho$ or $\Delta p$ instead of $\Delta \bar{\theta}$. 
In refs.~\cite{Giese:2020rtr,Giese:2020znk}
these alternative definitions of $\alpha$ have been
shown to over- or underestimate the GW signal, which is why we refrain from using them.
In the limit of
$c_{\text{s},i}^2 \to 1/3$, the pseudo-trace reduces to the trace of the
energy-momentum tensor, $\theta = g_{\mu\nu} T^{\mu\nu} = \rho - 3 p$,
and the definition of $\alpha$ becomes equivalent to
$\alpha = (\Delta V - \frac{1}{4} T \partial_T \Delta V) / \rho_\text{f}^\text{rad}$.

\begin{figure}[t]
  \centering
  \includegraphics[width=\textwidth]{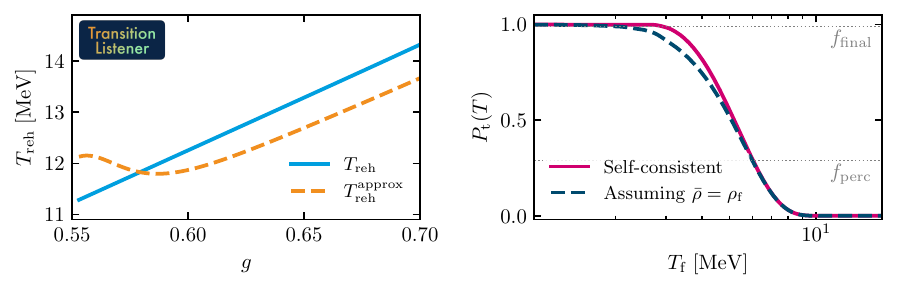}
  \caption{\textit{Left:} Reheating temperature and its approximation 
  $T_\text{perc}^\text{approx} =T_\text{perc} \ba{1 + \alpha}^{1/4}$
  as a function of the gauge coupling $g$
  in the conformal $U(1)^\prime$ model ($y = 0.01$, $v = 140 \, \text{MeV}$).
  \textit{Right:} Comparison of the self-consistent and approximate
  ($\bar{\rho} = \rho_\text{f}$)
  solution of the percolation integral
  $P_\text{t}(T) = 1 - \exp \bc{- I[H[P_\text{t}]](T)}$ for an
  ultraslow transition with $\ba{\beta/H}_{RH} = 3.5$ featured
  in the dark Abelian Higgs model ($\lambda = 0.018$,
  $g = 1$, $v = 100 \, \text{MeV}$).
}
  \label{fig:TrehDoneProperly}
\end{figure}

Using the transition strength $\alpha$, an often found approximation
for the reheating temperature reads
$T_\text{reh}^\text{approx} = T_\text{perc} (1 + \alpha)^{1/4}$,
see for instance refs.~\cite{Basler:2024aaf, Costa:2025pew}, which base
their GW computation on this approximation.
The left panel of fig.~\ref{fig:TrehDoneProperly} compares the reheating temperature
approximation with
our full computation based on eq.~\eqref{eq:truevacenergy} as a function 
of the gauge coupling $g$ in the conformal $U(1)^\prime$ model defined in
appendix~\ref{sec:models}. In general we find an $\mathcal{O}(10\%)$ discrepancy
due to this approximation, which translates directly to the predicted
GW signal through the amount of expected redshift. In the right panel we
further show the impact of solving the percolation integral
$P_\text{t}(T) = 1 - \exp \bc{- I[H[P_\text{t}]](T)}$ self-consistently instead
of neglecting the influence 
of the true vacuum bubbles on the Hubble expansion for 
an ultraslow transition featured in the dark Abelian Higgs model. The
influence on the percolation temperature is comparably small, but the final
temperature is clearly affected. Computing $P_\text{t}$ using the approximation
$\bar{\rho} \approx \rho_\text{f}$ hence bears
the risk of treating viable, extremely slow phase transitions as ruled out,
even though a final temperature could indeed be found when treating
the effect of $P_\text{t}$ on the Hubble rate self-consistently.

\subsection{Hydrodynamics and wall velocity}
\label{subsec:hydro}

The GW signal can be sourced by several mechanisms, most importantly by the bulk motion of
fluid during the phase transition (i.e., sound waves and turbulence), but also by bubble wall
collisions. In order to estimate the contribution of each mechanism to the
total signal, we compute their individual energy budgets. The most important quantity
going into their computation is the bubble wall velocity $v_\text{w}$, which appeared
already earlier, when we introduced the percolation integral in
eq.~\eqref{eq:percol-integral}. Computing the bubble wall velocity is itself a notoriously
difficult problem which we only try to address partially in this work. Even though
numerical tools like \texttt{WallGo}~\cite{Ekstedt:2024fyq} allow
the computation of the wall velocity in general models, the numerical cost of this is
still prohibitive for large-scale parameter scans. Most importantly, however, it is also
not strictly necessary to accurately compute $v_\text{w}$ in the strongly supercooled
cases we are most interested in: For a sufficiently large amount of supercooling, say
$\alpha > \mathcal{O}(1)$, the bubble wall velocity is expected to approach the speed of light,
$v_\text{w} \to 1$. In order to verify this, we employ the local thermal equilibrium (LTE)
approximation, which can be shown to be an upper bound on the wall velocity,
$v_\text{w} \lesssim v_\text{w}^\text{LTE}$~\cite{Ai:2023see}.
More important for the precise computation of the energy budget in strongly supercooled
cases is the question whether the bubble walls reach a terminal velocity before they
collide, or whether they accelerate up to very large Lorentz boost factors (between the
wall and the surrounding fluid) up until the point when they collide. Typically the latter
case requires very specific model building, i.e., negligible next-to-leading-order
friction, for instance due to extreme supercooling $\alpha > \mathcal{O}(10^{10})$.
For these cases, we follow
refs.~\cite{Espinosa:2010hh,Ellis:2019oqb} and compute the energy stored in bubble walls
based on the profile of a single bubble. In this case the efficiency factor, defined as
the fraction of the energy in the bubble walls compared to the released vacuum energy
$E_{V} = 4\pi \Delta V R_{*}^{3}/3$, reads
\begin{align}\label{eq:coll}
  \kappa_{\mathrm{col}} = \frac{E_{\mathrm{wall}}}{E_{V}} =
  \begin{cases}
    1 - \frac{\alpha_{\infty}}{\alpha}\, \quad &\text{if} \quad
    \gamma_{*} < \gamma_{\mathrm{eq}} \quad \text{(runaway walls)} \\
    \frac{\gamma_{\mathrm{eq}}}{\gamma_{*}}
    \left( 1 - \frac{\alpha_{\mathrm{\infty}}}{\alpha}
    \left( \frac{\gamma_{\mathrm{eq}}}{\gamma_{*}} \right)^{2}\right)\,
    \quad &\text{if} \quad  \gamma_{\mathrm{eq}} \le \gamma_{*}
    \quad \text{(walls with large terminal $v_\text{w}$)}  \\
    0\, \quad &\text{else (non-relativistic walls)}\, ,
\end{cases}
\end{align}
where $\gamma_\star= 2R_{\text{sep}} / (3R_0)$ is the maximally
possible bubble wall boost factor at the
time of collision. Here, $R_* \simeq R_{\text{sep}}$ ($R_0$) are
the bubble radii at the time of the bubble
collisions (nucleation). We evaluate $R_{0}$ as the radial
coordinate of the bubble profile
where $\phi(R_{0}) = \ba{\phi_{0} - \phi_{\mathrm{f}}}/2$,
where $\phi_{0}$
denotes the release point, i.e.~the field value at the
centre of the critical bubble.
The derived quantity
\begin{equation}\label{eq:coll2}
  \gamma_{\mathrm{eq}} = \frac{\alpha - \alpha_{\infty}}{\alpha_{\mathrm{eq}}}
  \quad \text{with} \quad
  \alpha_\infty=\frac{T^2}{24\,\rho_{\text{rad}}^{\text{false}}}
  \sum_i n_i\,\Delta m_i^2 \,  \quad \text{and} \quad
  \alpha_{\text{eq}}=\frac{T^3}{\rho_{\text{rad}}^{\text{false}}}
\sum_{\text{gauge}} n_i\,g_i^2\,\Delta m_i
\end{equation}
is the Lorentz factor reached by a bubble wall with terminal velocity, which has passed the
Bödeker-Moore leading-order friction~\cite{Bodeker:2009qy} ($\alpha_\infty$)
and is then stopped due to the next-to-leading-order friction from gauge boson
splitting radiation ($\alpha_{\mathrm{eq}}$). The sums run over all particles
(gauge bosons) $i$ which gather a positive mass square difference $\Delta m_i^2$
during the transition, carrying an intrinsic number of degrees of freedom $n_i$
and a gauge coupling $g_i$.

In most cases of phenomenological interest (strongly supercooled transitions involving
gauge bosons), $\gamma_\text{eq} \ll \gamma_*$ and $\alpha \gg \alpha_\infty$, meaning that
$\kappa_\text{col} \to 0$, even though the bubble walls might be relativistic at the time of
their collision. For these cases, this a-posteriori validates our initial
assumption that the energy density in bubble walls can be neglected in
eq.~\eqref{eq:rhobar}. The rest of the released vacuum energy goes into the bulk motion of
the plasma and dissipated heat. The first of these acts as the source for sound waves
after the bubble wall collisions. The efficiency factor of the energy transfer to sound
waves is defined as the fraction of energy in the bulk motion $E_{\mathrm{fl}}$ compared
to the vacuum energy released $E_{V}$ and can be computed from the enthalpy $w(\xi)$ and
velocity profile $v(\xi)$ of the bubble:
\begin{align}\label{eq:sw}
  \kappa_{\mathrm{sw}} = \frac{E_{\mathrm{fl}}}{E_{V}}
  = \frac{3}{v_{\mathrm{w}}^{3}} \int\dd \xi\,\xi^{2} \frac{v^{2} w}{1 - v^{2}}\,.
\end{align}
In order to evaluate this integral, we use the fitting functions
listed in the appendix of ref.~\cite{Espinosa:2010hh}.
Importantly, in the limit of strong FOPTs with $v_\text{w} \to 1$ and $\alpha \gg 1$,
the efficiency factor approaches $\kappa_\text{sw} \to 1$.

The decaying
sound waves leave behind turbulence in the plasma, which can also act as an
efficient GW source, which is however still subject to sizeable uncertainties
due to the practical difficulties of simulating turbulence~\cite{RoperPol:2019wvy}.
We take the standard heuristic
approach of approximating the efficiency factor for turbulence
as some fraction $\epsilon_{\mathrm{turb}}$ of the sound wave efficiency
factor~\cite{Caprini:2024hue}
\begin{align}\label{eq:turb}
  \kappa_{\mathrm{turb}} = \epsilon_{\mathrm{turb}} \kappa_{\mathrm{sw}}\,.
\end{align}
In practice, we set the turbulence efficiency factor to $\epsilon_{\mathrm{turb}} = 0.1$.

\subsection{Gravitational wave spectrum}
\label{subsec:grav-wave-spectr}

Simulating the nucleation of bubbles, their expansion and the backreaction of the plasma
in response to their growth and collision has become a mature field of research by now.
Still, simulating this process remains computationally demanding and is unfeasible for
scans over model parameter spaces. Luckily, there now exists a wealth of
different analytical and semi-analytical
approximations for each of the three aforementioned contributions to the total GW spectrum.
An overview of these approximations can be found in
refs.~\cite{Kamionkowski:1993fg, Kosowsky:1992vn, Kosowsky:1991ua,  Hindmarsh:2013xza,
Giblin:2014qia, Hindmarsh:2015qta, Hindmarsh:2015qta, Hindmarsh:2016lnk, Jinno:2017fby,
Hindmarsh:2017gnf, Konstandin:2017sat, Cutting:2018tjt, Caprini:2019egz, RoperPol:2019wvy,
Cutting:2019zws,  Hindmarsh:2019phv, Hindmarsh:2020hop, Cutting:2020nla}. This long list of
viable approximations is constantly growing due to the advances of numerical schemes being
developed for more specialised cases, see for instance refs.~\cite{Jinno:2022mie,
RoperPol:2023dzg,Caprini:2024gyk, Caprini:2026nnk} for the most recent developments. In this work
we follow the LISA working group and consider the GW spectra recommended in
ref.~\cite{Caprini:2024hue} for each of the contributions, which in our notation read
\begin{align} \label{eq:GWspec}
  h^2 \Omega_{\text{GW}}(f) = \mathcal{R}h^2 \times \begin{cases}
     A_{\mathrm{col}} K_{\mathrm{col}}^{2} (RH)^2
  S_{\mathrm{col}}(f) \quad & \text{for bubble wall collisions,}\\
A_{\mathrm{sw}} K_\text{sw}^2 \ba{RH}  \mathcal{Y}_{\mathrm{sw}} 
S_{\mathrm{sw}}(f) \quad & \text{for sound waves, and}\\
A_{\mathrm{turb}}\,K_{\mathrm{turb}}^2\,(RH)^3
S_{\mathrm{turb}}(f)\quad & \text{for turbulence.}
  \end{cases}
\end{align}
Each contribution includes a prefactor
\begin{align}
  \mathcal{R}h^2 &= \ba{\frac{a_\text{reh}}{a_0}}^2 \ba{\frac{H_\text{reh}}{H_0}}^2 h^2
  = \frac{\Omega_{\gamma}h^2}{D^{4/3}} \left(\frac{h_0}{h_{\text{eff}}^{\reh}}\right)^{4/3}
\frac{g_{\text{eff}}^{\reh}}{g_0}\,,
\end{align}
describing the amplitude redshift, where
$H_0 = 100  h \, \text{km} / (\text{s} \, \text{Mpc})$
is the Hubble constant, $\Omega_{\gamma}h^2 = 2.473 \times10^{-5}$ is the energy density in
radiation, and $g_{\gamma} = 2$, $h_{0} = 3.91$ are the numbers of relativistic degrees
of freedom, each evaluated today~\cite{Planck:2018vyg}. The energy and entropy degrees
of freedom in the broken phase at the time of reheating (i.e., the time of percolation) are
denoted by $g_\text{eff}^{\mathrm{reh}}$ and $h_\text{eff}^{\mathrm{reh}}$,
respectively. We further allow for an
additional dilution of the GW signal at later times due to a phase of entropy injection
through the dilution factor $D = S_\text{SM}^\text{0}/ S_\text{tot}^\text{init}$
with $S_\text{SM}^0$ being
the SM comoving entropy density today and $S_\text{tot}^\text{init}$ being the total
comoving entropy density just before the transition, cf.~ref.~\cite{Ertas:2021xeh}.
We ignore other effects from modified redshift histories, like potential
modulations of the infrared slope of the GW spectrum due to a
phase of early matter domination~\cite{Hook:2020phx,
Franciolini:2023wjm}, as these only have an impact on frequencies
far away from the peak of the GW signal.
For a standard $\Lambda$CDM redshift history following the transition, all frequencies
obtain the same amount of redshift. A helpful quantity is the inverse horizon size at the
time of the transition, redshifted to today, i.e.
\begin{align}\label{eq:freq-redshift}
  H_{*,0} = \frac{a_*}{a_0} H_* = \frac{11.2 \, \mathrm{nHz}}{D^{1/3}}
  \left( \frac{T_{\mathrm{reh}}}{100 \, \mathrm{MeV}} \right)
  \left( \frac{g_\text{eff}^{\mathrm{reh}}}{10} \right)^{1/2}
  \left( \frac{10}{h_\text{eff}^{\mathrm{reh}}} \right)^{1/3}\,,
\end{align}
which allows a direct comparison with the relevant length scale $L$ of the transition,
fulfilling $L \lesssim H_{*}^{-1}$ due to causality.

The normalisation constants in eq.~\eqref{eq:GWspec} read\footnote{In ref.~\cite{Caprini:2024hue},
the bubble wall collision spectra are expressed in terms of $\beta/H$. As we prefer to use $RH$ for the
aforementioned reasons and for consistent notation, we set the
prefactor to $A_\text{col} = 0.05 \times \ba{\beta R_\text{sep}}^2$,
where $\beta R_\text{sep} = (8 \pi / f_\text{perc})^{1/3} \max(v_\text{w},
c_\text{s})$, see eq.~\eqref{eq:betaH_RH}.
We further absorbed an explicit factor 3 appearing in
ref.~\cite{Caprini:2024hue} into $A_{\mathrm{turb}}$.}
$A_{\mathrm{col}} \approx 0.98$, $A_{\mathrm{sw}} = 0.11$ and
$A_{\mathrm{turb}} = 0.255$. The corresponding kinetic energy fractions can be
computed from the transition strength $\alpha$ and the efficiency factors,
\begin{equation}
  K_{\mathrm{col}} =\kappa_{\mathrm{col}}\frac{\alpha}{1+\alpha},\qquad
  K_{\text{sw}}=0.6\,\kappa_{\text{sw}}\frac{\alpha}{1+\alpha}\, ,\qquad \text{and}\qquad
  K_{\text{turb}}=0.6\,\kappa_{\text{turb}}\frac{\alpha}{1+\alpha} \, ,
\end{equation}
where the numerical factor 0.6 accounts for the reduced bulk kinetic energy
found in recent simulations compared to single bubble expansion,
cf.~refs.~\cite{Hindmarsh:2015qta,Cutting:2019zws}. The spectral shapes
are given by broken powerlaws following
\begin{align}
  S_{\col}(f)
    = \left( \frac{f}{f_{\col}} \right)^{2.4}
       \left[
         \frac{1}{2}
         + \frac{1}{2}
         \left( \frac{f}{f_{\col}} \right)^{1.2}
       \right]^{-4}
    \qquad
    \text{with}\quad
    f_{\col} = 0.49 \frac{H_{*,0}}{RH}
    \label{eq:Scol}
\end{align}
for the collision signal,
\begin{align}
  S_{\sw}(f)
    &= N
       \left( \frac{f}{f_{\sw,1}} \right)^3
       \left[ 1 + \left( \frac{f}{f_{\sw,1}} \right)^2 \right]^{-1}
       \left[ 1 + \left( \frac{f}{f_{\sw,2}} \right)^4 \right]^{-1} 
    \label{eq:Ssw}
  \\
    &\text{with}\qquad
    f_{\sw,1} = 0.2 \frac{H_{*,0}}{RH} \, , \qquad
    f_{\sw,2} = \frac{0.5}{\Delta_{\mathrm{w}}} \frac{H_{*,0}}{RH} \, , \qquad \text{and} \qquad
    \Delta_{\mathrm{w}} = \frac{|v_{\mathrm{w}} - c_{\mathrm{s}}|}{\max(v_{\mathrm{w}}, c_{\mathrm{s}})}
    \notag
\end{align}
for the sound wave signal, and
\begin{align}
  S_{\turb}(f)
    &= \left( \frac{f}{f_{\turb,1}} \right)^3
       \left[
         1 + \left( \frac{f}{f_{\turb,2}} \right)^{2.15}
       \right]^{-7.9}
       \begin{cases}
         \ln^2\!\left( 1 + \dfrac{H_{*,0}}{2\pi f_{\turb,3}} \right) 
         & \text{for}\ f \le f_{\turb,3} \\
         \ln^2\!\left( 1 + \dfrac{H_{*,0}}{2\pi f} \right) 
         & \text{for}\ f > f_{\turb,3}
       \end{cases}
    \label{eq:Sturb}
  \\
    &\text{with}\qquad
    f_{\turb,1} = H_{*,0} \, , \qquad
    f_{\turb,2} = 2.2 \frac{H_{*,0}}{RH} \, , \qquad \text{and} \qquad
    f_{\turb,3} = \frac{\sqrt{3 K_{\turb}}}{4} \frac{H_{*,0}}{RH} 
    \notag
\end{align}
for the turbulence signal. The normalisation and the lifetime factor
for the soundwave source are given by
\begin{align} \label{eq:swlifetime}
N = \frac{4}{\pi}
        \frac{
          f_{\sw,1}\bigl(f_{\sw,1}^4 + f_{\sw,2}^4\bigr)
        }{
          f_{\sw,2}^3
          \bigl(
            \sqrt{2}\,f_{\sw,1}^2
            - 2 f_{\sw,1} f_{\sw,2}
            + \sqrt{2}\,f_{\sw,2}^2
          \bigr)
        }  \qquad \text{and} \qquad \mathcal{Y}_{\sw} = \min\!
        \left( 1, \frac{2 RH}{\sqrt{3 K_{\sw}}} \right) \, .
\end{align}

\subsection{Observability of cosmological GW backgrounds}
\label{subsec:observability}

A given GW background from a FOPT could potentially be observed in the near future with
upcoming GW observatories. Excitingly, it could be that PTAs have already found
signals from a cosmological phase transition~\cite{Antoniadis:2022pcn,NANOGrav:2023gor,
EPTA:2023fyk,Xu:2023wog,Reardon:2023gzh,Miles:2024seg}. As of now, however, the only
relevant\footnote{Also the non-observations of CMB $B$-mode polarisations provide additional
constraints, which we however do not take into account here, as the corresponding signals
would need to be emitted at very late times, when other constraints become more relevant. We
further do not take $\mu$-distortion constraints~\cite{Ramberg:2022irf}, curvature
perturbation constraints~\cite{Liu:2022lvz} or BBN homogenisation
constraints~\cite{Bagherian:2025puf} into account, as these typically do not dominate
or still demand further investigation.} constraints on cosmological GW backgrounds stem
from pulsar timing arrays (PTAs), $\Delta N_\text{eff}$ constraints from BBN and the
CMB observations, and, to a lesser extent, from LIGO-Virgo-KAGRA non-observations.
In order to access the observability of a GW background and its feasibility given
cosmological constraints, we hence implement the following three measures:

\paragraph{Signal-to-noise ratios.} For a given GW background, the signal-to-noise ratio (SNR)
is a key metric for assessing its detectability. The SNR is defined as the ratio of the signal
power to the noise power, integrated over the detector's frequency band and weighted by
the observation time, 
\begin{equation}
  {\rm SNR}^2=2\,t_{\text{obs}}\int \mathrm{d}f\,
  \left[\frac{h^2\Omega_{\text{GW}}(f)}{h^2\Omega_{\text{noise}}(f)}\right]^2,
\end{equation}
The factor 2 accounts for cases where a cross-correlation can be performed when a
multi-detector network is used; in single-detector scenarios, the prefactor is 1
instead~\cite{Breitbach:2018ddu}. If the SNR exceeds a certain threshold $\text{SNR}_\text{thr}$,
the signal is considered detectable. In our work, we consider the spectral noise curves
for the following list of \textbf{current} and \textit{planned/proposed} GW observatories,
based strongly on ref.~\cite{Breitbach:2018ddu},
\begin{itemize}
  \item \textit{SKA} (5, 10, and 20 years of detection data)~\cite{Janssen:2014dka,Weltman:2018zrl},
  $\text{SNR}_\text{thr} = 4$, $t_\text{obs} = 5, 10, 20 \, \text{yr}$,
  \item \textbf{EPTA (18 years)}~\cite{EPTA:2015qep,EPTA:2016ndq},
  $\text{SNR}_\text{thr} = 1.19$, $t_\text{obs} = 18 \, \text{yr}$,
  \item \textbf{NANOGrav (11 and 15 years)}~\cite{NANOGRAV:2018hou},
  $\text{SNR}_\text{thr} = 0.697$, $t_\text{obs} = 11, 15 \, \text{yr}$,
  \item \(\mu\)\textit{Ares}~\cite{Sesana:2019vho},
  $\text{SNR}_\text{thr} = 10$, $t_\text{obs} = 7 \, \text{yr}$,
  \item \textit{LISA}~\cite{LISA:2017pwj},
  $\text{SNR}_\text{thr} = 10$, $t_\text{obs} = 4 \, \text{yr}$,
  \item \textit{B-DECIGO, DECIGO}~\cite{Seto:2001qf,Sato:2017dkf},
  $\text{SNR}_\text{thr} = 8, 10$, $t_\text{obs} = 4, 4 \, \text{yr}$,
  \item \textit{BBO}~\cite{Crowder:2005nr},
  $\text{SNR}_\text{thr} = 10$, $t_\text{obs} = 4 \, \text{yr}$,
  \item \textit{ET}~\cite{Sathyaprakash:2012jk},
  $\text{SNR}_\text{thr} = 5$, $t_\text{obs} = 5 \, \text{yr}$, and 
  \item \textbf{HLV (O2)}, \textit{HLVK (design)}~\cite{LIGOScientific:2014pky,VIRGO:2014yos,Kokeyama:2020lqf},
  $\text{SNR}_\text{thr} = 1$, $t_\text{obs} = 1 \, \text{yr}$.
\end{itemize}
The evaluation of these measures thus allows us to assess whether a given GW background
is detectable by some observatory or should have already been found by now, in case of the
already existing observatories.

\paragraph{PTA likelihoods.} To further address the compatibility of a given GW signal
in the nHz range with the observed PTA signal, we compute likelihoods with the
\texttt{PTArcade} tool~\cite{Mitridate:2023oar}, using the \texttt{Ceffyl} refitting
techniques presented in ref.~\cite{Lamb:2023jls}. In practice, we further smooth out
this likelihood beyond its original range of validity, towards a wider range of GW
amplitudes and frequencies, such that it can also be used efficiently within a sampler,
cf.~ref.~\cite{Bringmann:2026xcx}. The currently publicly available PTA likelihoods
are based on the NANOGrav 12.5\,yr~\cite{NANOGrav:2020bcs} and 15\,yr~\cite{NANOGrav:2023gor}
data sets, as well as the second IPTA data release~\cite{Antoniadis:2022pcn}.

\paragraph{Cosmological constraints.} Finally, we compute the GW contribution to the
effective number of relativistic species,
\begin{equation}
\Delta N_{\text{eff}}^{\text{GW}}=\frac{8}{7}\left(\frac{11}{4}\right)^{4/3}
\frac{1}{\Omega_\gamma h^2}\int_{f_{\text{min}}}^{f_{\text{max}}}
h^2\Omega_{\text{GW}}(f) \, \mathrm{d} \log f \, ,
\end{equation}
using $f_{\text{min}}=1.5\times 10^{-11}\,\text{Hz}$ and $f_{\text{max}}=10^{10}\,\text{Hz}$,
see ref.~\cite{Maggiore:2018sht}. This allows for a quick comparison with the bounds obtained
from BBN and CMB observations. For a typical power-law spectrum for instance, this bound
constraints the GW amplitude below $h^2 \Omega_{\text{GW}}(f_\text{peak}) \lesssim 10^{-6}$,
depending on the specific spectral shape and choice of cosmological parameters.

\subsection{Benchmark points and implemented models}
\label{subsec:benchmark-points}

In appendix~\ref{sec:models} we define several models that are implemented in
\texttt{TransitionListener v2.0}. Here, we want to introduce 
benchmark points from these models, spanning vastly different energy scales.
In table~\ref{tab:benchmarks} we list the 
benchmark points and their corresponding model parameters.

\begin{figure}[t]
  \centering
  \includegraphics[width=\linewidth]{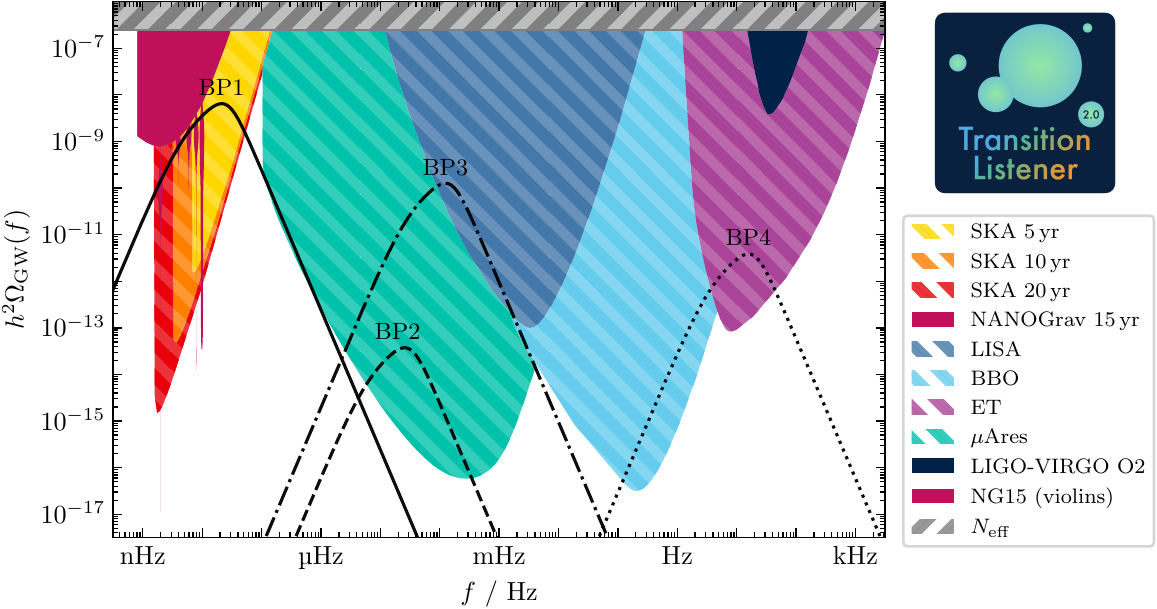}
  \caption{Gravitational wave spectra for the four benchmark points listed in
    table~\ref{tab:benchmarks}. The dashed sensitivity curves are expected
    powerlaw-integrated sensitivities, whereas
    the NANOGrav 15\,yr and LIGO-VIRGO O2 ones are based on
    observations. The thin violins in the nHz region stem from the NANOGrav 15\,yr
    data release. The grey shaded
    bar indicates the excluded region due to $\Delta N_\text{eff}^\text{GW}$
    constraints from BBN and CMB~\cite{Planck:2018vyg,Yeh:2022heq}.}
  \label{fig:gw-benchmarks}
\end{figure}

{\renewcommand{\arraystretch}{1.15}
\setlength{\tabcolsep}{3pt}
\begin{table}[t]
\noindent
\resizebox{\linewidth}{!}{%
\begin{tabular}{l l l p{0.44\linewidth}|c c c}
\toprule
BP & Model & Config file & Model parameters & $\alpha$ &
$(\beta/H)_{RH}$ & $T_{\rm reh}\,[\mathrm{GeV}]$ \\
\midrule

1 & Conformal dark $U(1)^\prime$ & \texttt{bp\_pta.yaml} &
\begin{tabular}[t]{@{}l@{}}
$g=0.7$, %g=0.692
$y=0.01$, %exact
$v=140\,\text{MeV}$. %exact
\end{tabular}
& $620$ %621.03
& $59$ %58.705
& $1.4\times10^{-2}$ %0.014154
\\

2 & Dark flipflop & \texttt{bp\_muares.yaml} &
\begin{tabular}[t]{@{}l@{}}
$\lambda_0=0.0051$, %0.005098
$\lambda_1=0.0021$, %0.002144
$\lambda_{12}=0.0031$, %0.003078
\\
$v=3.7\,\text{GeV}$, %3.728330741601088
$y=0.97$, %0.972319
$\gamma=0.75$. %0.7532
\end{tabular}
& $0.18$ %0.18407
& $1400$ %1402.8
& $5.9\times10^{-1}$ %0.59272
\\

3 & 2HDM (Type-I) & \texttt{bp\_lisa.yaml} &
\begin{tabular}[t]{@{}l@{}}
$\lambda_{1}=0.29$, %0.28894
$\lambda_{2}=0.26$, %0.26237
$\lambda_{3}=5.81$, %5.80958300958301
\\
$\lambda_{4}=-2.2$, %-2.17494
$\lambda_{5}=-2.2$, %-2.24170
$\tan\beta=21$, %21.3949
\\
$m_{12}^2=1.6\times10^{3}\,\text{GeV}^2$, %1573.171
$v=246\,\text{GeV}$. %246.22
\end{tabular}
& $0.60$ %0.5955000609132117
& $91$ %90.68692762240545
& $4.4\times10^{1}$ %43.869695874548206
\\

4 & Dark $U(1)^\prime$ & \texttt{bp\_et.yaml} &
\begin{tabular}[t]{@{}l@{}}
$\tilde{g}=2.65$, %exact
$\lambda=1.5\cdot10^{-3}$, %exact
$v=1\cdot10^{7}\,\text{GeV}$. %exact
\end{tabular}
& $1.4$ %1.3903
& $810$ %814.46
& $5.6\times10^{5}$ %559320.0
\\

\bottomrule
\end{tabular}%
}
\caption{Benchmark points used in fig.~\ref{fig:gw-benchmarks}. Unrounded
  values for the input parameters can be found in the listed \texttt{.yaml} files.}
\label{tab:benchmarks}
\end{table}
}

The resulting GW spectra for these benchmark points are shown in fig.~\ref{fig:gw-benchmarks}
together with the aforementioned $\Delta N_\text{eff}^\text{GW}$ bounds, the PTA data
(visualised as violins) and the powerlaw-integrated sensitivity curves for a selection
of representative GW observatories.
Let us briefly discuss each of the GW spectra here. Benchmark point 1, stemming from a
conformal dark sector model was studied in detail in ref.~\cite{Balan:2025uke}:
If a huge amount of vacuum energy density ($\alpha = 620$) is
released during the phase transition and the universe reheats to a temperature of
$T_\text{reh} = 14 \, \text{MeV}$, a good fit to the PTA data can be obtained. In
ref.~\cite{Balan:2025uke} it was shown that the same model, in a part of its parameter space
can further explain the observed DM abundance. Benchmark point 2, on the other hand, is based on
a dark flipflop model, first discussed in ref.~\cite{Bringmann:2026xcx}, and represents a
different approach to generating GWs: Here, no new gauge group is introduced, but instead
two SM singlets are considered. When one of the two singlets acquires a vacuum expectation
value through a second-order transition, it generates a potential barrier for the other
singlet field, which eventually will tunnel to its true vacuum afterwards. The shown benchmark point
corresponds to a comparatively weak phase transition ($\alpha = 0.18$) happening below the GeV
scale and could be tested by the proposed $\mu$Ares space-based
interferometer. Similarly, benchmark point 3 from the 2HDM showcases another possible
scenario for GW production: In this case the SM is extended by another Higgs doublet,
generating a potential barrier during the electroweak symmetry breaking. The benchmark transition
is strongly first-order ($\alpha = 0.60$), with bubble sizes at the few-percent level of the
horizon size at that time ($RH = 0.049$). The proposed point would result in a GW signal in
the LISA frequency band. Its observation would provide a unique test of the underlying
electroweak theory. Lastly, benchmark point 4, stemming from an Abelian dark Higgs
extension of the SM predicts a strong GW signal in the ET band, testable by
the next generation of ground-based interferometers. The identification of a FOPT signal
at these frequencies would correspond to the presence of strong dynamics at the $10^6$ GeV
scale, which is out of reach for conventional collider searches for new physics, see also
ref.~\cite{Ertas:2021xeh} for a detailed study of the dark Abelian Higgs model parameter space.

Each of these benchmark points is available as a \texttt{.yaml} file, allowing to be
reproduced easily, see sec.~\ref{sec:scan-and-plotting} for a tutorial. We cordially invite the reader
to reproduce these GW spectra on their own, using a one-line call of our code, as
presented in the following section.

\section{Program description and implementation}
\label{sec:3}

In this section we summarise the internal structure and functionality of
\texttt{TransitionListener v2.0} (abbreviated \texttt{TL}). We give a compact overview of
the numerical pipeline and the design choices that enable robust, end-to-end calculations:
from phase tracing and multi-field tunnelling to the computation of the true‑vacuum
fraction with its backreaction on the Hubble rate and the self-consistent treatment of
reheating. For each module we discuss the implemented algorithms
and the key accuracy controls.

This section is structured as follows: In the first subsection, we give a brief guide on
the installation of the code, followed by a first look on its overall module structure. We
then continue with a discussion of these modules. Starting from the
implementation of a new model in sec.~\ref{sec:model-definition}, we go through the chain
of individual modules which have to be called in order to produce the final observability
results. Sec.~\ref{sec:phase-tracing} then discusses the underlying algorithm behind the
phase tracing in a given temperature-dependent effective potential. The following
sec.~\ref{sec:bubble-nucleation-calc} describes the evaluation of the
transition rates between phases
found in the previous step. The algorithm to compute the nucleation history
and percolation temperature is discussed in sec.~\ref{sec:false-vacuum-fraction}. Finally, in
sec.~\ref{sec:gw-observability-comp} we discuss the computation of thermodynamical quantities,
the GW signal and its observability. We conclude this section with a brief note on error handling,
which we expand on in more detail in Appendix~\ref{sec:error_codes}.

\subsection{Installation instructions and structure of the repository}
\label{sec:installation}

In this subsection we first want to briefly describe the different installation options
for \texttt{TL}, before we then come to the description of the different modules the
program consists of.

\paragraph{Installation.}
\texttt{TL} can be obtained from the GitHub repository
\begin{center}
  \href{https://github.com/tasicarl/TransitionListener}{https://github.com/tasicarl/TransitionListener}
\end{center}
and can be installed in a virtual \texttt{conda} environment as follows\footnote{The \texttt{ptarcade} package on MacOS with Apple silicon processors requires the option \texttt{--platform osx-64}.}
\begin{lstlisting}[language=bash]
  conda create -n TL -c conda-forge python=3.10 ptarcade ultranest tqdm mpi4py sympy
  conda activate TL
  git clone https://github.com/tasicarl/TransitionListener.git
  cd TransitionListener
  pip install -e .
\end{lstlisting}
The editable (\texttt{-e}) install keeps the package in sync with local changes, for instance when
modifying model files or adding a new feature like an updated GW spectral template. The
public documentation at
\href{https://www.tasillo.de/TransitionListener}{https://tasillo.de/TransitionListener}
provides a quick-start guide, examples and the API references generated from the source
tree. We further want to encourage users of our code to report bugs and ask physics
questions using the issue tracker on \texttt{github}, where we plan to release future
versions of our code.

\begin{figure}[t]
  \centering
  \includegraphics[width=1\linewidth]{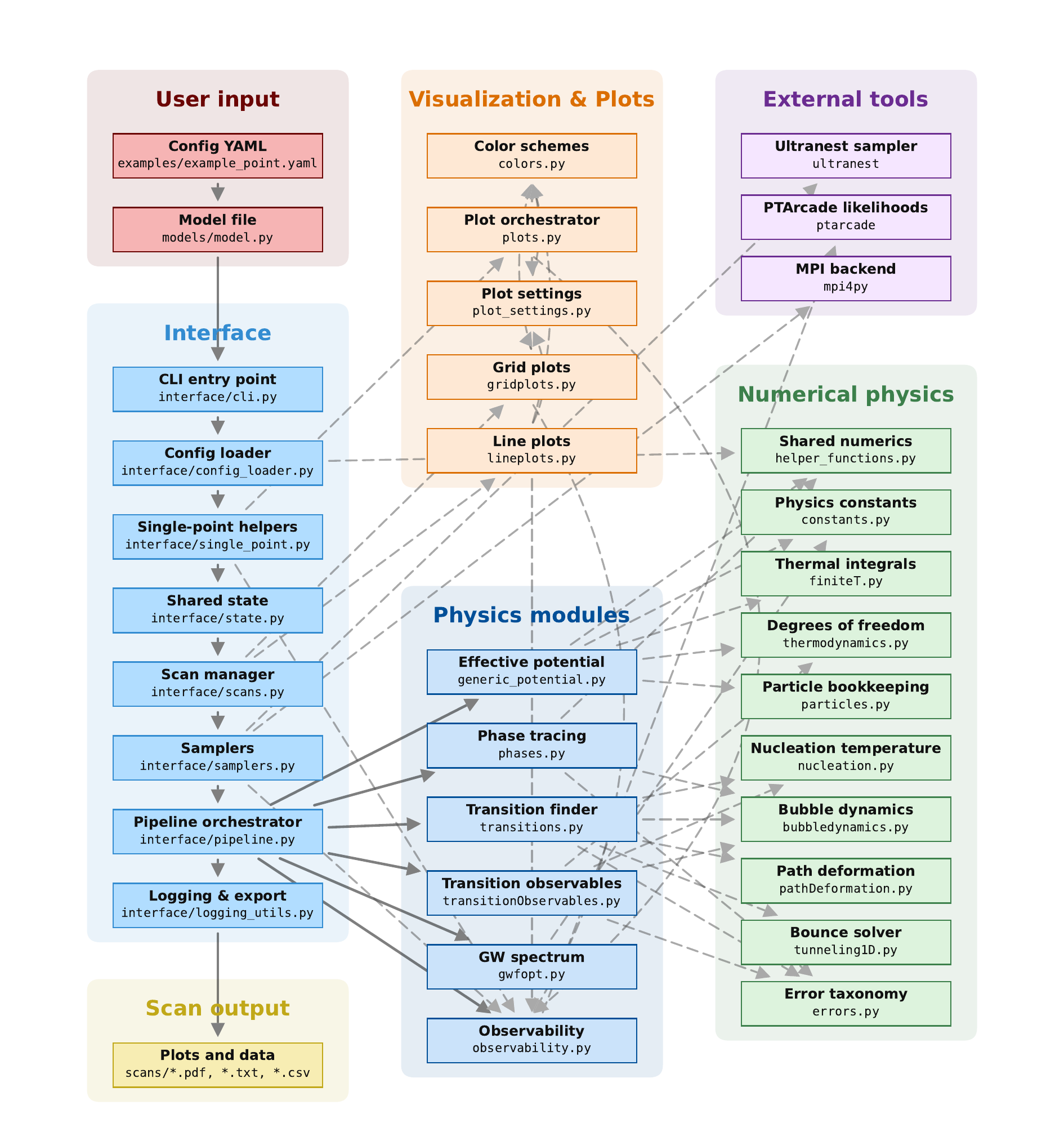}
  \caption{Module dependencies inside \texttt{TransitionListener}.}
    \label{fig:module_graph}
\end{figure}

\paragraph{Structure of the code.}
Version 2.0 of \texttt{TL} builds onto the original release from
2021~\cite{Ertas:2021xeh}, which itself was heavily influenced by the phase-tracing,
path-deformation and tunnelling foundations of
\texttt{CosmoTransitions}~\cite{Wainwright:2011kj}. In fig.~\ref{fig:module_graph} a
visual guide for the new structure of \texttt{TL} can be found, with arrows indicating the
typical module dependencies.

Overall, the different modules can be structured into several groups, indicated by
different colours and blocks in fig.~\ref{fig:module_graph}: The modules visualised in
light blue serve as a central interface to all other modules: A command line interface,
relevant for the \texttt{tl} command is introduced in \texttt{interface/cli.py}. A usual
call of \texttt{TL} consists of calling a\texttt{.yaml} configuration file (depicted as a
red block in fig.~\ref{fig:module_graph}) that contains the modelname and parameter
specifications, e.g.
\begin{lstlisting}
  tl --config config.yaml
\end{lstlisting}
This will then trigger the \texttt{config\_loader.py} to read the requested operations on
the defined model, be it computations on a single point in parameter space
(\texttt{single\_point.py}) or larger scans (\texttt{scans.py}), potentially involving
Monte Carlo methods available through the \texttt{samplers.py} module. If a GW signal
prediction for a single point of a given parameter space is wanted, several available plots can be
produced, each one stemming from a different part of the full end-to-end computation, see
sec.~\ref{sec:scan-and-plotting} for examples. Eventually, all the
output generated by \texttt{TL}, including log
files and plots, is handled through the \texttt{logging\_utils.py} module. By default,
these are saved into a \texttt{/scans/} subdirectory, relative to the installation point
of the code.

The main physics engine of \texttt{TL} is visualised in dark blue in
fig.~\ref{fig:module_graph}: The pipeline orchestrator, found in
\texttt{interface/pipeline.py} calls the individual modules necessary to answer a physics
question. Methods for the automated computation of effective potentials and derived
quantities like their gradients or the energy density are implemented in
\texttt{generic\_potential.py}. User-defined modules located in the folder
\texttt{/models/} inherit methods from \texttt{generic\_potential.py} and can also
overload them, if a given model does not fit the default implementation. An example is the
conformal $U(1)^\prime$ model described in Appendix~\ref{sec:models},
which requires a non-standard renormalisation
procedure. Each model file defines the field content, the counterterms and a
\texttt{MassSpectrum} object, which defines the field-dependent masses of all physical
states present in the theory. In case of SM particles being present in the theory, as for
instance in the 2HDM, the model file also allows the user to exclude these from the computation of
the background evolution described by the effective degrees of freedom
$g_\text{eff}^\text{ext}$ and $h_\text{eff}^\text{ext}$, cf.~eq.~\eqref{eq:rho}. Helper
functions (visualised in green in fig.~\ref{fig:module_graph}) for the bookkeeping of
different types of particles are implemented in \texttt{particles.py}, degrees of freedom
are computed via the methods implemented in \texttt{thermodynamics.py}, the $J_\text{b}$
and $J_\text{f}$ integrals are found in \texttt{finiteT.py}, and general physical
constants, like SM particle masses or the Planck mass are centrally managed in
\texttt{constants.py}.

Having defined an effective potential, the phase tracing is performed in
\texttt{phases.py}, which will hand back a \texttt{Phases} object to the pipeline
orchestrator. This object is then handed over to the \texttt{transition.py} module which
evaluates the possibility of all the different transitions between the found phases. In
particular, this requires the \texttt{bubbledynamics.py} helper module, which performs the
computation of the true-vacuum fraction, based on bounce action computations performed in
\texttt{tunneling1D.py}, after a viable tunnelling path has been found in
\texttt{pathDeformation.py}. If a first-order phase transition has been found, macroscopic
thermodynamic quantities like $\alpha$, $RH$, $T_\perc$ and $T_\reh$ are then computed in
\texttt{transitionObservables.py}, which itself returns a \texttt{TransitionObservables}
object. Within the pipeline orchestrator, this object is then handed to the
\texttt{gwfopt.py} modules which computes the GW spectrum following eq.~\eqref{eq:GWspec}.
If several transitions are present in a given model, the strongest FOPT is used to infer
the GW signal. The \texttt{FOPTspectrum} object created by this is then used to infer the
observability and possible cosmological viability of a given signal within the
\texttt{observability.py} module.

In the case of discrete scans over parts of a given model parameter space, the plot
orchestrator \texttt{plots.py} delegates tasks to the appropriate plotting modules,
including \texttt{gridplots.py} and \texttt{lineplots.py}. These use colour schemes and
plotting defaults defined in \texttt{colors.py} and \texttt{plot\_settings.py} (all
visualised in orange in fig.~\ref{fig:module_graph}). External tools like \texttt{MPI},
allowing the easy parallelisation of large model parameter space scans, the
\texttt{ultranest} sampler, or the \texttt{PTArcade} likelihood are optional dependencies
of the code, broadening the general functionality of \texttt{TL}. If at any point of the
computation an error occurs, it is logged and handled by the \texttt{logging\_utils.py}
module. The \texttt{errors.py} file contains a list of possible numerical and physical
errors which can occur when running \texttt{TL}. These include for instance the possibility
of the eternal inflation scenario, when a transition does not succed, or the case
of insufficient numerical accuracies
to make precise statements about the emitted GW signal. A list of all
error cases can be found in Appendix~\ref{sec:error_codes}.

  % PUT THIS ON WEBSITE!
  % \begin{table}[t]
  %   \centering
  %   \footnotesize
  %   \begin{tabular}{p{0.23\linewidth}p{0.18\linewidth}p{0.52\linewidth}}
  %     \toprule
  %     Object & Location & Purpose / key fields \\
  %     \midrule
  %     \texttt{PhaseInfo} & \texttt{phases.py} & Spline of a minqimum: $(X,T,dX/dT)$, \texttt{tck} spline, $T_{\min}$/$T_{\max}$, parent/child links, second-order links, \texttt{mirrorPhase}. \\
  %     \texttt{PhaseSegment} & \texttt{phases.py} & Piecewise segment of a phase history: entry/exit $T$ and a callable wrapper for $\phi(T)$. \\
  %     \texttt{TransitionHistory} & \texttt{phases.py} & Ordered list of \texttt{PhaseSegment} objects for the full thermal evolution. \\
  %     \texttt{TransitionInfo} & \texttt{transitions.py} & Candidate transition with $T_{\crit}$, $T_{\nuc}$, $T_{\perc}$, phases, action, and cached instanton data. \\
  %     \texttt{TransitionContext} & \texttt{transitionObservables.py} & Bundle of per-transition inputs (phases, configs, splines) used to compute derived observables. \\
  %     \texttt{PercolationResult} & \texttt{transitionObservables.py} & Percolation outputs and splines ($P$, $S_3/T$, $H$, $T_{\text{bro}}$). \\
  %     \bottomrule
  %   \end{tabular}
  %   \caption{Core data objects used during phase tracing and transition evaluation.}
  %   \label{tab:phase_dataclasses}
  % \end{table}

\subsection{Model definition}
\label{sec:model-definition}

\texttt{TL} allows for the semi-automatic computation of effective potentials, including
quantum and thermal corrections, once a theory and its mass spectrum are specified. A
selection of predefined models is available in the \texttt{models} directory. These
include several single- and multi-field dark sector models as well as the 2HDM, which can
be used out-of-the-box, or extended to more general cases. A discussion of these models
can be found in Appendix~\ref{sec:models}. Here we introduce the definition of a generic model file on a
technical level. To do so, we will describe the \texttt{templatePotential.py} template and
explain how to implement new models.

A generic model file starts with the definition of model parameters in the form of a
\texttt{python} dictionary, including a name of the parameter (including a \LaTeX version
for display in plots), a default value and bounds, see code snippet~\ref{lst:model-params}.
\texttt{TL} will automatically return an error if the parameters are out of bounds or if
they are not initialised properly. Further, the number of field dimensions \texttt{Ndim}
of the potential must be specified.
\begin{lstlisting}[language=python,label=lst:model-params,
caption=Any model file in \texttt{TL} begins with the definition of its
parameters in a dictionary.]
class TemplatePotential(generic_potential):
    model_parameters = {
        "parameter1": {
            "default": 1.0,
            "plotname": r"$\lambda$",
            "min": 0,
            "max": np.inf},
        # ...
    }

    def init(self, inp, verbose=True):
        self.Ndim = 1  # Specify the number of field dimensions
        self.mp = self.set_modelparams(inp)
        self.parameter1 = self.mp["parameter1"]["value"]
        # ...
\end{lstlisting}
Next, the tree-level potential \texttt{V0} and the counterterm potential \texttt{Vct} have to be
implemented as shown in code snippet~\ref{lst:potential}. For demonstrational
purposes, we show the implementation of a dummy tree-level and counterterm potential of the form 
\begin{align}
  V_\text{tree}(\phi_1) = - \frac{\mu_1^2}{2} \phi_1^2 + \frac{\lambda_1}{4} \phi_1^4 \, ,
  \qquad V_\text{ct}(\phi_1) = - \frac{\delta\mu_1^2}{2} \phi_1^2 + \frac{\delta\lambda_1}{4} \phi_1^4 \, .
\end{align}
Currently, there is no automated way to generate the counterterms for a generic model
implemented in \texttt{TL} and counterterms have to be implemented manually for each new
model. The model files delivered in this code release however include implementations for
several common models, including single- and multi-field models, which can be easily
adapted. In the case of the 2HDM, we implemented the counterterm solver in the same way
as \texttt{BSMPTv3}, to allow for a direct comparison between the two codes,
cf.~sec.~\ref{sec:scan-and-plotting}.

\begin{lstlisting}[language=python,label=lst:potential,caption=Example of a tree-level and
counterterm potential implementation.]
def V0(self, X: np.ndarray) -> np.ndarray:
    """ This method defines the tree-level potential."""
    X = np.asanyarray(X)

    phi1 = X[..., 0]
    # For Ndims > 1, the other field directions are exctracted as
    # phi2 = X[..., 1]
    # phi3 = X[..., 2]

    # Compute the contributions to the tree-level potential with the
    # model parameters
    V = - self.mu2_1*(phi1**2.)/2. + self.l_1*(phi1**4.)/4.

    # Potentially other field directions contribute, e.g.:
    # V += - self.mu2_2 * phi2**2/2 + self.l_2 * phi2**4/4.

    return V

def Vct(self, X: np.ndarray) -> np.ndarray:
    """ The counterterm lagrangian is the same as the tree-level Lagrangian but
    with masses and couplings replaced by counterterm values (i.e. here
    mu2 -> dmu2 and l -> dl).
    """
    X = np.array(X)
    phi1 = X[..., 0]

    # Compute the contribution from the counterterms
    V = - self.dmu2_1*(phi1**2.)/2. + self.dl_1*(phi1**4.)/4.
    # Potentially other field directions contribute, e.g.:
    # V += - self.dmu2_2 * phi2**2/2 + self.dl_2 * phi2**4/4.

    return V
\end{lstlisting}
Lastly the field-dependent mass spectrum of the theory has to be implemented, from which
the quantum and thermal corrections in the effective potential can be are computed. This
is done by specifying the (physical) \texttt{Scalar}s, the \texttt{Goldstone} bosons, the
\texttt{GaugeBosons} and \texttt{Fermions} of the theory
(see code snippet~\ref{lst:mass-spectrum}). Here it
is convenient to split the gauge bosons into their transverse and longitudinal part, since
only the latter receive a thermal mass. Additionally, a mass function has to be implemented
for the bosons and fermions, which returns the masses in the same order in which the
particles are listed.
\begin{lstlisting}[language=python,label=lst:mass-spectrum,
caption=Construction of the mass spectrum for fermions and bosons.]
    def build_mass_spectrum(self) -> MassSpectrum:
        scalar_particles = [
            Scalar(name="m_scalar1", latex_name=r"$m_{\phi_1}$",
                   dof=1, is_SM=False),
            Goldstone(name="m_goldstone1", latex_name=r"$m_{\varphi_1}$",
                      dof=1, is_SM=False)]
        gauge_particles = [
            GaugeBoson(name="m_gb_T", latex_name=r"$m_{GB,T}$",
                       dof=2, gauge_coupling=self.g, is_SM=False),
            GaugeBoson(name="m_gb_L", latex_name=r"$m_{GB,L}$",
                       dof=1, gauge_coupling=self.g, is_SM=False)]
        fermion_particles = [
            Fermion(name="m_fermion", latex_name=r"$m_\chi$",
                    dof=4, is_SM=False)]

        def boson_mass_function(X, T):
            # return bosonic mass squares in the same order
            # as scalar_particles/gauge_particles specified above
            ...

        def fermion_mass_function(X):
            # return fermionic mass squares in the same order
            # as fermion_particles specified above
            ...

        return MassSpectrum(
            scalars=scalar_particles,
            gaugeBosons=gauge_particles,
            fermions=fermion_particles,
            boson_massSq_fn=boson_mass_function,
            fermion_massSq_fn=fermion_mass_function,
        )
\end{lstlisting}

Furthermore, the model file has to set the renormalisation scale used in the computation
of the effective potential. Optionally, the user can also override the default accuracy
settings in the function \texttt{setConfigParameters} for all the different algorithms, described in
detail in the following of this section and listed in appendix~\ref{sec:error_codes}.

To allow for a consistent and numerically stable computation, the potential and field values
are by default rescaled by a conversion factor
\begin{align}
  \texttt{conversionFactor} \equiv C = \frac{v_{\text{GeV}}}{\texttt{internal\_scale}}\,,
\end{align}
which is used to convert between physical quantities and internal units. The reason for this is the
difficulty in maintaining numerical stability across a wide range of energy scales, which
often enough appear in powers of four, as in the effective potential. Exponentials, as in
the bubble nucleation rate in eq.~\eqref{eq:nucrate}, further amplify this potential
hazard to numerical stability. The user-defined \texttt{internal\_scale} (in internal
units) sets the scale at which the dynamics of the phase transition should be resolved.

Normalising a dimensionful quantity used in the model definition, for instance a vacuum
expectation value $v_{\text{GeV}}$ in GeV, to \texttt{internal\_scale=1000} will
internally normalise the effective potential such that the endpoint of tunnelling will be
at around 1000 internal units, regardless if $v_{\text{GeV}} = 1 \, \text{MeV}$ or
$v_{\text{GeV}} = 10^{10} \, \text{GeV}$. As a result, quantities can be converted from
one to the other unit system by using
\begin{align}
  Q_{\text{GeV}} = Q\,C^{d}\,\text{GeV}^{d}\,,
\end{align}
where $d$ is the mass dimension of $Q$. For example, $T=T_{\text{GeV}}/C$,
$m=m_{\text{GeV}}/C$ and $\rho=\rho_{\text{GeV}}/C^4$. In the following of this section, all
quantities are understood to be in internal units unless they are explicitly labelled in
GeV (or frequencies in Hz). The default settings are optimised and tested to work well
with an internal reference scale \texttt{internal\_scale=1000}.

\subsection{Phase tracing}
\label{sec:phase-tracing}

The starting point for studying phase transitions, once an effective potential is defined,
is to identify the different phases of the system and the temperature interval in which
they exist. To do so, we
introduced the \texttt{Phases} class, defined within \texttt{phases.py}, which wraps and
extends the original \texttt{CosmoTransitions} tracing algorithms~\cite{Wainwright:2011kj}. We
will review it here for completeness.

Starting from a minimum temperature, which we by default set to the CMB temperature
$T_\text{min} = T_0 = 238 \, \mu \text{eV}$, the minima of the user-defined effective
potential $V$ are specified by a list \texttt{X0}, which has to be defined in the model file.
Any minimum satisfies $\partial V(\bm{\phi}_{\mathrm{min}})/\partial \phi_{i} = 0$
which, using the chain
rule, can be differentiated with respect to temperature to yield an evolution equation of
the position of the minima in the individual field dimensions,
\begin{equation}\label{eq:phase-prediction}
\frac{\partial^2 V}{\partial \phi_i \partial \phi_j}\,\frac{\mathrm{d}\phi_{\mathrm{min},j}}{\mathrm{d}T}
+ \frac{\partial^2 V}{\partial \phi_i \partial T}=0 \quad
\Rightarrow \quad \td{\phi_{\text{min},i}}{T} = - H^{-1}_{ij} \partial_{T}\partial_{\phi_{j}} V\,.
\end{equation}
Here, $H_{ij}^{-1}$ is the inverse of the Hessian matrix of the effective potential.
The phase tracing starts by solving these evolution equations, starting from a minimum at
$T_{\mathrm{min}}$ and tracing it up in temperature, until the maximal temperature
$T_\text{max}$ is reached, which by default is defined as \texttt{Tmax\_factor=2.5} times the
\texttt{internal\_scale}. Eq.~(\ref{eq:phase-prediction}) is used to obtain a guess for
the location of the minimum after a temperature step $\Delta T$. The precise minimum is then
determined by a gradient-free minimiser and the difference between the linear prediction
and the resulting point is used to adapt the step size of the following point. The end of
a phase is indicated by a (near-) singular Hessian $H_{ij}$. When this happens, the routine
searches for neighbouring minima and appends them to the list of minima to trace. To ensure
the full phase structure is found, all found phases 
are eventually also traced down in temperature. After this, redundant
minima are merged by comparing overlaps in field space and second-order transitions are
identified for phases that terminate in saddle points. A summary of the different accuracy
parameters going into the phase tracing routines, together with their default values and a
description of their function in the code, can be found in
table~\ref{tab:tracingconf_defaults} in Appendix~\ref{sec:error_codes}.

As of now, \texttt{TL} does not allow for the detailed treatment of models with several
discrete symmetries in multiple field dimensions, which would require a more sophisticated
approach for treating the tunnelling between mirror phases~\cite{Athron:2024xrh}.
Currently, the user can
manually specify the symmetry structure of the theory, by excluding phase tracing in
specific regions of field space, for instance regions with $\phi < - 5$ (in internal units)
for the case of $\mathbb{Z}_2$ symmetries. That way, the found phases are limited to the specified
regions. If more than one discrete symmetry is present, the code may erroneously classify
phase transitions as impossible, because it only considers transitions within a given
fraction of the total field space\footnote{Intriguingly, this kind of studies are needed for
assessing the production of domain walls, which themselves can complicate the details of
Hubble expansion through their contribution to the total energy density and their decay,
potentially also contributing to the total GW signal of a given phase transitions. It will
therefore be an obvious future goal to extend \texttt{TL} such that it can also reliably
deal with these cases.}.

After the phase tracing, a tree of the possible transition histories leading to the
$T_\text{min}$ phase(s) is constructed. To do so, first the deepest high-temperature
minimum $\bm{\phi}_{\mathrm{high}}$ is identified as the global minimum at the maximal
temperature. A transition to another phase $\bm{\phi}_{\mathrm{low}}$ becomes possible if the
two phases coexists across a temperature interval and the low-temperature phase becomes
energetically favoured, i.e. when
$\Delta V(T) = V_{\tot}(\bm{\phi}_{\text{high}},T) - V_{\tot}(\bm{\phi}_{\text{low}},T) > 0$. A link
between each low-temperature phase fulfilling that criterion is added to the tree. Then,
for each child-phase, this procedure repeats until the $T_\text{min}$ phase(s) are
reached. This tree of possible transitions will then allow the study of possible
first-order transitions in the following step.

\subsection{Computation of the bubble nucleation rate}
\label{sec:bubble-nucleation-calc}

Starting with the highest-temperature phase link, the code iterates over phase pairs and
evaluates the transition rates between them in a range of temperatures using a combination
of path deformation and shooting algorithms. The algorithm used in our work was first
presented and discussed in ref.~\cite{Wainwright:2011kj}, after a qualitatively similar
algorithm was proposed in ref.~\cite{Cline:1999wi}. Here we want to briefly describe the
method for a given phase pair and for a given effective potential $V$.

In order to obtain $\Gamma(T)$ following eq.~\eqref{eq:nucrate}, the bounce action needs to be
computed, which itself requires the tunnelling path in field space, along which the
transition will take place. This tunnelling path starts at the false vacuum and stops at a
release point, which is often close to the true vacuum; the bounce itself describes the
equations of motion of the scalar fields between the false vacuum and the release point,
from where on the fields roll down to reach the true vacuum state. For a given temperature
at which we want to compute the bounce action, we thus iterate over different paths in
order to find the one which minimises the bounce action.

We start by parameterising the tunnelling path $\bm{\phi}(x)$ as a path through field space,
where $x$ is a parameter indicating the position along the path. For simplicity, we
require $x$ to be normalised such that $|\dd \bm{\phi} / \dd x| = 1$, i.e.~that one unit of
$x$ corresponds to one unit of path length in field space. This reparameterisation allows
us to split the bounce eq.~\eqref{eq:bounce-equation} into a part parallel and
a part perpendicular to the tunnelling path,
\begin{align}\label{eq:eom-bounce-equation}
  \frac{\dd^2x}{\dd r^2} + \frac{2}{r} \frac{\dd x}{\dd r} = \frac{\partial}{ \partial x}V(\bm{\phi}(x), T) \qquad \text{and} \qquad
  \frac{\dd^{2} \bm{\phi}}{\dd x^2} \left( \frac{\dd x}{\dd r} \right)^2 = \bm{\nabla}_{\perp} V(\bm{\phi}, T)\,,
\end{align}
see for instance ref.~\cite{Athron:2023xlk} for a derivation. Starting with an initial
guess for the path (a straight line), the bounce equation is solved using a shooting
algorithm along this path, yielding $\bm{\phi}(x(r))$. The algorithm then computes the normal
force
\begin{align}\label{eq:bounce-normal-force}
  \bm{N} = \frac{\dd^{2}\bm{\phi}}{\dd x^{2}} \left(\frac{\dd x}{\dd r}\right)^{2} - \bm{\nabla}_{\perp}V(\bm{\phi}, T)
\end{align}
and deforms the path $\bm{\phi}(x)$ until the normal force vanishes,
$\bm{N} = 0$. This path is then again fed into the shooting
algorithm to infer the function $x(r)$, describing how the path evolves with
respect to the radial coordinate. These two steps are
iterated until the path converges. The solver then returns the bubble
profile $\bm{\phi}(r)=\bm{\phi}(x(r))$ as well as the bounce
action $S_3$ computed following eq.~\eqref{eq:bounce-action-O3}.

The accuracy of the shooting part of the algorithm is governed by
two parameters: \texttt{xtol} controls the tolerance on the field
release point $\bm{\phi}_{0}$ from which the bounce is shot, while
\texttt{phitol} bounds the integration error when solving the
parallel component of the bounce eq.~\eqref{eq:eom-bounce-equation}.
The path deformation has its own pair
of parameters. Its convergence is decided by \texttt{fRatioConv}:
the algorithm terminates once the largest normal force along the
path, divided by the largest potential gradient, has fallen below
this threshold. The size of the deformation applied at each
iteration is set by \texttt{startstep}; if it is chosen too large,
the deformation overshoots and amplifies deviations from the
correct path rather than damping them. The full settings that
determine the accuracy of the bounce solver
and path-deformation algorithm can be found in
table~\ref{tab:tunneling_params} in Appendix~\ref{sec:error_codes}.

Given the bounce action $S_3$ at a given temperature, we can
evaluate the bubble nucleation rate between the false and true vacuum
using eq.~\eqref{eq:nucrate}, which describes the probability of
nucleating a bubble in a given volume within a certain time
interval. The relevant volume and time scales for cosmological
phase transitions are given by the Hubble parameter,
which itself depends on the presence of the true vacuum.
An algorithmic solution to this problem is the topic of the
following subsection.

\subsection{Computation of the true-vacuum fraction}
\label{sec:false-vacuum-fraction}

We now come to one of the core parts of our improvements with
respect to existing codes
and the original \texttt{CosmoTransitions} module~\cite{Wainwright:2011kj}:
Eq.~\eqref{eq:false-vacuum-time} identifies the false-vacuum fraction
$P_\text{f}(t) = \exp[-I(t)]$ with the exponential of minus the percolation integral
$I(t)$, so that the true-vacuum fraction reads
$P_\text{t}(T) = 1 - \exp\{- I[H[P_\text{t}]](T)\}$. This implicit equation is
notoriously difficult to solve numerically:
$I$ is a double integral over the cosmic expansion history, which itself depends on the
transition history via $P_\text{t}$. The situation is further complicated by the
time-temperature relation depending on the equation of state in the false vacuum and by
the large numerical cost of evaluating the bubble nucleation rate: 
A single multi-field bounce evaluation through the expensive shooting
and path-deformation algorithms described above can take up to $1\,\text{min}$ on a
modern CPU, such that computing $\Gamma(T)$ on a fine grid is computationally prohibitive.
An efficient approach which
minimises the number of bounce evaluations is hence needed in order to compute the
percolation integral with sufficient accuracy and in order to perform large-scale scans
over model parameter spaces. Since the percolation temperature is not known a priori, the
upper boundary of the integration domain is not fixed, adding further to the described
difficulty.

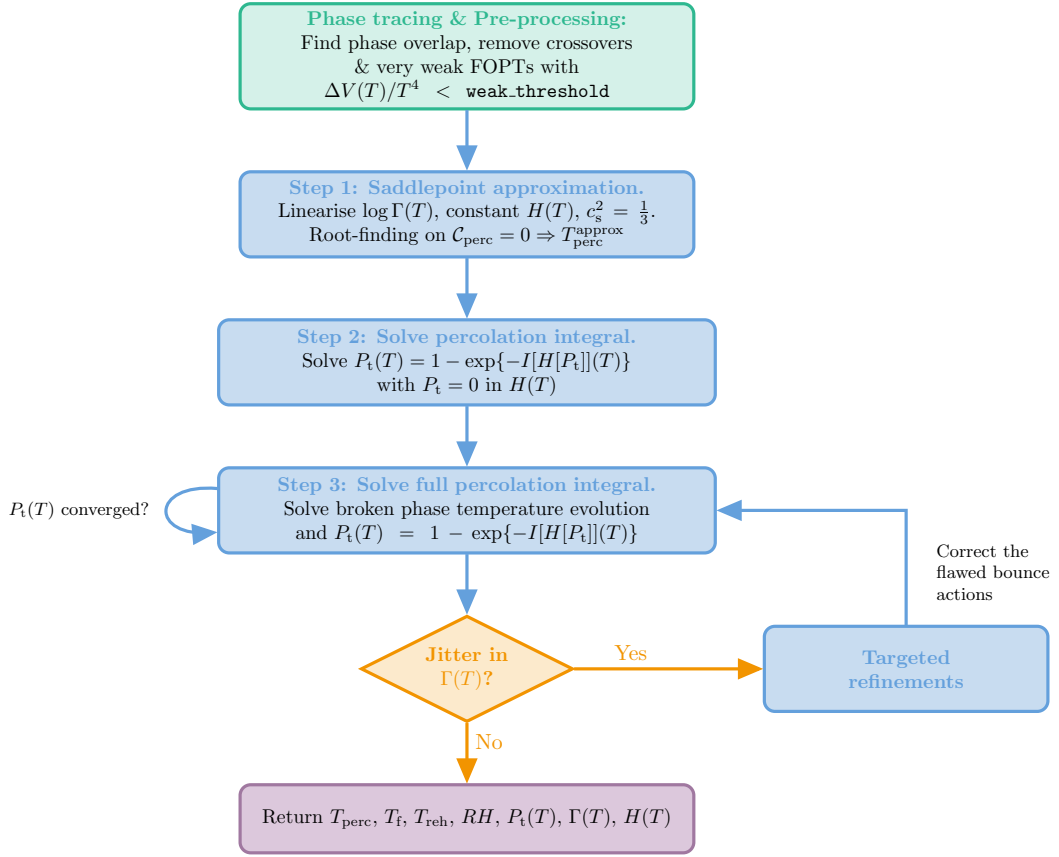
\begin{figure}[t]
  \centering

  \begin{tikzpicture}[node distance=0.6cm, auto, scale=0.75, every node/.style={scale=0.75}]

  % Nodes
  \node (preprocess) [preprocessing] {
      \textbf{\color{greenbox}Phase tracing \& Pre-processing:} \\
      Find phase overlap, remove crossovers \\
      \& very weak FOPTs with $\Delta V(T) / T^4 < \texttt{weak\_threshold}$
  };

  \node (step1) [process, below=0.8cm of preprocess] {
      \textbf{\color{bluebox}Step 1: Saddlepoint approximation.} \\
      Linearise $\log \Gamma(T)$, constant $H(T)$, $c_\text{s}^2 = \frac{1}{3}$. \\
      Root-finding on $\mbox{$\mathcal{C}_{\text{perc}} = 0 \Rightarrow T_{\text{perc}}^{\text{approx}}$}$
  };

  \node (step2) [process, below=0.8cm of step1, text width=8.5cm] {
      \textbf{\color{bluebox}Step 2: Solve percolation integral.} \\
      Solve $\mbox{$P_\text{t}(T) = 1 - \exp\{-I[H[P_\text{t}]](T)\}$}$\\
      with $\mbox{$P_\text{t} = 0$}$ in $H(T)$
  };

  \node (step3) [process, below=0.8cm of step2, text width=8.5cm] {
      \textbf{\color{bluebox}Step 3: Solve full percolation integral.} \\
      Solve broken phase temperature evolution \\
      and $P_\text{t}(T) = 1 - \exp\{-I[H[P_\text{t}]](T)\}$
  };

  \node (jitter) [decision, below=0.8cm of step3] {
      \textbf{\color{orangediamond}Jitter in} \\
      \textbf{\color{orangediamond}$\Gamma(T)$?}
  };

  \node (refinements) [process, right=2.5cm of jitter, text width=4.5cm, minimum width=5cm] {
      \textbf{\color{bluebox}Targeted} \\
      \textbf{\color{bluebox}refinements}
  };

  \node (return) [terminator, below=0.8cm of jitter] {
      Return $T_{\text{perc}}$, $T_\text{f}$, $T_{\text{reh}}$, $RH$, $P_\text{t}(T)$, $\Gamma(T)$, $H(T)$
  };

  % Arrows
  \draw [arrowblue] (preprocess) -- (step1);

  \draw [arrowblue] (step1) -- (step2);
  \draw [arrowblue] (step2) -- (step3);
  \draw [arrowblue] (step3) -- (jitter);

  \draw [arroworange] (jitter.east) -- node[pos=0.3, above] {\color{orangediamond}Yes} (refinements.west);

  \draw [arrowblue] (refinements.north) |- (step3.east);

  \draw [arroworange] (jitter) -- node[pos=0.3, right] {\color{orangediamond}No} (return);

  % Loop arrows
  \draw [looparrow] (step3.175) .. controls +(175:1.2cm) and +(-175:1.2cm) .. (step3.-175);

  % Standalone text nodes (not attached to arrows)
  \node [right=0.3cm of refinements.north, font=\footnotesize, text width=2.5cm,
  anchor=south west, yshift=0.3cm] {Correct the flawed bounce actions};

  \node [left=0.8cm of step3, text width=3.8cm, font=\footnotesize, align=right] (cond3) {
      $P_\text{t}(T)$ converged?
  };

  \end{tikzpicture}

  \caption{Flowchart showing the structure of the percolation algorithm.}
  \label{fig:flowchart-percolation}
\end{figure}

In practice, we solve the integral using a mixture of numerical ODE solvers (instead of
performing a double integral) and Picard sweeps (in order to iteratively solve the self-consistency
equation on $P_\text{t}$) once an approximate solution to the problem has been found.
Fig.~\ref{fig:flowchart-percolation} illustrates the steps involved in this approach.
The remainder of this subsection will describe each of the illustrated steps.

\paragraph{Phase tracing and pre-processing.} Before starting the percolation algorithm,
we perform phase tracing and pre-select relevant transitions: often one is only interested
in phase transitions that emit phenomenologically relevant GW signals. Phase transitions
that occur only over a very short timescale and release little energy are identified and
discarded if the potential difference $\Delta V(T)/T^{4} < \texttt{weak\_threshold}$ in the
temperature range of coexistence between the two phases. By default this value is set to
\texttt{weak\_threshold} = $10^{-4}$, other values can be set by the user. Internally
these transitions are then treated as smooth second-order transitions happening at the critical
temperature.

\paragraph{Step 1: Saddlepoint approximated $T_\perc$.}
The first step of the percolation algorithm uses the saddlepoint approximation
of the percolation integral $I(T)$ to find an
initial estimate for $T_\perc$. We expand $\ln \Gamma(T)$ up to the linear term,
cf.~eq.~\eqref{eq:beta}, approximate $H(T) = H(T_{\perc})$ as constant, set
$c_\text{s}^2 = 1/3$, and push the upper limit of the percolation integral from
$T_\text{max}$ to infinity. This allows an analytic evaluation of the percolation integral,
\begin{align} \label{eq:percIntegralApprox}
  I(T) \simeq \frac{4 \pi v_\text{w}^3}{3} \frac{\Gamma(T)}{H^4(T)} \int_0^\infty
  \diff u \, u^3 \mathrm{e}^{- u \, b(T)}
           = \frac{8 \pi v_\text{w}^3}{b^4(T)} \frac{\Gamma(T)}{H^4(T)}\,.
\end{align}
Each evaluation of $I(T)$ therefore only requires a single evaluation
of the nucleation rate at $T$ and one more at a
neighbouring temperature to estimate the logarithmic slope
$b(T) = - T \,\Delta \log \Gamma / \Delta T$. In the limit of $\Delta T \to 0$ and at
$T = T_\perc$ one recovers $b(T_\perc) = \beta/H$, cf.~eq.~\eqref{eq:beta}. Equating
eq.~\eqref{eq:percIntegralApprox} with
$I(T_\perc) \overset{!}{=} -\ln\bc{1 - f_\text{perc}} \simeq 0.342$ then yields the simple
criterion
\begin{align}
  \mathcal{C}_\perc(T) = \frac{1}{b(T)} \ba{\frac{\Gamma(T)}{H^4(T)}}^{1/4}
  \ba{\frac{8 \pi v_\text{w}^3}{I_\text{perc}}}^{1/4} - 1\,,
\end{align}
which vanishes at $T = T_\perc^\text{approx}$. We use this to infer a first
guess for the percolation temperature, noting that it becomes increasingly inaccurate for the most
interesting cases of slowly varying $\Gamma(T)$, i.e.~whenever $\beta/H$ (and hence
$b(T)$) is small or even negative. All bounce action evaluations performed in this and the following
steps are cached in memory and reused in subsequent stages.

\paragraph{Step 2: ODE solutions to the simplified percolation integral.}
The second step solves the percolation integral with the approximation $P_\text{t} = 0$ in the
Hubble rate, i.e.~by approximating $\bar{\rho} \approx \rho_\text{f}$. Currently
two methods are implemented to solve the percolation integral in
eq.~\eqref{eq:percol-integral}. The first one
($\texttt{algorithm\_mode}=\texttt{fixed\_step\_size}$) starts by
searching for an approximation $T_\text{nuc}^\text{approx}$ of the nucleation temperature 
using a criterion $\mathcal{C}_\text{nuc}(T) = \log \Gamma/H^4$, similar to $\mathcal{C}_\text{perc}$.
Using $T_\text{nuc}^\text{approx}$ as a starting point, it
simply evaluates the nucleation rate on a grid with
fixed step size $\Delta T$, determined by the distance between $T_{\mathrm{nuc}}^\mathrm{approx}$ and
$T_{\mathrm{perc}}^{\mathrm{approx}}$. At each step $T_i$, 
the percolation integral is solved until $I(T)$ is sufficiently large, such that
$P_\text{t} = 0.99$ is reached, meaning the final temperature $T_\text{f}$ has been found.
This method is very robust. As it, however,
relies on a fixed step size $\Delta T$ it can become computationally expensive
for transitions with large temperature hierarchies or if the nucleation temperature
cannot be inferred.

The second method (\texttt{adaptive\_step\_size}) reformulates the
percolation integral into a system of coupled ordinary
differential equations and integrates them with a variable temperature step, depending
on the size of the integrand.
Eq.~\eqref{eq:percol-integral} can be expressed as $I(T) = \frac{4\pi}{3} v_\text{w}^3 J_3(T)$,
where
\begin{align}
  J_n(T) \equiv \int_{T}^{T_{\text{max}}} \diff T^\prime \, \gamma(T^\prime) R(T, T^\prime)^n \,
  , \quad \text{with} \qquad \qquad  \qquad \qquad \qquad  \\
  \gamma(T)  \equiv \frac{a^3(T) \Gamma(T)}{3 c_\text{s}^2(T) H(T) T}\,,\quad
  R(T, T^\prime) \equiv \int_{T}^{T^\prime} \diff \tilde{T} \, \nu(\tilde{T})\,,\quad \text{and} \quad
  \nu(T) \equiv \frac{1}{3 c_\text{s}^2(T) H(T) T a(T)}\,.
\end{align}
Using the Leibniz integral rule and observing that $\partial_T R(T,T^\prime) = - \nu(T)$, the derivative of $J_n$ reads
\begin{align}
  \frac{\partial J_n(T)}{\partial T} = \begin{cases}
    - \gamma(T) \quad & \text{if } n = 0\,,\\
    - n\,\nu(T)\,J_{n-1}(T) \quad & \text{if } n \ge 1\,.
  \end{cases}
\end{align}
The function $J_{3}(T)$ and hence $I(T)$ can then be obtained numerically
by integrating the above system of ODEs
with the initial conditions $J_n(T_{\text{max}}) = 0$, corresponding to a vanishing bubble
nucleation rate at $T_\text{max}$, using an adaptive step size solver. This method is faster on average, but
more prone to numerical instabilities if the bounce action is noisy.

\paragraph{Step 3: Solution of the full percolation integral.} In the previous steps, we
set $P_\text{t} = 0$ inside the Hubble rate. In this last step we also compute
the evolution of the true-vacuum energy density $\rho_\text{t}$
following eq.~\eqref{eq:truevacenergy}, 
based on the true-vacuum fraction from the previous step. By
iterating over this step (called a Picard sweep), a self-consistent solution
to $P_\text{t} = 1 - \exp \bc{- I[H[P_\text{t}]](T)}$ is found. We stop the iteration when 
the relative changes of the percolation temperature, the final temperature and the mean
bubble separation meet the user-defined accuracy goals: $\delta T_\perc < \texttt{acc\_tperc}$,
$\delta T_\text{f} < \texttt{acc\_tfinal}$ and $\delta RH < \texttt{acc\_rh}$. In 
practice we use percent-level accuracies, such that the solution of the full
percolation integral, including the previous steps typically requires less than
100 bounce action evaluations.

\paragraph{Bounce action accuracy control.}
In rare cases the default path-deformation or tunnelling accuracies are not sufficient to
produce a smooth bounce action profile, leading to jumps in $S_{3}(T)$. The algorithm
therefore runs a smoothness test on $\Gamma(T)/H^{4}(T)$ after step~3. It fits a quadratic
polynomial through the evaluations of $\log_{10}(\Gamma/H^4)$ and checks if the deviation of
any point exceeds the threshold \texttt{jitter\_GH4\_threshold}, by default set to one
order of magnitude. If so, the solution is flagged as potentially unreliable and, if
\texttt{jitter\_rescue} is enabled, it recomputes the bounce action with higher accuracies.
Eventually, the obtained percolation temperature $T_\perc$, the reheating temperature
$T_\text{reh}$ and the final temperature $T_\text{final}$,
as well as the nucleation temperature $T_\text{nuc}$ (based on the full
expression in eq.~\eqref{eq:bubblenumberNuc}) and the mean bubble separation $RH$
are returned.

\subsection{Thermodynamics, GW spectra and observability}
\label{sec:gw-observability-comp}

Once the function \texttt{calcPercAndEvolve} has returned the converged set of milestone
temperatures listed above, four steps of the computation of
observables and their detectability remain, following the description from
sec.~\ref{sec:2}. These steps only depend on a small number of optional user input
collected in the settings \texttt{GWConf}; most of them are switches to base the
computation of the GW signal on one or the other convention.

\paragraph{Thermo- and hydrodynamic quantities.}
The module \texttt{transitionObservables.py} computes the thermo- and hydrodynamic
quantities that enter the GW templates at the percolation temperature. The transition
strength $\alpha$ is obtained from the pseudo-trace of the energy-momentum tensor using the
symmetric- and broken-phase sound speeds returned by the hydrodynamics module,
cf.~eq.~\eqref{eq:pt-strength}. The mean bubble separation $RH$ follows straight from the
stored splines, solving the integral in eq.~\eqref{eq:mean-bubble-separation}. The inverse
duration $\beta/H$ is evaluated in two independent ways, following eq.~\eqref{eq:transition-speed}
and eq.~\eqref{eq:betaH_RH}. We refer to
them as \texttt{betaH\_S3} and \texttt{betaH\_RH} inside the code, respectively.
As a sanity check, we cross-check the two quantities internally, as they
should be equivalent in the limit of fast transitions, see also appendix
\ref{app:betaH-RH} and fig.~\ref{fig:bubble-separation}. The default value of the switch
\texttt{GWConf.use\_mean\_bubble\_separation=True} sets the GW signal computation to use $RH$
instead of $\ba{\beta/H}_{S_3}$.
By default, \texttt{sound\_speed="compute"} is active, such that the speed
of sound in both phases is computed numerically. Alternatively, the user can set an
arbitrary value, with $0 < c_\text{s} \leq 1$ in both phases.

The reheating temperature $T_\reh$ follows from local energy
conservation across the transition, see the discussion below eq.~\eqref{eq:truevacenergy},
where the field-independent energy contributions in both phases are fixed through the effective
degrees of freedom\footnote{In practice, we use splines based on tabulated data for the degrees 
of freedom in the SM~\cite{Husdal:2016haj, Saikawa:2018rcs}. If the phase transition also gives mass
to SM particles, we only use the splines for temperatures up to half the lowest, nonzero SM particle
mass from the model file. In the 2HDM model file for instance, the $\tau$ lepton is the lightest
particle present in the model file. For temperatures below $m_\tau/2 = 0.9 \, \text{GeV}$,
we hence use the SM spline for describing the degrees of freedom in the plasma. For temperatures
above $m_\tau/2$, also additional contributions from the SM and BSM particles described in the model file are
taken into account.} $g_\text{eff}$ and $h_\text{eff}$. 

The efficiency coefficients $\kappa_\text{col}$, $\kappa_\text{sw}$, and
$\kappa_\text{turb}$ are evaluated within
\texttt{hydrodynamics.py} from $\alpha$ and $v_\text{w}$ based on the computations described in
sec.~\ref{subsec:hydro}. Alternatively the user can control $v_\text{w}$ through changing
\texttt{wall\_velocity="LTE"} to any number between 0 and 1. By default, we
deactivate contributions from bubble wall collisions through \texttt{bw\_collisions="off"}.
The bubble wall collision switch instead can take the values
\texttt{"full"} (meaning $\kappa_\phi =1$) or \texttt{"NLO"}, in which case eq.~\eqref{eq:coll} is
used to infer the bubble wall contributions to the GW signal.
The relative efficiency $\epsilon_\text{turb}$ of turbulence, introduced in
eq.~\eqref{eq:turb}, is controlled by \texttt{epsilon\_turbulence}.

\paragraph{GW spectra and the observability module.}
Within \texttt{gwfopt.py}, the thermodynamic observables and the hydrodynamic variables are
used to assemble the total GW spectrum following eq.~\eqref{eq:GWspec},
\begin{equation}
  h^2\Omega_\text{GW}(f) \;=\; \sum_{i\in\{\text{col},\,\text{sw},\,\text{turb}\}}
  h^2\Omega_{\text{GW},i}(f)\,.
\end{equation}
The result is stored as a \texttt{FOPTspectrum} object.
Within \texttt{observability.py}, the \texttt{FOPTspectrum} is then finally evaluated
against the three observability measures introduced in sec.~\ref{subsec:observability}:
For every detector listed there the SNR and the corresponding boolean
\texttt{detectable\_by\_<detector>} flag (triggered when
$\text{SNR}>\text{SNR}_\text{thr}$) are computed. Further, PTA
log-likelihoods are returned for the NANOGrav 12.5\,yr, and NANOGrav 15\,yr data sets,
as well as the
second IPTA data release\footnote{We return three separate log-likelihoods
for each observatory. As described in ref.~\cite{Bringmann:2026xcx}, the \texttt{ceffyl}
likelihoods used within \texttt{PTArcade} are not reliable for signals far from the PTA signal
region. In order to help sampling algorithms, we smoothly extend the likelihood to a larger range of
frequencies and amplitudes, thus allowing samplers to be guided towards relevant regions of parameter
space. The three likelihoods are referred to as \texttt{PTArcade\_lnL}, \texttt{smoothened\_lnL} and
\texttt{mock\_lnL}, the latter being the lightweight surrogate introduced in ref.~\cite{Bringmann:2026xcx}.},
as well as the contribution of the GW background to $\Delta N_\text{eff}^\text{GW}$. Additionally we save
the spectral peak amplitude and frequency, as well as other follow-up diagnostics when performing parameter scans.

\paragraph{Output scheme and diagnostic flags.}
All derived quantities, i.e.~percolation-stage observables, hydrodynamic
outputs, peak spectral amplitudes and frequencies, SNRs, PTA likelihoods and
$\Delta N_\text{eff}^\text{GW}$, are written to a single standardised
\texttt{txt} table, which is also printed to terminal when using the command
line interface. Further, a series of diagnostic flags
is printed. Among these, \texttt{betaH\_small} and
\texttt{betaH\_very\_small} warn when $\ba{\beta/H}_{RH}$ drops
below 10 and 3, respectively, indicating that the spectral templates are
being extrapolated (far) outside their tested regime. The flag
\texttt{betaH\_mismatch} records a disagreement between
$\ba{\beta/H}_{S_3}$ and $\ba{\beta/H}_{RH}$
if they are more than one order of magnitude apart. 
\texttt{betaH\_nonfinite} signals that the logarithmic derivative of
$S_3/T$ could not be evaluated at $T_\perc$. The flag
\texttt{nucleationRate\_nonexponential} is raised when $\ba{\beta/H}_{S_3} < 0$ is detected.
The warnings \texttt{spline\_tnuc\_unavailable}/\texttt{\_not\_reached}/\texttt{\_failed}
track any odd behaviour found in the computation of $T_\text{nuc}$ from the
percolation splines via $N(T_\text{nuc}) = 1$. If solving this criterion
using Brent's method returned no value, returned an
out-of-range value, or failed outright, respectively, a warning is raised.
Finally, \texttt{not\_T0\_global\_min} is raised when the global minimum 
does not coincide with the found potential minimum at $T_\text{min}$, i.e.~if
the tracked potential is metastable until today. This check is relevant
for instance in computations involving the electroweak phase transition,
which certainly has to have happened until today.

\subsection{Error handling}
\label{sec:percolation-errors}

Any large-scale scan over model parameter spaces has to deal with the fact that the
computation described in section~\ref{sec:false-vacuum-fraction} can fail for physical as
well as numerical reasons, and that distinguishing between the two is essential: a
genuinely trapped or eternally inflating vacuum must not be confused with a merely poorly
sampled bounce, and vice versa. \texttt{TransitionListener} therefore raises a dedicated
exception for each characteristic error and propagates it to the output files.
Table~\ref{tab:errorcodes} in appendix~\ref{sec:error_codes} lists the implemented error
codes.

\section{Scans and plotting methods}
\label{sec:scan-and-plotting}

\texttt{TransitionListener} comes with several scan methods to explore the parameter space of a model. In
this section we discuss how to set up and run a scan using a configuration file and
illustrate each scan method with one example of the implemented models (see
appendix~\ref{sec:models}). An example of how to use the code instead in a \texttt{python} script can
be found in appendix~\ref{sec:python-usage}.

\subsection{Command line interface}
\label{sec:label}

The code provides a command line interface that automatises the computation steps
described in sec.~\ref{sec:3}. It requires to set up a configuration file in the \texttt{.yaml}
format, in which the user specifies a model file which defines the potential and the
particle content of a theory, the scan method and the parameter
values or ranges. Additionally the output format and path as well as descriptions for the
plots can to be specified. In code snippet~\ref{lst:config-yaml} we show a basic example for how
to compute the phase transition signal of a single benchmark point in the dark flipflop model.

\begin{lstlisting}[language=yaml,deletekeywords={[2]{y,v_GeV}},label=lst:config-yaml,
  caption={Anatomy of a configuration file. Here, we fix the input parameters for the
  dark flipflop model to benchmark point 2 from table~\ref{tab:benchmarks}.}]
# Scan settings ====================================
Modelfile:
  models/TL_dark_flipflop.py

Potential:
  DarkFlipFlop

Scan:
  SinglePoint

Parameters:
  lambda0: 0.005098
  lambda1: 0.002144
  lambda12: 0.003078
  v_GeV: 3.728330741601088
  y: 0.972319
  gamma: 0.7532

# ==================================================
# output configuration
# ==================================================
timeout: 300  # Timeout in seconds for the individual TL run, -1 for no timeout
format: txt  # either "txt" or "hdf5"
output_path: scans/exampleScan
description: Example
plot_description: Flipflop benchmark point

additional_plots:
  # Create additional plots
\end{lstlisting}
The scan can be run from the command line as shown below. The option \texttt{-c
  "config.yaml"} sets the location of the scan configuration file and the option
\texttt{-j N} the number of cores to use in multi-point scans.
\begin{lstlisting}[language=bash,label=lst:tl-command-line,
% caption=Use of the command line tool to run a scan with a configuration file \texttt{example.yaml}. With the option
% \texttt{-j N} the number of cores are specified.
]
$ tl -c "example.yaml" -j N
\end{lstlisting}

\subsection{Single point}

\begin{figure}[t]
  \includegraphics[width=\textwidth]{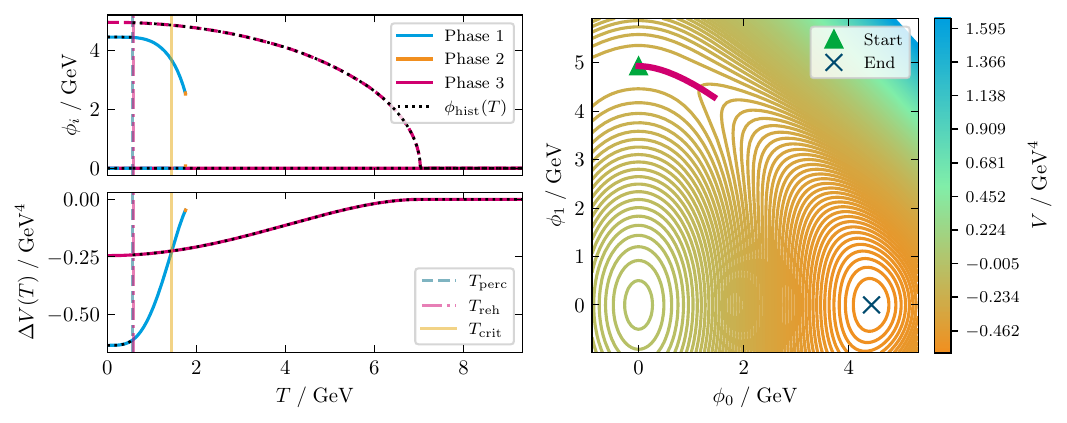}
  \caption{Compilation of the output plots using the \texttt{phases} and \texttt{profileV}
    plotting options for the dark flipflop benchmark point 2 from
    table~\ref{tab:benchmarks}. \textit{Left panels:} Phase evolution $\bm{\phi}(T)$
    and $\Delta V(T) = V(\bm{\phi}_\mathrm{min}(T),T)-V(0,T)$ across the relevant transition
    history. In the upper panel the $\phi_0$ ($\phi_1$) component is
    visualised by a solid (dashed) line. \textit{Right panel:} Potential slice at
    $T=T_{\text{nuc}}$ with the tunnelling path overlaid. From the end of the pink
    line the field rolls to its end point (cross).}
  \label{fig:plot-potential-phases}
\end{figure}

\begin{figure}[t]
  \centering
  \includegraphics[width=0.85\textwidth]{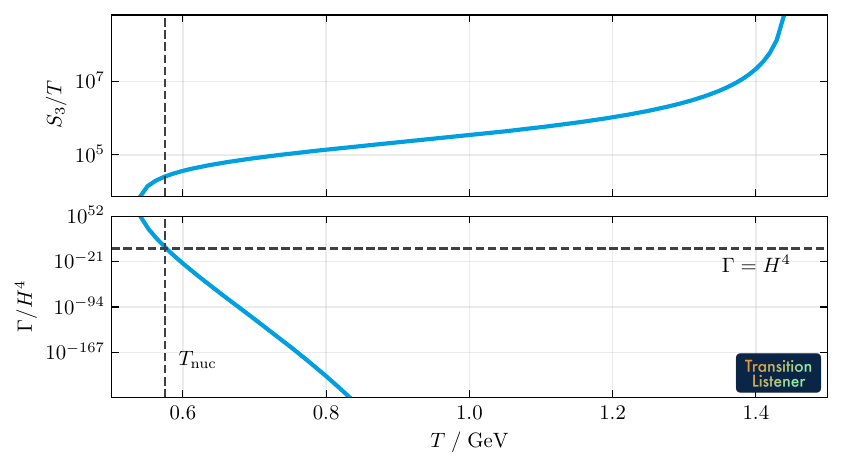}
  \caption{Output of the \texttt{action} plot option for the dark flipflop
  benchmark point 2 from table~\ref{tab:benchmarks}.
  \textit{Upper panel:} Bounce action $S_{3}/T$ in dependence of temperature.
  \textit{Lower panel:} Corresponding bubble nucleation rate $\Gamma$ in units of $H^4$.}
  \label{fig:plot-action}
\end{figure}

\begin{figure}[t]
  \centering
  \includegraphics[width=0.85\linewidth]{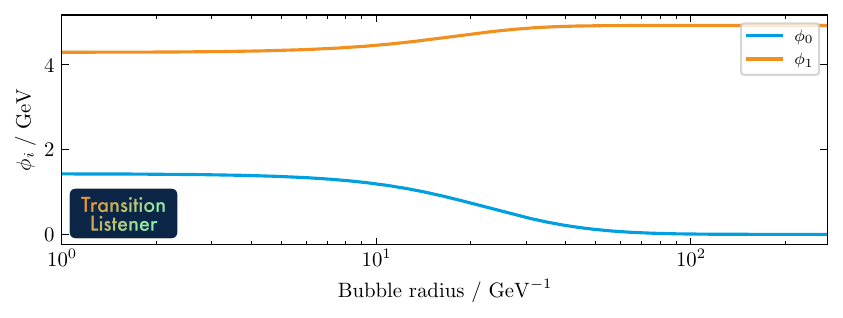}
  \caption{Bubble profile at nucleation for the dark flipflop benchmark point 2 from
  table~\ref{tab:benchmarks}, generated using the \texttt{profile} plot option.
  Inside the bubble, at $r=0$, the fields approach their release point. At
  $r \to \infty$ (outside the bubble), the field values approach their
  false-vacuum values.}
  \label{fig:plot-bubble-profile}
\end{figure}

\begin{figure}[t]
  \centering
  \includegraphics[width=0.9\linewidth]{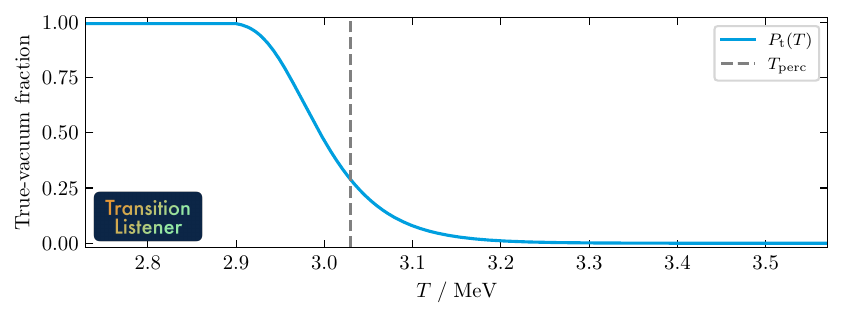}
  \caption{True-vacuum fraction evolution of the conformal dark $U(1)^\prime$
  benchmark point 1 from table~\ref{tab:benchmarks}
  ($g = 0.7$, $v = 0.14\,$GeV and $y = 0.01$), generated using the
  \texttt{percolation} plot option.}
  \label{fig:plot-percolation}
\end{figure}

The first scan method is the \texttt{SinglePoint} mode, which computes the phase transition
GW signal for a fixed set of input parameters, as shown in the example configuration file in code
snippet~\ref{lst:config-yaml} above. Additionally, it provides the possibility to directly plot
several intermediate results of the computation. We describe them below in code
snippet~\ref{lst:plotting}. In the following we showcase these different plotting
routines. Figures~\ref{fig:plot-potential-phases}--\ref{fig:plot-bubble-profile}
use the dark flipflop benchmark point 2 from table~\ref{tab:benchmarks}, while
fig.~\ref{fig:plot-percolation} uses the conformal benchmark point 1.
\begin{lstlisting}[language=yaml,morekeywords={[2]{potential,profileV,phases,action,profile,percolation,gw_spectrum}},deletekeywords={[2]{N}},label=lst:plotting,
caption={Configuration for creating additional plots in the single-point
dark flipflop benchmark example used in
figs.~\ref{fig:plot-potential-phases}--\ref{fig:plot-bubble-profile}.
}]
additional_plots:
  potential:
    plot?: True
    T_GeV: 5.7467e-01
    phi_ranges_GeV: [0, 5, 0, 6]
    n: 100
  profileV:
    plot?: True
    field_index_1: 0
    field_index_2: 1
  phases:
    plot?: True
    include_transitions?: True
  action:
    plot?: True
    Tmin_GeV: 0.5
    Tmax_GeV: 1.5
    phase_indices: [P1, P3]
    n: 100
  profile:
    plot?: True
  percolation:
    plot?: True
  gw_spectrum:
    plot?: True
\end{lstlisting}

\paragraph{Phases and potential.} The first option \texttt{potential} allows the user to plot
the effective potential along chosen field directions at a fixed temperature. The code
evaluates $V_{\tot}(\bm{\phi},T)$ on the user-specified grid. Similarly the option
\texttt{profileV} plots the effective potential at the nucleation temperature, including
the tunnelling path as shown in the right panel of fig.~\ref{fig:plot-potential-phases} for
the example of the dark flipflop benchmark point 2 (see table~\ref{tab:benchmarks}). The
green triangle indicates the false (meta-stable) vacuum from where the field starts to tunnel along
the pink path to the release point (end of the path), from where it then rolls to the stable true vacuum
indicated by a cross.

The left panels show the output of the \texttt{phases} plotting option. The upper left
panel visualises the temperature-dependent field values of all traced minima and
optionally indicates the critical, percolation and reheating temperature through vertical
lines. In the lower left panel, the value of the effective potential (relative to
$V(0, T)$) for each phase is shown. The dotted black line indicates the phase that the
universe is in at a given temperature. Phase 2, present in a tiny temperature interval,
stems from the finite step sizes used in phase tracing. Since the energy released in the
transition from phase 2 to phase 1 is virtually 0
(see lower left panel), it is treated as a crossover transition in the code.

\paragraph{Action and nucleation rate.}

The result of the \texttt{action} plot option is shown in fig.~\ref{fig:plot-action}. It
evaluates the action $S_{3}(T)$ at \texttt{n} points between the specified minimal and
maximal temperature as shown in the upper panel. Additionally it computes the nucleation
rate in units of $H^4$ (in the $P_\text{t}=0$ approximation) and plots it in the lower panel. A
horizontal dashed line indicates the approximate nucleation criterion $\Gamma = H^4$.
Here we again use the dark flipflop benchmark point 2 from
table~\ref{tab:benchmarks}. The two relevant vacua become degenerate at
$T_\text{c} \simeq 1.45\,\text{GeV}$, where the bounce action diverges and the
nucleation rate vanishes. At $T_\text{nuc} \simeq 0.57\,\text{GeV}$ the
nucleation rate crosses the threshold, marking the onset of the transition.

\paragraph{Bubble Profile.}
Whenever the tunnelling solver succeeds, \texttt{TL} stores the bounce profile
$\bm{\phi}(r)$ sampled along the radial coordinate. The bubble profile plot \texttt{profile}
displays the profiles of the individual field components along the radial coordinate
at the nucleation temperature as shown in fig.~\ref{fig:plot-bubble-profile}.
This bubble profile corresponds to the
tunnelling path in field space shown in the right panel of fig.~\ref{fig:plot-potential-phases}.
For $r \to 0$, the fields approach their release point $\bm{\phi}_{0}$, indicated by the end of
the pink line in fig.~\ref{fig:plot-potential-phases}; for $r \to \infty$ (outside the bubble),
the field values approach the false-vacuum values, which are
indicated by a green triangle in fig.~\ref{fig:plot-potential-phases}.

\paragraph{Evolution of the true-vacuum fraction.} The evolution of the
true-vacuum fraction can be visualised with the \texttt{percolation} option.
The resulting plot shows $P_{\mathrm{t}}(T)$ between the nucleation
temperature and the completion temperature. In contrast to
figs.~\ref{fig:plot-potential-phases}--\ref{fig:plot-bubble-profile},
fig.~\ref{fig:plot-percolation} uses the conformal dark $U(1)^\prime$
benchmark point 1 from table~\ref{tab:benchmarks}, with
$g = 0.7$, $v = 0.14\,$GeV and $y = 0.01$. The grey dashed line indicates the
percolation temperature, when $P_\text{t}(T) = f_\text{perc}$.

\paragraph{GW spectrum.}
The \texttt{gw\_spectrum} option evaluates the total GW spectrum and its individual source components
and overlays them with the detector sensitivity curves described in sec.~\ref{subsec:observability}.
Fig.~\ref{fig:gw-benchmarks} shows the results for the benchmark points
listed in table~\ref{tab:benchmarks}.

\subsection{Line scan}

A line scan provides the possibility to study the dependence of the model predictions
in dependence of a single parameter. It allows varying a single model parameter while keeping the
others fixed. The user has to specify the parameter to scan over, its range
$[\texttt{min},\texttt{max}]$, the number of support points $N$ for the scan,
whether the sampling is linear
or logarithmic, as well as the values of the fixed parameters. An example for this is shown
in code snippet~\ref{lst:config-linescan} for the conformal dark $U(1)^\prime$ model. 

\begin{lstlisting}[caption={Parameter configuration of a line scan for the conformal $U(1)^\prime$ model.
  This computes the GW signal for 30 points, with $g \in {[0.5, 1]}$ on a linear grid and fixed values of
  $y = 0.01$ and $v = 0.14\,$GeV.},language=YAML,label=lst:config-linescan]
Parameters:
  line_param:
    g:
      scale: lin
      range: [0.5, 1]
  other_params:
    y: 0.01
    v_GeV: 0.14
  N: 30
\end{lstlisting}

\begin{figure}[t]
  \centering
  \includegraphics[width=0.98\linewidth]{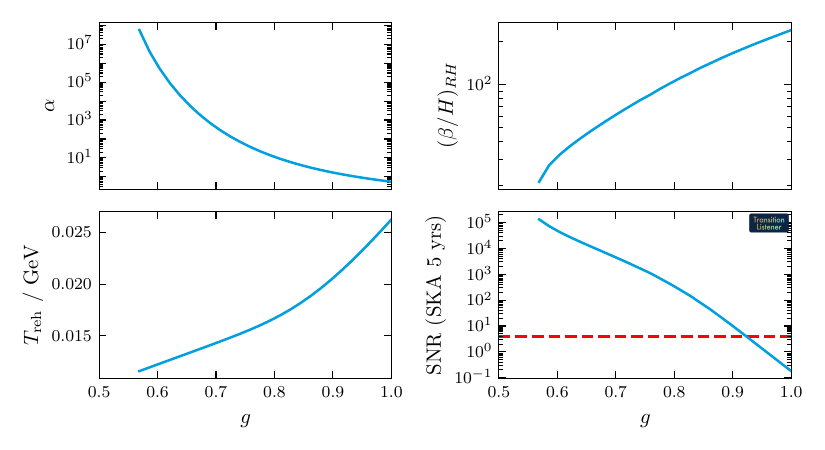}
  \caption{Overview plot of the example line scan based on the
  conformal model near the benchmark point from table~\ref{tab:benchmarks},
  scanning $g$ at fixed $y = 0.01$ and $v = 140 \, \text{MeV}$.
  The fourth panel shows the expected SNR after 5 years of data taking at SKA.
  The red dashed line indicates the detection threshold.}
  \label{fig:line-scan}
\end{figure}

A line scan produces a large set of output files, including the thermodynamic
and hydrodynamic quantities and other intermediate results, plotted against the scanned parameter.
In addition, an overview plot is created, which shows $\alpha$, $\beta/H$,
$T_{\reh}$ and an expected SNR in dependence of the scanned parameter in a
single plot. Fig.~\ref{fig:line-scan} shows the produced overview plot for
the benchmark point 1 from table~\ref{tab:benchmarks}, when varying the
dark gauge coupling $g$. Towards smaller
$g$, the transition becomes more supercooled, up to the point where the universe is
eventually trapped in the false vacuum for $g \lesssim 0.55$. As the
supercooling increases, the transition strength $\alpha$ rises, the 
transition happens later at slightly smaller $T_\text{reh}$ and
$(\beta/H)_{RH}$ drops to values below 100. Correspondingly, the SNR curve
(here taken to be for SKA after 5 years of data taking)
grows far into the testable region, indicating that this model
parameter slice can be tested up to $g \lesssim 0.9$.
For reproducibility, the configuration file to create this
plot can be found in \texttt{examples/example\_line.yaml}.

\subsection{Grid scan}

Grid scans vary two parameters simultaneously while keeping the other parameters fixed.
Analogously to the line scan, the user has to specify the two varying parameters, their
range $[\texttt{min},\texttt{max}]$, the number of steps $N$ and whether the sampling is
linear or logarithmic. This constructs the Cartesian product of the two parameters,
resulting in $N^{2}$ total points. The values of the fixed parameters are specified as before. An
example for this is shown in code snippet~\ref{lst:config-grid} for the dark Abelian Higgs model.

\begin{lstlisting}[language=yaml,deletekeywords={[2]{v_GeV}},label=lst:config-grid, caption=Example of a grid scan configuration.]
Scan:
  GridScan

Parameters:
  grid_params:
    x:
      name: l
      scale: log
      range: [-4, -2]
    y:
      name: v_GeV
      scale: log
      range: [6, 10]
  other_params:
    g_tilde: 2.69
  N: 30
\end{lstlisting}

\begin{figure}[t]
  \centering
  \includegraphics[width=0.98\linewidth]{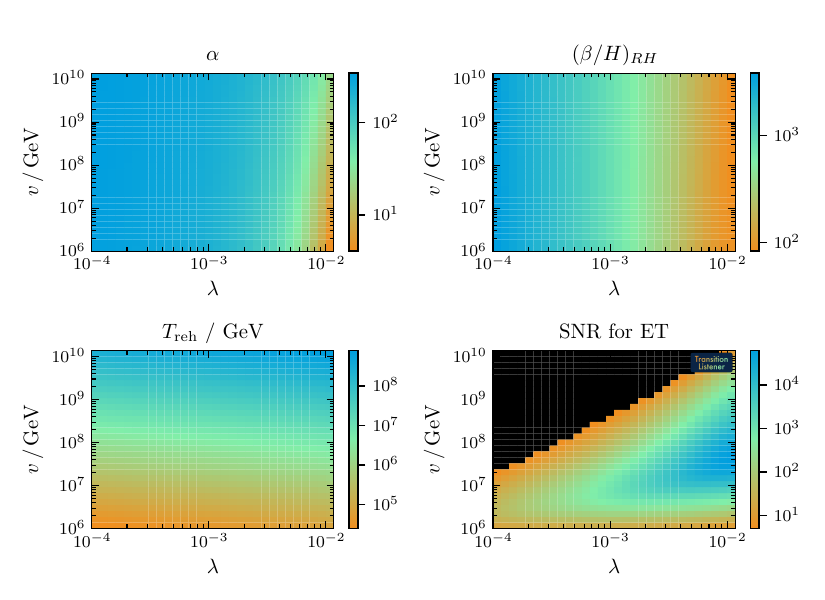}
  \caption{Overview plot of the example grid scan for the Abelian dark Higgs model from appendix~\ref{sec:models}
  scanning over $\lambda\in[10^{-4},10^{-2}]$ and $v\in[10^{6},10^{10}]\,\mathrm{GeV}$ at
  fixed $g\cdot\lambda^{-1/4}=2.69$. The fourth panel shows predicted SNR at the Einstein Telescope.}
  \label{fig:grid-scan}
\end{figure}

Figure~\ref{fig:grid-scan} shows the resulting \texttt{overview\_plot.pdf} of an example grid scan over the
dark Higgs self coupling $\lambda \in [10^{-4}, 10^{-2}]$ and the vev $v \in [10^{6}, 10^{10}]$\,GeV with the gauge coupling fixed 
via the condition $g\cdot\lambda^{-1/4} = 2.69$, see ref.~\cite{Bringmann:2026xcx} for the reasoning behind that scaling.
This overview plot as well as figures for all other computed observables in dependence of the grid parameters can be reproduced using the configuration file
\texttt{examples/example\_grid.yaml}.

\subsection{Random scan and nested sampling}

The random scan method draws parameter points from uniform or logarithmic priors
specified by the user. For Bayesian inference on the PTA likelihoods,
\texttt{TL} interfaces with \texttt{UltraNest}'s
\texttt{ReactiveNestedSampler}~\cite{Buchner:2021cql}. In both cases, the
\texttt{.yaml} file lists the scanned
parameters and their priors. An example configuration file for the conformal
$U(1)^\prime$ model is shown in code snippet~\ref{lst:random-scan-conformal}.
This randomly draws \texttt{N=10000} points from the
priors $g \in [0.5, 1]$, $y \in [10^{-3}, 1]$ and
$v \in [10^{-2}, 10]\,\text{GeV}$, where the latter two are specified in logarithmic scale.

\begin{lstlisting}[language=yaml,label=lst:random-scan-conformal,caption={Example of a random scan for
the conformal dark $U(1)^\prime$ model. The \texttt{UltranestScan} follows the same structure, but
does not use the \texttt{N} parameter as it samples until convergence of the Bayesian posterior
distribution is reached.}]
Scan:
  RandomScan # Or: UltranestScan

Parameters:
  scan_params:
    g:
      scale: lin
      range: [0.5, 1]
    y:
      scale: log
      range: [1e-3, 1e0]
    v_GeV:
      scale: log
      range: [1e-2, 1e1]
  other_params:
    # no other params
  N: 10000 # Only used for RandomScan
\end{lstlisting}

For both the random scan and the nested sampling methods, the scan output is
continuously written to the file
\texttt{output\_table.csv}. \texttt{MPI} environments are supported by
tagging every \texttt{.csv} file with
the local rank of each worker process to avoid concurrent writes. In
appendix~\ref{sec:error_codes} we list all the parameters
available as derived observables which are saved as default output. 

\begin{figure}[t]
   \centering
   \includegraphics[width=0.9\linewidth]{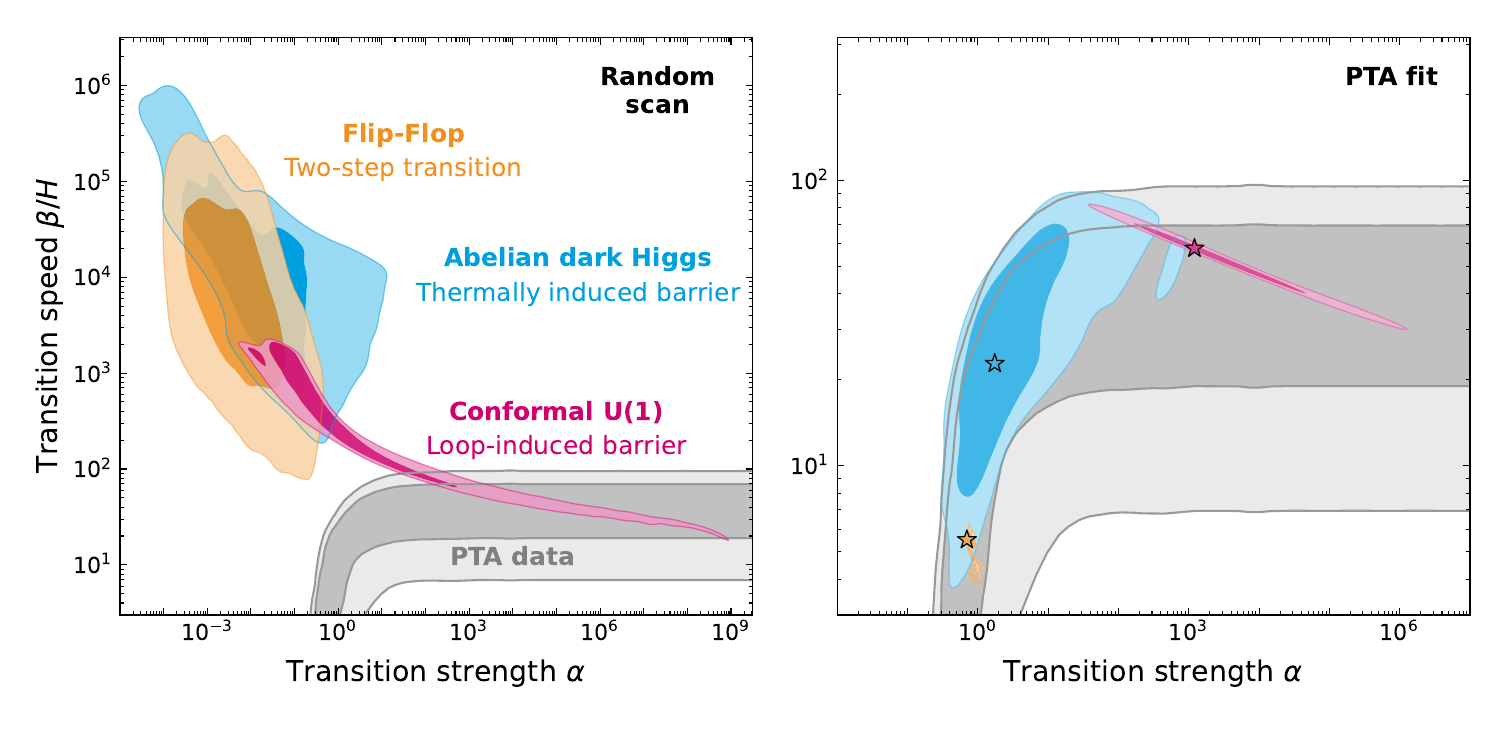}
   \caption{Posterior distributions in the $\alpha$--$(\beta/H)_{RH}$ plane
   for the three dark sector models described in appendix~\ref{sec:models}.
   \textit{Left:} Generic model predictions generated based on flat and log
   priors using \texttt{TransitionListener}'s random scan method.
   \textit{Right:} Parameter regions favoured by the individual models
   when requiring a fit to the NANOGrav 15~yr data set, produced using the \texttt{UltraNest}
   backend in \texttt{TransitionListener}. This figure was presented first
   in a separate study; we refer to ref.~\cite{Bringmann:2026xcx} for further details.}
   \label{fig:comparisonRandomFit}
 \end{figure}

In fig.~\ref{fig:comparisonRandomFit} we compare the 
conformal dark $U(1)^\prime$ model, the dark flipflop model as well as the dark Abelian Higgs model
(see Appendix~\ref{sec:models}) in their capability of explaining the observed PTA
signal. To do so, we compare the resulting transition strength and speed distributions of points
generated using a random scan (\textit{left panel}) and using nested sampling (\textit{right panel})
based on the NANOGrav 15\,yr data set~\cite{NANOGrav:2023gor}. The used methodology as well as the 
underlying parameter ranges have been described by us in a separate publication~\cite{Bringmann:2026xcx}.

\section{Comparison with existing codes (BSMPT)}
\label{sec:6}

The ecosystem of phase-transition tools is split between specialist packages that target a
single subtask (e.g.\ vacuum stability, dimensional reduction, bounce solvers or GW
templates) and all-rounder codes that provide end-to-end pipelines. Specialist tools
include \texttt{vevacious} and \texttt{DRalgo} for effective potentials
\cite{Camargo-Molina:2013qva,Ekstedt:2022bff}, \texttt{AnyBubble},
\texttt{BubbleProfiler}, \texttt{FindBounce}, \texttt{SimpleBounce}, \texttt{OptiBounce}
and \texttt{BubbleDet} for tunnelling
\cite{Masoumi:2016wot,Athron:2019nbd,Guada:2020xnz,Sato:2019wpo,Bardsley:2021lmq,Ekstedt:2023sqc},
\texttt{PTPlot} and \texttt{PTTools} for GW
spectra~\cite{Caprini:2019egz,Caprini:2015zlo}, and \texttt{WallGo} for bubble
wall velocities~\cite{Ekstedt:2024fyq}. Below we focus on the end-to-end tools
most closely related to \texttt{TransitionListener}. Table~\ref{tab:comparison} summarises
the capabilities of these tools and compares them with each other. It includes the code
\texttt{CosmoTransitions}~\cite{Wainwright:2011kj} on which \texttt{TransitionListener} is
built and which first implemented the path-deformation algorithm, the tool
\texttt{BSMPT}~\cite{Basler:2018cwe,Basler:2020nrq,Basler:2024aaf} which focuses on the
electroweak phase transition and extended Higgs sectors,
\texttt{PhaseTracer2}~\cite{Athron:2024xrh}, the Mathematica tool
\texttt{PT2GWFinder}~\cite{Brdar:2025gyo}, and finally \texttt{ELENA}~\cite{Costa:2025pew}
that implements the tunnelling potential to efficiently compute the bounce action for 1D
potentials. We note that \texttt{TransitionListener} is currently the only code available
which is capable of performing computations which self-consistently solve
$P_\text{t} = 1 - \exp \bc{- I[H[P_\text{t}]](T)}$. Other strengths are the capability of
performing computations in more than one field dimension, for transitions happening not
necessarily at the electroweak scale, with a generic time-temperature relation, including
the $T_\text{reh}$ computation using energy conservation instead of a proxy scaling, and
using the LTE bubble wall velocity. On the usability side,
\texttt{TransitionListener} additionally offers a user-friendly interface and
seamless integration with samplers and scanning tools.

{\renewcommand{\arraystretch}{1.15}
\setlength{\tabcolsep}{3pt}
\begin{table}[t]
\noindent
\resizebox{\linewidth}{!}{%
\begin{tabular}{lccccccc|ccccccc}
\toprule
 & \multicolumn{7}{c|}{Features} & \multicolumn{7}{c}{Observables} \\
\cmidrule(lr){2-8}\cmidrule(lr){9-15}
Software
& Lang & MF & TTR & Meth & Reh & Scans & Self-cons.
& $T_*$ & $T_\text{reh}$ & $T_\text{f}$ & $v_\text{w}$ & $\alpha$ & $\ba{\beta/H}_{S_3}$ & $RH$ \\
\midrule

\texttt{TransitionListener v2}
& Py & \fancyc & gen & P+S & \fancyc & \fancyc & \fancyc
& $T_{\rm perc}$ & \fancyc & \fancyc & LTE
& $\frac{\bar{\theta}_{\mathrm{f}}-\bar{\theta}_{\mathrm{t}}}{3w_{\mathrm{f}}(T_{\mathrm{perc}})}$
& \fancyc & \fancyc \\

\texttt{CosmoTransitions}
& Py & \fancyc & -- & P+S & \fancyx & \fancyx & \fancyx
& -- & -- & -- & -- & -- & -- & -- \\

\texttt{BSMPT v3}
& C++ & \fancyc & bag & P+S & (\fancyc) & \fancyc & \fancyx
& $T_{\rm perc}$ & \fancyc & \fancyc & LTE
& $\frac{\theta_{\mathrm{f}}-\theta_{\mathrm{t}}}{4 \rho_{\mathrm{rad}}(T_{\mathrm{perc}})}$
& \fancyc & \fancyc \\

\texttt{PhaseTracer2}
& C++ & \fancyc & -- & P+S & \fancyx & (\fancyc) & \fancyx
& $T_{\rm nuc}$ & \fancyx & \fancyx & input
& $\frac{\theta_{\mathrm{f}}-\theta_{\mathrm{t}}}{\rho_{\mathrm{rad}}(T_{\mathrm{nuc}})}$
& \fancyc & \fancyx \\

\texttt{PT2GWFinder}
& Ma & \fancyx & bag & poly & \fancyx & (\fancyc) & \fancyx
& $T_{\rm perc}$ & \fancyx & \fancyx & input
& $\frac{\theta_{\mathrm{f}}-\theta_{\mathrm{t}}}{4\rho_{\mathrm{rad}}(T_{\mathrm{perc}})}$
& \fancyc & \fancyx \\

\texttt{ELENA}
& Py & \fancyx & gen & tun & (\fancyc) & (\fancyc) & \fancyx
& $T_{\rm perc}$ & \fancyc & \fancyc & $1$
& $\frac{\bar{\theta}_{\mathrm{f}}-\bar{\theta}_{\mathrm{t}}}{3w_{\mathrm{f}}(T_{\mathrm{perc}})}$
& \fancyc & \fancyc \\

\bottomrule
\end{tabular}%
}
\caption{
Comparison of software tools for phase transitions.
Abbreviations: Lang = language (Py = Python, Ma = Mathematica);
MF = multi-field; TTR = time--temperature relation (gen = general, bag = bag model);
Meth = method (P+S = path deformation + shooting,
poly = polygonal bounces, tun = tunnelling potential);
Reh = reheating; Scans = parameter-scan support; Self-cons. = self-consistent
treatment of the Hubble rate and $P_\text{t}$.
Bracketed checkmarks indicate support under additional assumptions or in a restricted sense;
for Reh, this means an instantaneous reheating approximation.
We further point out that BSMPT does not require $v_\text{w} = v_\text{w}^\text{LTE}$
and $T_* = T_\text{perc}$, but also allows other choices.}
\label{tab:comparison}
\end{table}
}

To benchmark \texttt{TransitionListener} against an established end-to-end tool, we
compare \texttt{TL v2.0.0} to \texttt{BSMPT v3.1.8} for the same 2HDM benchmark point
already used in fig.~\ref{fig:gw-benchmarks} and listed in table~\ref{tab:benchmarks},
however scanning over $\lambda_3$ from 5.5 to 6.0. For better comparability we adopted
the same treatment of effective degrees of freedom in the \texttt{TL} model
implementation as \texttt{BSMPTv3} (see appendix of \cite{Basler:2024aaf}) instead of
computing them directly from the effective potential. Further, we fix $v_\text{w} = 1.0$ in
\texttt{BSMPT} instead of $v_\text{w} = 0.95$ employed there by default. Additionally, we
use a minimally modified version of \texttt{BSMPT}, which also returns $RH$ instead of only
$(\beta/H)_{S_3}$ in the output \texttt{.tsv} files, allowing us to reconstruct
$(\beta/H)_{RH}$ for a fair comparison of the two codes. The main
scan uses the default \texttt{TL} precision and is complemented by an increased-precision
edge scan near the strongly supercooled region, with \texttt{converge\_0=1} and
\texttt{fRatioConv=5e-3}. For each point we store $T_\text{crit}$,
$T_\text{nuc}$, $T_\perc$, $T_{\rm reh}$, $\alpha$, $(\beta/H)_{RH}$ (inferred from $RH$ using
eq.~\eqref{eq:RH-betaH}), the expected SNR at LISA, and runtime, and then compare
point-by-point. In the $\lambda_3$ scan shown below we deliberately do \textit{not} enforce the
light Higgs state to have a 125\,GeV mass at every point (which is however in place at
$\lambda_3 = 5.81$, where the largest SNR is obtained), in order to isolate and illustrate
numerical stability when varying only one model parameter.

\begin{figure}[t]
  \centering
  \includegraphics[width=\linewidth]{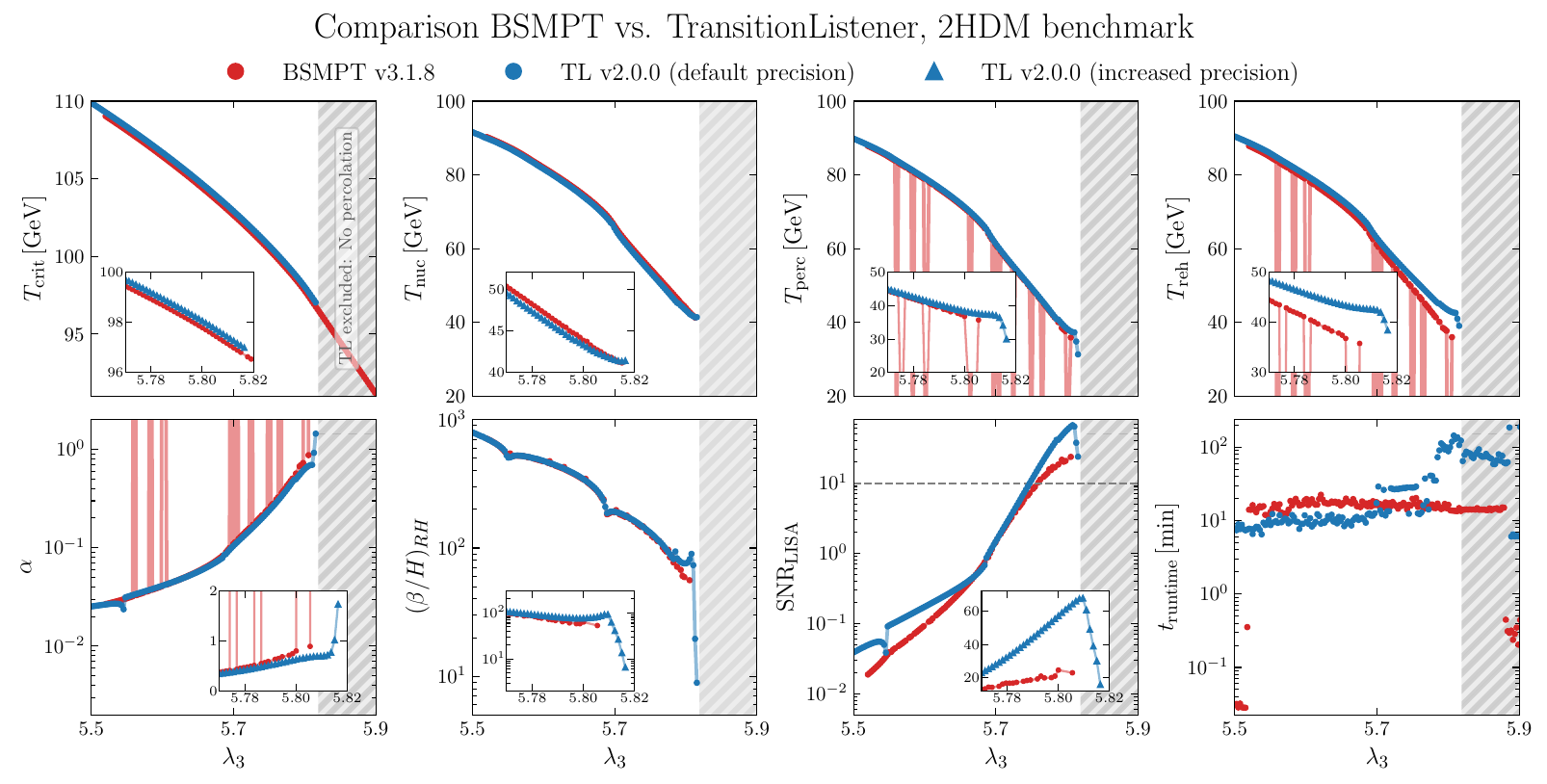}
  \caption{Comparison of a 2HDM benchmark scan between
  \texttt{TransitionListener v2.0.0} and \texttt{BSMPT v3.1.8}. The shaded region marks the \texttt{TL}
  exclusion boundary where vacuum trapping occurs. Requiring a higher tunnelling precision in \texttt{TL}
  (shown in insets) leads to visually vanishing noise levels. The discrepancies
  between both codes are discussed in the main text.}
  \label{fig:comparison_2hdm_obs}
\end{figure}

\begin{figure}[t]
  \centering
  \includegraphics[width=\linewidth]{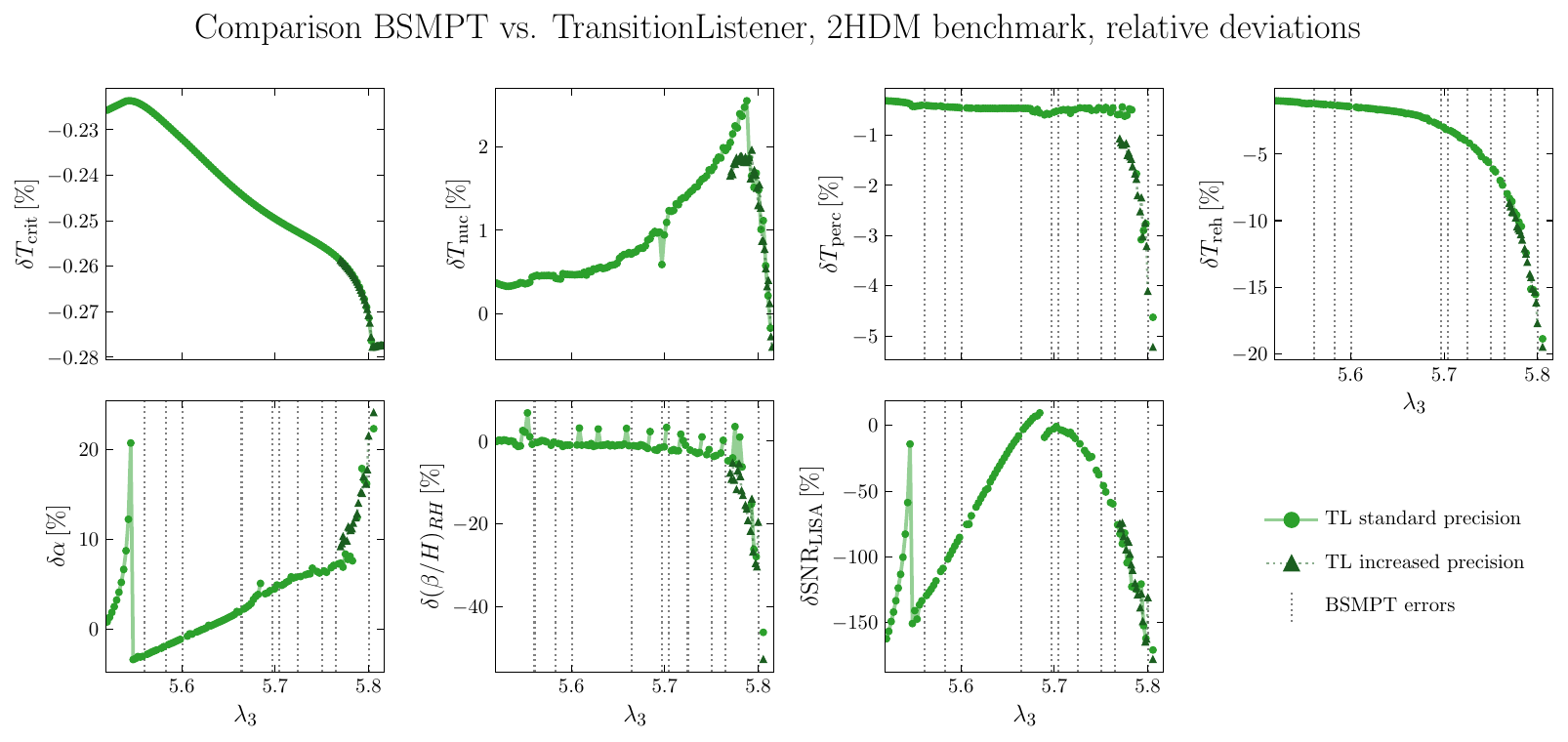}
  \caption{Relative differences between the \texttt{BSMPT} and \texttt{TL} curves shown in
  fig.~\ref{fig:comparison_2hdm_obs}. Dotted vertical lines
  indicate points which we manually identified as unflagged \texttt{BSMPT} errors,
  which we excluded from this comparison. Note that \texttt{BSMPT} does not automatically 
  flag these points as invalid or numerically questionable.}
  \label{fig:comparison_2hdm_rel}
\end{figure}

The comparison in Figs.~\ref{fig:comparison_2hdm_obs} and \ref{fig:comparison_2hdm_rel} is
overall encouraging: both codes show comparable behaviour over most of the scan range, and
the GW signal strengthens monotonically with increasing $\lambda_3$, up until
$\lambda_3 = 5.81$. Beyond that, $(\beta/H)_{RH}$ drops so low that the signal leaves the LISA
frequency band (see SNR panel). At values of $\lambda_3> 5.82$, \texttt{TL} reports vacuum
trapping, whereas \texttt{BSMPT} has no directly analogous warning flag. The steps visible
at $\lambda_3 = 5.55$ and $5.69$ in $\alpha$ and $\beta/H$ are due to $V_\text{daisy}$ obtaining
imaginary parts. We note that \texttt{TL} does not store $T_\text{crit}$ for transitions
that do not complete successfully.

The critical temperature $T_\text{crit}$ agrees at the per-mille level between both codes.
The residual differences are consistent with limited significant digits in the counterterm
inputs. This serves as a consistency check to probe whether the effective potentials which
the two codes use are actually equivalent. Quantitatively, $T_\text{perc}$ typically
agrees at about the $0.5\%$ level and degrades to around $5\%$ for the most supercooled
points, i.e.~the points which LISA will be able to probe best. The reheating temperature
$T_{\rm reh}$ shows larger differences: \texttt{TL} uses local energy conservation, while
the widely used approximation $T_{\rm reh} \approx T_{\perc}(1+\alpha)^{1/4}$ in \texttt{BSMPT} can
induce shifts up to $\sim20\%$, which map directly into the predicted GW peak frequency. For
transitions with $\alpha \gg 1$, this approximation becomes increasingly inaccurate, resulting in
correspondingly large uncertainties in the GW peak frequency. The strength parameter
$\alpha$ is typically larger in \texttt{BSMPT} by up to $20\%$, depending on the specific point
in the parameter space: We identified two effects that contribute to this. The first one
is the definition of $\alpha$, where \texttt{TL} goes beyond the bag equation of state used in
\texttt{BSMPT}. The second one\footnote{We informed the \texttt{BSMPT} authors about these
  mismatches. They confirmed the underlying errors, created a corresponding
  \href{https://github.com/BSMPT/BSMPT/issues/300}{GitHub issue}, and fixed them in
  version 3.2.1. Future versions of this work will incorporate comparisons with an updated
  version of \texttt{BSMPT}.} comes from a bug in \texttt{BSMPT} in the temperature
derivative of the daisy potential, which was missing a term. We find that this alone
contributes an $\mathcal{O}(10\%)$ difference in $\alpha$. The most difficult to compute quantity is
$(\beta/H)_{RH}$: With default settings, it is numerically noisy in both codes. For
$\lambda_{3} \ll 5.8 $, i.e.~for sufficiently weak transitions, the two codes agree, whereas for
stronger supercooling we observe a growing discrepancy of up to 40\%. In these cases
($\beta/H \lesssim 100$), \texttt{TL} in standard precision becomes numerically less reliable, but
still returns points in regions where \texttt{BSMPT} fails to converge. With increased
\texttt{TL} precision (\texttt{converge\_0=1}, \texttt{fRatioConv=5e-3}), the fine-tuned
region can still be resolved down to $\beta/H=\mathcal{O}(5)$. The LISA SNR predictions are overall
comparable, with discrepancies up to roughly a factor of two near the edge. The underlying
reasons for that are the aforementioned shift in the peak frequency (due to generally
larger $T_\text{reh}$ predicted by \texttt{TL}), slightly different LISA noise budgets, as
well as different assumptions on the LISA observation period and the soundwave lifetime
factor $\mathcal{Y}_\text{sw}$ in eq.~\eqref{eq:swlifetime} (compare with eq.~(3.77) in
ref.~\cite{Basler:2024aaf}). Specifically, in \texttt{BSMPT} noise curves from
ref.~\cite{LISA_Science_Requirements:2018} are used, with an assumed observation period of
3 years, while we follow the SNR computations presented in \cite{Breitbach:2018ddu} with a
4 year observation period.

The scan shows several points with numerical instabilities in the results from
\texttt{BSMPT}, visible as $T_\text{perc}=-1\, \text{GeV}$, negative $T_{\rm reh}$ and
$\alpha\sim10^7$. Further, for $\lambda_3$ at the borders of the scanned region, BSMPT fails to
construct an effective potential. Runtime is nevertheless broadly comparable between the
two codes up to moderate supercooling, with \texttt{BSMPT} staying constant in
runtime (roughly 12\,min per point), regardless of the phase transition's properties,
whereas \texttt{TL} shows a more significant increase in runtime from around 9\,min to
about three hours for large supercooling. The latter is due to \texttt{TL} performing more
percolation loops until a specific accuracy target is met.

In summary, we conclude that while runtime still poses severe limitations for
parameter space studies, the comparison with \texttt{BSMPTv3} validates our calculations
at low supercooling, where common approximations valid for weak and sufficiently fast
transitions hold. In the strong
supercooling regime, however, where a more careful
treatment of the reheating dynamics is required, we observe relevant discrepancies
between the two codes, also regarding numerical stability. Having two alternative
implementations provides a valuable cross-check, which will be especially
relevant in coming studies of the strong supercooling regime, 
which will be probed first by upcoming GW
observatories.

\section{Summary}
\label{sec:conclusion}

We have presented \texttt{TransitionListener v2.0}, a comprehensive Python framework
that advances the state of the art in automated calculations for cosmological
first-order phase transitions and their gravitational wave signatures. With
pulsar timing arrays (PTAs) reporting evidence for stochastic GW backgrounds and
LISA on the horizon, translating BSM models into quantitative, testable
predictions has become both urgent and technically demanding.

Version 2 represents a substantial evolution beyond the original code, driven by the
recognition that conventional approximations -- radiation domination, the usual
$\dot{T}=-HT$ time-temperature relationship, $\rho \simeq \Delta V$ thermodynamics, reheating scaling
like $T_\text{reh} \propto (1+ \alpha)^{1/4}$, and fixed nucleation criteria like
$S_3/T \simeq 140$ -- break down where GW signals are strongest and best testable. The
framework now implements a self-consistent treatment of the true-vacuum fraction evolution,
iteratively solving for $P_\text{t}(T)$ while accounting for backreaction on the Hubble
rate, depending itself on the spatially averaged energy density including vacuum
contributions. Reheating is handled using local energy conservation during the transition;
the mean bubble separation $R_{\text{sep}}$ is computed directly from the nucleation
history, enabling faithful mapping to simulation-calibrated GW spectral templates where
estimates based on $(\beta/H)_{S_3}$ break down. Bubble wall velocities are bounded using local
thermal equilibrium matching conditions, and all thermodynamic quantities are derived from
the full effective potential with proper accounting of field-independent radiation terms.

In a comparison with the established code \texttt{BSMPTv3} we show that these
improvements prove critical in the regime of strong supercooling.
We find excellent agreement of the
predicted phase transition parameters $\alpha$ and $(\beta/H)_{RH}$
for intermediate transition strengths of the electroweak phase transition in the real 2HDM.
In the strongly supercooled regime ($\alpha\gg 1$, $\beta/H\lesssim100$),
which upcoming GW detectors will be able to probe first,
the necessity to go beyond the simplifying assumptions mentioned above becomes apparent,
cf.~fig.~\ref{fig:comparison_2hdm_obs}. In the presented comparison we find that
these approximations introduce $\mathcal{O}(1)$ uncertainties in the predicted SNRs,
thus affecting directly the expected observability of strong electroweak transitions with LISA.

Beyond single-point precision, the code enables systematic parameter space
exploration through multiple scan modes (line, grid, random, and nested sampling
via \texttt{UltraNest}~\cite{Buchner:2021cql}), with uniform \texttt{.csv}
output and direct integration
of PTA likelihoods through \texttt{PTArcade}~\cite{Mitridate:2023oar} and
\texttt{Ceffyl}~\cite{Lamb:2023jls}. Built-in
detector sensitivities span twelve orders of magnitude in frequency. The modular
architecture will facilitate the future evolution of the code, accommodating
new hydrodynamic prescriptions, refined GW templates, faster bounce
action evaluations, or dimensionally reduced effective potentials.
We have implemented benchmark models spanning dark sector extensions, extended Higgs
sectors, and conformal scenarios, validated in previous
works~\cite{Bringmann:2026xcx,Balan:2025uke,Bringmann:2023iuz}.
All numerical tolerances are exposed and
documented, allowing users to balance precision against computational cost. The result is
a tool suitable both for precision phenomenological studies and for large-scale inference
that coming GW data releases will increasingly demand.

With gravitational wave astronomy now transitioning from discovery to precision
measurement, \texttt{TransitionListener v2.0} provides the community with a
robust, validated pipeline for connecting fundamental theory to observable
signals. Whether probing TeV-scale electroweak physics, sub-GeV dark sectors, or
exotic multi-step transitions, the framework delivers the numerical accuracy and
physical consistency needed to extract genuine physics from the gravitational
wave backgrounds we are now beginning to observe.

\section*{Acknowledgments}
We thank Felix Kahlhoefer and Fatih Ertas for their assistance during and since the
development of the first version of \texttt{TransitionListener}. We are grateful to our
collaborators Torsten Bringmann, Frederik Depta, Thomas Konstandin, and Kai Schmidt-Hoberg
for their feedback on preliminary results, which significantly improved the code. We
further thank Safa Helal for $\beta$-testing an early version and for developing an SMEFT
model file, to be presented in a future publication. We also thank André Pousette for
valuable input on the use of BSMPT in the 2HDM, as well as finding benchmark point 3
from table~\ref{tab:benchmarks}. We acknowledge helpful discussions with
Lisa Biermann, Francesco Costa, Andreas Ekstedt, Rikard Enberg, Maciek Kierkla, William
Lamb, Michele Lucente, Henda Mansour, Andrea Mitridate, and Jorinde van de Vis.
Computations for this work were performed on the local SOM clusters at the Instituto de
Física Corpuscular (IFIC), including \texttt{Graviton}. JM acknowledges funding by the
Deutsche Forschungsgemeinschaft (DFG) through Grant No. 396021762 – TRR 257. CT
acknowledges support from a Feodor Lynen Fellowship of the Alexander von Humboldt
Foundation, as well as a EuCAPT travel grant that contributed to the development of this
work. CT is supported by the Spanish National Grant PID2022-137268NA-C55 and Generalitat
Valenciana through the grant CIPROM/22/69.

\appendix

\section{Models}
\label{sec:models}

\texttt{TransitionListener} comes with several implemented models that cover single-field,
multi-field and conformal scenarios. Each model is located in \texttt{models/} and
inherits the methods from the base class \texttt{generic\_potential}. In this section we
define the tree-level potentials as well as the mass spectra of the implemented models.
These include a dark photon model, a conformal dark sector,  a model with two dark singlets and
the CP-conserving Two-Higgs Doublet model (2HDM).

\subsection{Dark Abelian Higgs model}

A simple dark photon model is implemented in \texttt{models/TL\_dark\_U1.py}. It contains
a complex scalar field $\Phi$ with self-coupling $\lambda$ and the gauge coupling $g$ to the dark
photon $A^{\prime}_{\mu}$, for more details see ref.~\cite{Ertas:2021xeh}.
The Lagrangian before symmetry breaking reads
\begin{align}\label{eq:lagrangian-dark-u1}
  \mathcal{L} = |D_{\mu} \Phi|^{2} - \frac{1}{4}A^{\prime}_{\mu\nu}A^{\prime\mu\nu} - V_\text{tree}(\Phi),
\end{align}
with the covariant derivative $D_{\mu} = \partial_{\mu} + \mathrm{i} gA_{\mu}^{\prime}$ and the field strength tensor
of the dark photon
$A_{\mu\nu}^{\prime} = \partial_{\mu}A^{\prime}_{\nu} - \partial_{\nu}A^{\prime}_{\mu}$. Upon the breaking of the dark
$U(1)^\prime$ symmetry, the scalar obtains a vev, such that $\Phi = (\phi + v + \mathrm{i}\varphi)/\sqrt{2}$ and the tree-level
potential is given by
\begin{align}\label{eq:potential-dark-u1}
V_0(\phi)&=-\frac{\mu^2}{2}\phi^2+\frac{\lambda}{4}\phi^4 ~~~ \text{ with } ~~~ \mu^2=\lambda v^2\,.
\end{align}
The spectrum in the broken phase consists of a dark Higgs $\phi$, a Goldstone boson $\varphi$, and a dark
photon with transverse $A^{\prime}_\text{T}$ and longitudinal $A^{\prime}_\text{L}$ polarisations. The
field-dependent masses are
\begin{align}
m_\phi^2&=-\mu^2+3\lambda \phi^2+\Pi_\phi,\qquad
m_{A^\prime_\text{T}}^2=g^2\phi ^2,\nonumber\\
m_\varphi^2&=-\mu^2+\lambda\phi^2+\Pi_\phi,\qquad
       \,\,m_{A^\prime_\text{L}}^2 = m_{A^\prime}^2+\Pi_{A^\prime_\text{L}},
\end{align}
with the hard thermal masses
\begin{align}
  \Pi_\phi&=\left(\frac{\lambda}{3}+\frac{g^2}{4}\right)T^2,\qquad ~~\Pi_{A^\prime_\text{L}}=\frac{g^2}{3}T^2\,.
\end{align}
The 1-loop corrections require to specify counterterms which are of the same form
as the tree-level potential in eq.~\eqref{eq:potential-dark-u1},
\begin{align}\label{eq:Vct-dark-u1}
  V_{\mathrm{ct}} = - \frac{\delta\mu^2}{2}\phi^{2} + \frac{\delta\lambda}{4}\phi^{4}\,.
\end{align}
The values of the counterterms are fixed by the renormalisation conditions, which we
implemented to fix the mass of the scalar and the vev to the tree-level values
\begin{align}
  \delta \mu^2 &= \left( \frac{3}{2\phi} \frac{\dd V_{\mathrm{cw}}}{\dd \phi} - \frac{1}{2}
  \frac{\dd^2 V_{\mathrm{cw}}}{\dd \phi^{2}}\right) \Big|_{\phi = v} && \text{and} &&
  \delta \lambda &= \left( \frac{1}{2\phi^3} \frac{\dd V_{\mathrm{cw}}}{\dd \phi} -
  \frac{1}{2\phi^2} \frac{\dd^2 V_{\mathrm{cw}}}{\dd \phi^2}\right) \Big|_{\phi = v}\,.
\end{align}

\subsection{Conformal dark $U(1)^\prime$ model}

The conformal model \texttt{models/TL\_conformal\_dark\_u1.py} realises a radiatively
broken symmetry of a dark Higgs. Furthermore it contains a dark photon and two
left-handed fermions, described by the Lagrangian
\begin{align}\label{eq:lagrangian-conformal}
  \mathcal{L} = |D_{\mu}\Phi |^2  - \frac{1}{4} F^{\prime}_{\mu\nu}F^{\prime\mu\nu}
  + \bar{\chi}_1 \mathrm{i}\slashed{D} \chi_1 + \bar{\chi}_{2} \mathrm{i} \slashed{D} \chi_2 - \frac{y}{2}
  \left(\Phi \bar{\chi}_1^c\chi_1 +  \Phi^* \bar{\chi}_{2}^c\chi_{2} + \text{h.c.}\right)\, - V_\text{tree}(\Phi)
\end{align}
The covariant derivatives read $D_{\mu} = \partial_{\mu} + \mathrm{i} g Q_{i}A^{\prime}_{\mu}$, where
$U(1)^\prime$ charges $Q_i$ are assigned such that the Lagrangian is gauge-invariant: $Q_{\Phi} = 1$,
$Q_{y_{1}} = -1/2$ and $Q_{y_{2}} = 1/2$. 
The tree-level potential exhibits a classical conformal symmetry 
\begin{align}
V_\text{tree}(\phi) &=\frac{\lambda}{4} \phi^{4}\,,
\end{align}
which is broken by the radiative Coleman-Weinberg term at the 1-loop level. We employ
renormalisation conditions ensuring that $m_{\phi}(\phi = 0) = 0$ and define the renormalised
self-coupling $\lambda$ at the renormalisation scale, chosen to be the vev.
This fixes the self-coupling of the Higgs through the other couplings by dimensional
transmutation~\cite{Balan:2025uke}
\begin{align}
  \lambda = \frac{11}{48\pi^2} \left( 10 \lambda^2 + 3 g^4 - y^4 \right)\,.
\end{align}
Therefore all model parameters can be expressed in terms of the gauge coupling $g$,
the Yukawa coupling $y$ and the
vev $v$. For this model it is convenient to combine the counterterms and 1-loop
corrections into one expression. In the model implementation the usual 1-loop term
\texttt{V1} is hence overwritten by a custom function which implements
\begin{align}
  V_{\text{1-loop}}(\phi) = \sum_a \pm  \frac{n_a}{64 \pi^2} \frac{m_a^4(v)}{v^4} \phi^4
 \left[\log \ba{\frac{\phi^2}{v^2}} - \frac{1}{2}\right] \, .
\end{align}
Here the sign is positive (negative) for bosons (fermions) and $n_{a}$ are the respective
internal degrees of
freedom. The sum runs over the particle spectrum, which includes a dark Higgs, a
Goldstone boson, transverse and longitudinal dark photon polarisations,
and the Dirac fermion. The field-dependent
masses are
\begin{align}
m_\phi^2=3\lambda \phi^2+\Pi_\phi \, , &&
m_\varphi^2=\lambda\phi^2+\Pi_\phi \, , && 
m_{A^\prime_{\text{T}}}^2=g^2 \phi^2 \, , && 
m_{A^\prime_{\text{L}}}^2=m_{A^\prime}^2+\Pi_{A^\prime_\text{L}} \,, &&
m_{\psi_{1,2}}^2=\frac{1}{2}y^2 \phi^2,
\end{align}
with the thermal masses
\begin{align}
\Pi_\phi&=\left(\frac{\lambda}{3}+\frac{g^2}{4}+\frac{y^2}{12}\right)T^2,\qquad
\Pi_{A^\prime_{\text{L}}}=\left(\frac{1}{3}+\frac{1}{12}\right)g^2T^2.
\end{align}

\subsection{Dark flipflop model}

The dark flipflop model (\texttt{models/TL\_dark\_flipflop.py}) provides a
potential which can realise a two-step phase transition, see ref.~\cite{Bringmann:2026xcx}.
The potential depends on two real scalar fields $\phi_{1}$ and $\phi_{2}$ and contains
a Majorana fermion $\psi=\psi^{c}$ which only couples to $\phi_{1}$:
\begin{align}\label{eq:flipflop-lagrangian}
  \mathcal{L} = \frac{1}{2}\left(\partial_{\mu}\phi_{1}\right)\left(\partial^{\mu}\phi_{1}\right) +
  \frac{1}{2}\left(\partial_{\mu}\phi_{2}\right)\left(\partial^{\mu}\phi_{2}\right) +
  \frac{1}{2} \mathrm{i} \bar{\psi}\slashed{\partial} \psi - \frac{y}{2} \phi_{1}\bar{\psi} \psi
  - V_\text{tree}(\phi_{1}, \phi_{2}) \, .
\end{align}
The tree-level potential is parameterised in the suggestive form
\begin{align}
  V_\text{tree} = \frac{\lambda_0}{4}(\phi_1^2+\gamma^2\phi_2^2-v^2)^2-\frac{\lambda_1}{2}v^2\phi_1^2
  +\frac{\lambda_{12}}{2}\phi_1^2\phi_2^2 \,,
\end{align}
with minima along the field axes $(\phi_{1}, \phi_{2}) = (\pm v_1, 0) =  (\pm \sqrt{1 + \lambda_{1}/\lambda_{0}\,v}, 0)$ and
$(\phi_1,\phi_{2}) = (0, \pm v/\gamma)$.
The entries of the scalar mass matrix read
\begin{align}
\mathcal{M}_{11}&=2\lambda_0\phi_1^2+\lambda_0(\phi_1^2+\gamma^2\phi_2^2-v^2)-
\lambda_1 v^2+\lambda_{12}\phi_2^2+\Pi_1, \nonumber\\
\mathcal{M}_{22}&=2\gamma^4\lambda_0\phi_2^2+\lambda_0\gamma^2(\phi_1^2+\gamma^2\phi_2^2-v^2)
+\lambda_{12}\phi_1^2+\Pi_2, \quad \text{and} \nonumber\\
\mathcal{M}_{12}&=2\gamma^2\lambda_0\phi_1\phi_2+2\lambda_{12}\phi_1\phi_2,
\end{align}
whereas the fermion has the field-dependent mass $m_{\psi} = y \phi_{1}/2$.
The thermal masses of the scalar fields read
\begin{align}
\Pi_1=\frac{y^2+2\left((3+\gamma^2)\lambda_0+\lambda_{12}\right)}{24}\,T^2 \, ,&& \text{and} &&
\Pi_2=\frac{\gamma^2\lambda_0+3\gamma^4\lambda_0+\lambda_{12}}{12}\,T^2\,.
\end{align}
The imposed renormalisation conditions keep the global vev and the masses of the
two scalars fixed after including the 1-loop corrections. This gives rise to the
counterterm potential 
\begin{align}\label{eq:Vct-flipflop}
  V_{\mathrm{ct}} = - \frac{\delta\mu_{1}^2}{2}\phi_{1}^2 - \frac{\delta\mu_2^2}{2}\phi_{2}^{2}
  + \frac{\delta\lambda_{1}}{4}\phi_{1}^4 \, .
\end{align}
The counterterms are fixed by the renormalisation conditions
\begin{align}
  \label{eq:counterterm-conditions-flipflop}
 \delta \lambda_{1} &= \frac{1}{2 v_1^3} \frac{\partial V_{\mathrm{CW}}}{\partial \phi_{1}}\Big |_{(v_{1}, 0)} -
            \frac{1}{2 v^2_{1}} \frac{\partial^2V_{\mathrm{CW}}}{\partial\phi^2_{2}}\Big |_{(v_{1}, 0)} \, , \\
 \delta\mu_{1}^{2} &= \frac{3}{2 v_{1}} \frac{\partial V_{\mathrm{CW}}}{\partial\phi_{2}}\Big |_{(v_{1}, 0)} -
                      \frac{1}{2} \frac{\partial^2V_{\mathrm{CW}}}{\partial\phi^2_{2}}\Big |_{(v_{1}, 0)}  \, ,
                      \qquad \text{and} \\ 
 \delta\mu_{2}^2 &= \frac{\partial^2V_{\mathrm{CW}}}{\partial\phi^2_{2}}\Big |_{(v_{1}, 0)} \, . 
\end{align}

\subsection{Two-Higgs Doublet model (2HDM)}

The CP-conserving 2HDM \cite{Lee:1973iz,Branco:2011iw} is implemented in
\texttt{models/TL\_2HDM.py} via the \texttt{R2HDM} class. The model encompasses two Higgs
doublets $\Phi_{a} = (\phi_{a}^{+}, \phi_{a}^{0})^{T}$ with $a = 1,2$, transforming under
$SU(2)_{L}\times U(1)_{Y}$. Assuming a softly broken $\mathbb{Z}_{2}$ symmetry
$(\Phi_{1}, \Phi_{2}) \to (-\Phi_{1}, \Phi_{2})$, the tree-level potential of the model is given by
\begin{align}
  \begin{split}
  V_0(\Phi_1,\Phi_2)
  &= m_{11}^2|\Phi_1|^2+m_{22}^2|\Phi_2|^2 - \left( m_{12}^2 \Phi_{1}^{\dagger}\Phi_{2} + \text{h.c.} \right)
    + \frac{\lambda_{1}}{2}|\Phi_{1}|^{4}  + \frac{\lambda_{2}}{2}|\Phi_{2}|^{4} \\
    & ~~~ ~~~+ \lambda_{3}|\Phi_{1}|^{2}|\Phi_{2}|^{2} + \lambda_{4}
    (\Phi_{1}^{\dagger}\Phi_{2})(\Phi_{2}^{\dagger}\Phi_{1})
      + \frac{\lambda_5}{2} \left(  (\Phi_1^{\dagger} \Phi_{2})^2 + \mathrm{h.c.} \right) \,.
      \end{split}
\end{align}
The CP-even components of the two Higgs doublets $\phi_{a}^{0}$ can be decomposed as
\begin{align}
  \Phi_{a} = \begin{pmatrix} 
\phi^{+}_a \\
\phi_{a}^{0}
 \end{pmatrix}
 =
  \begin{pmatrix} 
\phi_{a}^{+} \\
\frac{1}{\sqrt{2}}\left(\phi_{a} + \varphi_{a} + \mathrm{i} \psi_{a}\right)
 \end{pmatrix} \, ,
\end{align}
where $\phi_{a}$ are static background fields that together give the electroweak vacuum
$\left\langle \phi_{1} \right\rangle^{2} + \left\langle \phi_{2} \right\rangle^{2} = v_1^2 + v_2^2 =
v_{\mathrm{EW}}^{2}$ at zero temperature. The implementation is restricted to CP-even
neutral background fields and sets possible intermediate charge- and CP-breaking minima to
zero $\left\langle \phi^{+}_{a} \right\rangle = \left\langle \psi_{a} \right\rangle = 0$.
This yields the tree-level potential of the background fields
\begin{align}
  V_0(\phi_1,\phi_2)
  &= \frac{1}{2}m_{11}^2\phi_1^2+\frac{1}{2}m_{22}^2\phi_2^2
  -m_{12}^2\phi_1\phi_2 +\frac{\lambda_1}{8}\phi_1^4+\frac{\lambda_2}{8}\phi_2^4
   +\frac{\lambda_{345}}{4}\phi_1^2\phi_2^2 \, ,
\end{align}
with $\lambda_{345}=\lambda_3+\lambda_4+\lambda_5$. The mass matrices of the CP-even
($\varphi_{a}$), CP-odd ($\psi_{a})$ and charged scalar ($\phi_{a}^{+}$) read
\begin{align}
\mathcal{M}^2_{\text{even}}&=
\begin{pmatrix}
m_{11}^2+\tfrac{3}{2}\lambda_1\phi_1^2+\tfrac{1}{2}\lambda_{345}\phi_2^2+\Pi_1
&-m_{12}^2+\lambda_{345}\phi_1\phi_2 \\
-m_{12}^2+\lambda_{345}\phi_1\phi_2
&m_{22}^2+\tfrac{3}{2}\lambda_2\phi_2^2+\tfrac{1}{2}\lambda_{345}\phi_1^2+\Pi_2
\end{pmatrix}, \nonumber\\
\mathcal{M}^2_{\text{odd}}&=
\begin{pmatrix}
m_{11}^2+\tfrac{1}{2}\lambda_1\phi_1^2+\tfrac{1}{2}\lambda_{34-5}\phi_2^2+\Pi_1
&-m_{12}^2+\lambda_5\phi_1\phi_2 \\
-m_{12}^2+\lambda_5\phi_1\phi_2
&m_{22}^2+\tfrac{1}{2}\lambda_2\phi_2^2+\tfrac{1}{2}\lambda_{34-5}\phi_1^2+\Pi_2
\end{pmatrix}, \nonumber\\
\mathcal{M}^2_{\text{ch}}&=
\begin{pmatrix}
m_{11}^2+\tfrac{1}{2}\lambda_1\phi_1^2+\tfrac{1}{2}\lambda_3\phi_2^2+\Pi_1
&-m_{12}^2+\tfrac{1}{2}\lambda_{45}\phi_1\phi_2 \\
-m_{12}^2+\tfrac{1}{2}\lambda_{45}\phi_1\phi_2
&m_{22}^2+\tfrac{1}{2}\lambda_2\phi_2^2+\tfrac{1}{2}\lambda_3\phi_1^2+\Pi_2
\end{pmatrix},
\end{align}
with $\lambda_{34-5}=\lambda_3+\lambda_4-\lambda_5$, 
$\lambda_{45}=\lambda_4+\lambda_5$ and the thermal masses 
\begin{align}\label{eq:thermal-masses-2hdm}
  \Pi_{1} &= \left( 12 \lambda_1 + 8\lambda_{3} + 4\lambda_{4} + 3(3 g_2^2 + g_1^2) \right)
  \frac{T^2}{48} + \Pi_{\mathrm{ferm},1}\,, \\
  \Pi_{2} &= \left( 12 \lambda_2 + 8\lambda_{3} + 4\lambda_{4} + 3(3 g_2^2 + g_1^2) + 12 y_{t}^2 \right)
  \frac{T^2}{48} + \Pi_{\mathrm{ferm},2}\,. \\
\end{align}
Here either $\Pi_{1}$ or $\Pi_{2}$ receive a contribution depending on whether the down-type
quarks couple to the first or the second Higgs doublet,
\begin{align}\label{eq:2hdm-thermal-masses-ferm}
  \Pi_{\mathrm{ferm},i}
        &= \begin{cases}
          12 y_{b}^2 \frac{T^2}{48} &\text{ if Yukawa type} = i, \\
          0  &\text{ else}.
          \end{cases}
\end{align}
% \begin{align}\label{eq:2hdm-thermal-masses-ferm}
%   \Pi_{\mathrm{ferm},i}
%         &= \begin{cases}
%           & \left( 24 y_{b}^2 + 8  y_{\tau}^2 \right) \frac{T^2}{48}~~ \text{ if Yukawa type} = i, \\
%           & 0 ~~~~~~~~~~~~~~~~~~~~~~~~~~~~~~~\text{ else}.
%           \end{cases}
% \end{align}
%
The mass matrices can be diagonalised to give the mass eigenstates after symmetry
breaking, which includes the light and heavy Higgs bosons $h$ and $H$, the pseudo scalar
$A$, the neutral and charged Goldstones $G^{0}$, $G^{\pm}$ as well as the charged scalars
$H^{\pm}$. These are related to the gauge eigenstates via
\begin{align}
  \begin{pmatrix} 
G^{\pm} \\
H^{\pm }
 \end{pmatrix} = R(\beta) \begin{pmatrix} 
\phi_1^{\pm } \\
\phi_2^{\pm }
 \end{pmatrix} , \quad
  \begin{pmatrix} 
G^0 \\
A
 \end{pmatrix} = R(\beta) \begin{pmatrix} 
\psi_1 \\
\psi_2
 \end{pmatrix}, \quad
  \begin{pmatrix} 
H \\
h
 \end{pmatrix} = R(\alpha) \begin{pmatrix} 
\varphi_{1} \\
\varphi_{2}
 \end{pmatrix} \, .
\end{align}
The rotation matrices $R$ are of the form
\begin{align}
  R(\beta) = \begin{pmatrix} 
\cos\beta   & \sin\beta \\
-\sin\beta  & \cos\beta
 \end{pmatrix}\, ,
\end{align}
where $\beta$ is fixed by the vev ratio $\tan\beta = v_2 / v_1$. The input
parameters are therefore the quartic couplings $\lambda_{i}$ for $i = 1,\dots,5$, the
soft-breaking mass $m_{12}^2$, $\tan\beta$ and the Yukawa type. CP conservation further restricts
all parameters to be real.
The spectrum of the transversal and longitudinal gauge boson components is
\begin{align}\label{eq:2hdm-gauge-masses}
 m^2_{W_\text{T}} &= \frac{g_2^2}{4}(\phi_1^2+\phi_2^2)\,, \qquad~~~~~~ m^2_{W_\text{L}} = 
 \frac{g_2^2}{4}(\phi_1^2+\phi_2^2) + \Pi_{W_\text{L}} \,.\\
 m_{B_\text{T}}^2 &= \frac{g_1^2+g_2^2}{4}(\phi_1^2+\phi_2^2)\,, \qquad m_{B_\text{L}}^2 =
 \frac{g_1^2+g_2^2}{4}(\phi_1^2+\phi_2^2) + \Pi_{B_\text{L}}\,.
\end{align}
With the thermal masses  $\Pi_{W_\text{L}} = 2 g_{2}^2T^2 $ and $\Pi_{B_\text{L}} = 2 g_{1}^2 T^2$.
From the quark and lepton sector we only account for the most important contributions to
the potential from the top, bottom and tau. Their field-dependent masses are
\begin{align}\label{eq:2hdm-mass-fermions}
   m_t^2 = y_t^2\phi_{i}^2 \, , \qquad m_b^2 = y_b^2\phi_{i}^2 \, , \qquad m_\tau^2 = y_\tau^2\phi_{i}^2 \, ,
\end{align}
where the Higgs doublet they couple to depends on the Yukawa type.

{\sloppy
The counterterms are computed using a dedicated solver in
\texttt{counterterms/}\allowbreak\texttt{twohdm\_solver.py}
For the CP-conserving 2HDM we determine one-loop counterterms in the on-shell
scheme matched to the BSMPT prescription~\cite{Basler:2018cwe}.
\par}
The counterterm potential is
\begin{align}
V_{\text{ct}} &=
\delta m_{11}^2\,\Phi_1^\dagger\Phi_1
+\delta m_{22}^2\,\Phi_2^\dagger\Phi_2
-\delta m_{12}^2\left(\Phi_1^\dagger\Phi_2+\Phi_2^\dagger\Phi_1\right)
\nonumber\\
&\quad+\frac{\delta\lambda_1}{2}\left(\Phi_1^\dagger\Phi_1\right)^2
+\frac{\delta\lambda_2}{2}\left(\Phi_2^\dagger\Phi_2\right)^2
+\delta\lambda_3\left(\Phi_1^\dagger\Phi_1\right)\left(\Phi_2^\dagger\Phi_2\right)
\nonumber\\
&\quad+\frac{\delta\lambda_5}{2}\left[\left(\Phi_1^\dagger\Phi_2\right)^2+
\left(\Phi_2^\dagger\Phi_1\right)^2\right].
\end{align}
The vector of counterterms reads
\begin{equation}
\delta=\left(\delta m_{11}^2,\delta m_{22}^2,\delta m_{12}^2,
\delta\lambda_1,\delta\lambda_2,\delta\lambda_3,\delta\lambda_5\right)^T,
\end{equation}
with $\delta\lambda_4=0$ fixed by the chosen renormalisation prescription.

Derivatives of the Coleman-Weinberg potential at the tree-level minima
are evaluated from cached curvature
tensors (\texttt{models/generated/2hdm\_curvature\_tensors.json}), generated via
\texttt{sympy} if needed. The seven renormalisation conditions are two
tadpole conditions and five conditions on the Hessian of the Coleman-Weinberg potential,
evaluated at $v=(v_1,v_2)$:
\begin{align}
\left.\partial_i\left(V_{\text{CW}}+V_{\text{ct}}\right)\right|_{v}&=0,
\quad i\in\{4,6\}, \nonumber\\
\left.\partial_i\partial_j\left(V_{\text{CW}}+V_{\text{ct}}\right)\right|_{v}&=0,
\quad (i,j)\in\{(4,4),(6,6),(4,6),(5,5),(0,0)\}.
\end{align}
Here, $i, j = 0, \dots, 7$ denote the indices of the fields in the vector
\begin{align}
(\text{Re} \, \phi_1^+, \text{Im} \, \phi_1^+, \text{Re} \, \phi_2^+,
\text{Im} \, \phi_2^+, \phi_1, \psi_1, \phi_2, \psi_2)^T.
\end{align}
We define the Coleman-Weinberg derivative vector entering the linear system as
\begin{equation}
\Delta_{\text{CW}}\equiv
\left(
\left.\partial_4 V_{\text{CW}}\right|_{v},
\left.\partial_6 V_{\text{CW}}\right|_{v},
\left.\partial_4^2 V_{\text{CW}}\right|_{v},
\left.\partial_6^2 V_{\text{CW}}\right|_{v},
\left.\partial_4\partial_6 V_{\text{CW}}\right|_{v},
\left.\partial_5^2 V_{\text{CW}}\right|_{v},
\left.\partial_0^2 V_{\text{CW}}\right|_{v}
\right)^T.
\end{equation}
These are assembled into a linear system
\begin{equation}
A(v_1,v_2)\,\delta=-\Delta_{\text{CW}},
\end{equation}
which is then solved numerically for the counterterms $\delta$.

\section{Energy density and effective degrees of freedom}
\label{app:effective_dof}

Here we provide details about the computation of the energy density, as well as the 
implementation of the effective degrees of freedom, introduced in eq.~\eqref{eq:rho}.

\paragraph{Radiation energy density.}
The 1-loop temperature-dependent corrections $V_T$ within the effective potential in
eq.~\eqref{eq:Vtot} are given in terms of the thermal functions $J_\text{b/f}$. Rewritten in
a more suggestive way, these read
\begin{align}
  J_{\mathrm{b/f}}\left( \frac{m^2}{T^2} \right) = \frac{1}{T^{3}} \int_{0}^{\infty} \dd p\, p^{2}
  \ln \left( 1 \mp \mathrm{e}^{-\sqrt{p^{2} + m^{2}}/T} \right) \, .
  % = \frac{1}{T^{3}} \int_{m}^{\infty} \dd E\, E^{2}
  % \ln \left( 1 \mp e^{-E/T} \right)
\end{align}
Noting that
\begin{align}
  \frac{\dd }{\dd p} \ln \left( 1 \mp \mathrm{e}^{-\sqrt{p^{2}+ m^{2}}/T} \right)
  = \pm \frac{p}{E T} \frac{1}{\mathrm{e}^{E/T} \mp 1} 
\end{align}
and using integration by parts, we can rewrite the thermal functions as
\begin{align}
  J_\text{b/f} \left( \frac{m^2}{T^2} \right)
  &= \mp  \frac{1}{T^{3}} \int_{0}^{\infty} \dd p \, \frac{p^{4}}{3 T E}
  \frac{1}{\mathrm{e}^{E/T} \mp 1}
  = \mp  \frac{2\pi^{2}}{T^{4}} \int \frac{\dd^{3}p}{(2\pi)^{3}}
  \frac{p^{2}}{3 E} \frac{1}{\mathrm{e}^{E/T} \mp 1} \, .
\end{align}
Using eq.~\eqref{eq:thermal-potential} we can easily see that the thermal
contributions describe the pressure of a bath of non-interacting particles
\begin{align}
  V_{T}(\bm{\phi}) =  - \sum_{a} n_{a} \int \frac{\dd^{3}p}{(2\pi)^{3}}
  \frac{p^{2}}{3 E} \frac{1}{\mathrm{e}^{E/T} \mp 1} = -p_{\mathrm{rad}}\,.
\end{align}
Similarly, the energy density calculated from the effective potential simply yields
the analogous expression, still assuming non-interacting particles
\begin{align}
  \rho_{\mathrm{rad}}
  &= V_{T} - T \frac{\partial V_{T}}{\partial T} 
  = \sum_{a}n_{a} \int \frac{\dd^3 p}{(2\pi)^{3}} \frac{E}{\mathrm{e}^{E/T} \pm 1}\,.
\end{align}
This can be reformulated in terms of the effective degrees of freedom $g_{\mathrm{eff}}$,
\begin{align}\label{eq:geff-def}
   \rho_{\mathrm{rad}} =  \frac{\pi^2}{30} \sum_{a} \underbrace{n_{a}
  \left(  \frac{15}{\pi^4} \int_{m_a/T}^{\infty} \dd u \frac{u^2 \sqrt{u^2 -
  (m_a/T)^2}}{\mathrm{e}^{u} \mp  1}\right)}_{\equiv g_{\mathrm{eff},a}(T)} T^4 
  = \frac{\pi^2}{30}  \sum_{a}g_{\mathrm{eff},a}(T) T^{4} \,,
\end{align}
where $g_{\mathrm{eff},a}(T)$ denotes the effective degrees of freedom for species $a$, carrying
a mass $m_a$, see also ref.~\cite{Husdal:2016haj} for equivalent formulations of the same quantity.
Within \texttt{TransitionListener} we implement the radiation energy density in terms of the
thermal functions and their derivatives, which are pre-computed and tabulated in terms of $z_a = m^2_a/T^2$,
\begin{align}
  \rho_{\mathrm{rad}} = -\sum_{a}n_{a} \left( \frac{3 T^4}{2\pi^{2}}  J_\text{b/f}\left( z_a \right) -
    \frac{2 m^2T^2}{2\pi^{2}} \frac{\dd}{\dd z_a} J_\text{b/f}(z_a)  \right)\,.
\end{align}
\paragraph{Counting the degrees of freedom.}
Note that the contribution from the of the gauge sector to the thermal
  corrections of the potential in the $R_{\xi}$ gauge has the following form \cite{Patel:2011th}
\begin{align}\label{eq:}
  V_{\mathrm{T}}^{\mathrm{gauge}} = \frac{T^4}{2\pi^2} \sum_{\mathrm{g}}
  \left[3 J_{b} \left( \frac{m_{g}^2}{T^2} \right) -
  J_{b} \left( \frac{\xi m_{g}^2}{T^2} \right) \right]\,.
\end{align}
In the Landau gauge $\xi = 0$ the second term gives a field-independent (pure radiation)
contribution which is often neglected. We include this term in the effective potential to
ensure the correct number of degrees of freedom from gauge and Goldstone
bosons in the true and false vacua when computing the pressure and energy densities.

\paragraph{Daisy energy density.}

The total energy density of a plasma does not only obtain contributions from the radiation term
of non-interacting particles, but also corrections from particle interactions, cf.~eq.~\eqref{eq:rho},
\begin{align}
  \rho_{\mathrm{int}}
  &= V_{\mathrm{daisy}} - T \frac{\partial V_{\mathrm{daisy}}}{\partial T}
  = \frac{T^2}{8\pi} \sum_{i \in B_\text{L}} \left( m^2_{i}(\phi) + \Pi_{i}(T)\right)^{1/2}
  \frac{\partial \Pi_i(T)}{\partial T} \, ,
\end{align}
where the sum runs over the scalars and longitudinal modes of the gauge bosons. This term
is of order $g^{3}$ in the couplings, and hence typically suppressed compared to the
radiation energy density $\rho_{\mathrm{rad}}$ derived above. We therefore neglect this
contribution in our computations.

\paragraph{Effective degrees of freedom.}
The number of effective degrees of freedom of a single species $a$ contributing to the total
energy density, the pressure and entropy density are defined through
\begin{align}
\rho_{a}=\frac{\pi^2}{30}\,g_{\text{eff},a}(T)\,T^4 \, , &&
p_{a}=\frac{\pi^2}{90}\,k_{\text{eff},a}(T)\,T^4 \, , && \text{and} &&
s_{a} = \frac{\rho_{a} + p_{a}}{T} = \frac{2\pi^{2}}{45} h_{\mathrm{eff},a}(T) T^{3}\,,
\end{align}
such that the respective quantities for the full thermal bath can be obtained by summing over $a$.
The individual functions $g_{\text{eff},a}(T)$, $k_{\text{eff},a}(T)$, and $h_{\mathrm{eff},a}(T)$
can be computed in terms of the thermal integrals $I_\rho^{\text{b}/\text{f}}$ and
$I_p^{\text{b}/\text{f}}$,
\begin{align}
  g_{\text{eff},a}(T) = g_a I_\rho^{\text{b}/\text{f}} \ba{\frac{m_a}{T}}\,, &&
  k_{\text{eff},a}(T) = g_a I_p^{\text{b}/\text{f}} \ba{\frac{m_a}{T}}\,, &&
  h_{\mathrm{eff},a}(T) = \frac{3 g_{\text{eff},a}(T) + k_{\text{eff},a}(T)}{4}\,,
\end{align}
where 
\begin{align}
I_\rho^{\text{b}/\text{f}}(x)=\frac{15}{\pi^4}\int_x^\infty
\frac{u^2\sqrt{u^2-x^2}}{\mathrm{e}^u\mp1}\,\mathrm{d}u \, , && \text{and} &&
I_p^{\text{b}/\text{f}}(x)=\frac{15}{\pi^4}\int_x^\infty
\frac{(u^2-x^2)^{3/2}}{\mathrm{e}^u\mp1}\,\mathrm{d}u \, .
\end{align}
We use tabulated values for these integrals within the \texttt{thermodynamics} module to
speed up numerical computation while preserving accuracy.

\section{Analytical relation between transition speed and bubble size}
\label{app:betaH-RH}
Here we derive an analytical relation between $R_{\mathrm{sep}}$
and $\beta$ at percolation, as first presented in
ref.~\cite{Megevand:2016lpr}. Approximating the nucleation
rate as an exponential around
the percolation time $t_\text{perc}$ and neglecting the cosmic
expansion, we obtain 
\begin{align}
  \Gamma(t) &\approx \Gamma_\perc \mathrm{e}^{-\beta(t- t_\perc)}\,,
  \qquad \text{and} \quad P_\text{f}(t) \approx
  \exp \left( - \frac{8\pi v^3_\text{w}}{\beta^4} \Gamma_\perc
  \mathrm{e}^{-\beta(t-t_\perc)}\right)\,.
\end{align}
Neglecting cosmic expansion in the bubble density as well, we obtain
\begin{align}\label{eq:bubble-density-approx}
  n(t) &\approx \int_{-\infty}^{t} \dd t^{\prime} \,  \Gamma(t^{\prime})
  P_\text{f}(t^{\prime}) = -\frac{\beta^3}{8\pi v_\text{w}^3}
  \int_{-\infty}^{t_\perc} \diff t \,  \td{}{t}
  \left( \exp \left( -\frac{8\pi v_\text{w}^{3}}{\beta^4}
    \Gamma_\perc \mathrm{e}^{-\beta (t-t_{\perc})} \right) \right) \nonumber \\
       &= \frac{\beta^3}{8\pi v_\text{w}^3}
       \left(1 -  \exp \left(-\frac{8\pi v^3_{w}}{\beta^{4}}\Gamma_{\perc}\right) \right)
         = \frac{\beta^3}{8\pi v_\text{w}^3} \left(1 -  P_\text{f}(t_{\perc})\right)
\end{align}
The mean bubble separation at percolation is hence related to the transition speed through
\begin{align}
  R_\text{sep}(t_{\perc}) = \frac{1}{n(t_\perc)^{1/3}}
  \approx \ba{\frac{8 \pi}{f_{\mathrm{perc}}}}^{1/3} \frac{v_\text{w}}{\beta} 
\end{align}
with the factor $f_{\mathrm{perc}} = 1 - P_\text{f}(t_{\mathrm{perc}}) \approx 0.29$ taking into
account the true-vacuum fraction.

\section{Using TransitionListener in a Python script}
\label{sec:python-usage}

In code snippet~\ref{lst:using-the-code} we show how to manually perform the steps from
initialising the potential to plotting the GW spectrum. First the potential is
initialised, and phases are traced using the \texttt{Phases} class, followed by viable
transitions being found using the \texttt{Transitions} class. The evolution of the
true-vacuum fraction and the percolation temperature are computed using the
\texttt{TransitionsObservables} which gives back the \texttt{percolationResults} object.
The instanton solution, the action and the nucleation rate can be computed with the
functions \texttt{bounceSolution} and \texttt{Gamma}. Lastly the GW spectrum is created
from which the SNR ratios are determined through the \texttt{Observability} class. More
details can be found in the documentation on
\begin{center}
  \href{https://tasillo.de/TransitionListener/index.html}{https://tasillo.de/TransitionListener/index.html}.
\end{center}

\begin{lstlisting}[language=python,label=lst:using-the-code,
caption={Alternative computation of the GW spectrum and other intermediate
results for the Abelian dark Higgs model without using the
\texttt{TransitionListener} interface class.}]
from models.TL_dark_U1 import specific_potential
from transitionlistener.phases import Phases
from transitionlistener.transitions import Transitions
from transitionlistener.transitionObservables import TransitionObservables
from transitionlistener.observability import Observability
from transitionlistener.gwfopt import FOPTspectrum
from transitionlistener.plots import plotGWSpectrum
from transitionlistener.pathDeformation import bounceSolution
from transitionlistener.bubbledynamics import Gamma

# Initialising the potential
# =====================================================
pot = specific_potential(inp={"g_tilde": 2.69, "l": 1.5e-3, "v_GeV": 1e7})

# Phases and possible transitions
# =====================================================
phases = Phases(pot, verbose=True)
transitions = Transitions(phases, pot, verbose=True)

# Percolation
# =====================================================
observables = TransitionObservables(pot, phases, transitions, verbose=True)
percolation = observables.percolationResults[0]
Tperc = percolation.Tperc
Tperc_GeV = Tperc * pot.conversionFactor

# Bounce action, bubble profiles and , nucleation rates
# ======================================================
x_start, x_end = phases[1].valAt(Tperc), phases[0].valAt(Tperc)
bSol = bounceSolution(pot, Tperc, x_start, x_end)
action = bSol.action
Phis = bSol.Phi
R = bSol.profile1D.R
nuclRate = Gamma(Tperc, action)

# Gravitational wave spectrum
# ======================================================
gwspectrum = FOPTspectrum(observables[0], verbose=False)
observability = Observability(gwspectrum, verbose=True, include_smbhb=False)
plotGWSpectrum(observables[0], showplot=True)
\end{lstlisting}

\newpage

\section{Tables of error codes, structure of the output and accuracy settings}
\label{sec:error_codes}

Table~\ref{tab:errorcodes} lists the error codes returned by \texttt{TransitionListenerw}
together with a short explanation of the underlying issue.  Tab.~\ref{tab:csv_columns}
lists the columns used within the \texttt{output\_table.csv} file automatically created 
during random and \texttt{UltraNest} scans. Tables~\ref{tab:tracingconf_defaults},
\ref{tab:tunneling_params}, \ref{tab:gwconf_defaults} and \ref{tab:percolationconf_defaults}
list the available configuration options for the phase tracing,
the bounce action computation, the solution of the percolation integral 
as well as the computation of the GW spectrum, respectively.

\begin{table}[ht]
  \centering
  \footnotesize
  \setlength{\tabcolsep}{4pt}
  \renewcommand{\arraystretch}{1.1}
  \begin{tabularx}{\linewidth}{@{}c l X@{}}
    \toprule
    Code & Exception & Typical meaning \\
    \midrule
    1   & \texttt{TachyonError}                     & Tachyonic mass at the $T=T_0$ vacuum;
    usually cured by adjusting the model parameters or the renormalisation conditions. \\
    2   & \texttt{NucleationError}                  & The nucleation criterion
    $\mathcal{C}_\text{nuc}(T_\text{nuc}) = 0$ could not be satisfied; surfaces only
    in the \texttt{fixed\_step\_size} workflow. \\
    3   & \texttt{WrongT0MinimumError}              & The last traced phase does not match
    the expected $T=T_0$ minimum. \\
    4   & \texttt{NoPhases}                         & Phase tracing failed altogether. \\
    5   & \texttt{OnlyOnePhase}                     & Only the high-temperature phase was
    found; no transitions can be constructed. \\
    6   & \texttt{NoTransitionFound}                & Multiple phases exist but no viable
    transitions exist. \\
    7   & \texttt{PercolationApproximation1Error}   & The saddlepoint approximation for
    percolation failed to converge, usually signaling a very flat $S_3/T$ curve. \\
    8   & \texttt{TooMuchSupercoolingError}         & $P_\text{t}(T_\text{min}) < f_\perc$
    at the coldest explored temperature: the transition did not reach percolation in the
    evaluated window. \\
    9   & \texttt{OnlySecondOrderTransitionsError}  & All traced transitions are continuous,
    no GWs are emitted. \\
    10  & \texttt{PercolationError}                 & The percolation integral could not be
    evaluated reliably: the step-2/3 Brent solve for $T_\perc$ did not bracket a root, the
    ODE step collapsed below floating-point spacing, or the action evaluation budget was exhausted. \\
    11  & \texttt{TunnelingError}                   & The path-deformation or overshoot/undershoot
    solver failed to converge at the requested temperature. \\
    12  & \texttt{InitPotentialError}               & The model parameters produce an ill-defined
    potential (e.g.~unbounded from below or violating input bounds). \\
    13  & \texttt{SplineError}                      & Construction of the deformation spline failed,
    often because the initial path collapses in multi-field space. \\
    14  & \texttt{WrongHighTPhaseError}             & The first transition did not originate from
    the high-temperature phase, indicating an inconsistent phase tree. \\
    15  & \texttt{EternalInflationError}            & $f_\perc \le P_\text{t}(T_\text{min}) < f_\text{final}$
    at the coldest explored temperature: percolation occurred but residual false-vacuum regions
    persist and continue to inflate. \\
    16  & \texttt{Timeout}                          & The per-point timeout (specified in the
    \texttt{.yaml} file) elapsed before the computation finished. \\
    17  & \texttt{ActionRateJitterError}            & Jitter in $\Gamma/H^4$ across the active
    percolation band detected, which could not be fixed automatically. \\
    999 & \texttt{UnexpectedError}                  & Catch-all code for uncategorised failures;
    consult \texttt{errormsg} for the Python traceback. \\
    \bottomrule
  \end{tabularx}
  \caption{Error codes exposed by the CLI and scan backends. Codes are defined in
  \texttt{transitionlistener/errors.py} and are
  propagated to \texttt{output\_table.csv} as well as external interfaces.}
  \label{tab:errorcodes}
\end{table}

\begin{table}
\centering
\footnotesize
\setlength{\tabcolsep}{4pt}
\renewcommand{\arraystretch}{1.1}
\begin{tabular}{p{0.32\linewidth}p{0.62\linewidth}}
\toprule
Column & Description \\
\midrule

\texttt{[model parameters]} & Model-specific input
parameters (names depend on the chosen model). \\

\texttt{smoothened\_lnL} & Smoothed PTA log-likelihood for the default PTA label. \\
\texttt{PTArcade\_lnL} & PTArcade log-likelihood for the default PTA label. \\
\texttt{mock\_lnL} & Mock PTA log-likelihood for the default PTA label. \\

\texttt{error} & Error code (0 for success, see Tab.~\ref{tab:errorcodes}). \\
\texttt{errormsg} & Exception message when \texttt{error} $\neq 0$. \\

\texttt{alpha}, \texttt{alpha\_inf}, \texttt{alpha\_eq} & Strength
parameters ($\alpha$ and hydrodynamic variants). \\

\texttt{betaH\_S3}, \texttt{betaH\_RH} & $\ba{\beta/H}_{S_3}$ and $\ba{\beta/H}_{RH}$. \\

\texttt{RH} & Mean bubble separation $RH$ at percolation. \\

\texttt{Treh\_SM\_GeV}, \texttt{Tperc\_SM\_GeV} & Reheating and
percolation temperatures in GeV. \\

\texttt{g\_eff\_tot\_reh}, \texttt{h\_eff\_tot\_reh} & Effective
energy/entropy d.o.f.\ at reheating. \\

\texttt{kappa\_phi}, \texttt{kappa\_sw}, \texttt{kappa\_turb} & Efficiency
factors for collisions, sound waves, turbulence. \\

\texttt{g0}, \texttt{h0} & Present-day effective d.o.f.\ used in redshift. \\

\texttt{v\_wall} & Wall velocity. \\
\texttt{D} & Late-time dilution factor. \\

\texttt{c\_s}, \texttt{c\_s\_sym}, \texttt{c\_s\_bro} & Sound speed and
phase-specific values. \\

\texttt{step}, \texttt{total\_steps} & Transition step index within a
multi-step history; total step number. \\

\texttt{Tnuc\_SM\_GeV}, \texttt{Tcrit\_SM\_GeV}, \texttt{Tf\_SM\_GeV} & Nucleation,
critical and completion temperatures (GeV). \\

\texttt{xi\_crit} & $\xi = v/T$ parameter at the critical temperature. \\

\texttt{WARNING:too\_weak\_to\_compute\_perc} & Flag for weak transitions (percolation skipped). \\
\texttt{WARNING:no\_perc\_splines} & Flag when percolation splines are unavailable. \\
\texttt{WARNING:betaH\_[very]\_small} & Flag when $\ba{\beta/H}_{RH} < 10$ [$\ba{\beta/H}_{RH} < 3$]. \\
\texttt{WARNING:betaH\_mismatch} & Flag when $\ba{\beta/H}_{S_3}$ and $\ba{\beta/H}_{RH}$ differ by more than a factor 10. \\
\texttt{WARNING:betaH\_nonfinite} & Flag when $\ba{\beta/H}_{S_3}$ or $\ba{\beta/H}_{RH}$ are non-finite. \\
\texttt{WARNING:nucleationRate\_nonexp} & Flag when $\ba{\beta/H}_{S_3} < 0$, indicating non-exponential growth of $\Gamma(T)$. \\
\texttt{WARNING:spline\_tnuc\_<missing>} & Flag when the nucleation temperature spline is unavailable or cannot be constructed. \\
\texttt{WARNING:not\_T0\_global\_min} & Flag when the reached $T=T_0$ minimum is not the global minimum. \\

\texttt{lnL\_<variant>\_<pta\_label>} & PTA likelihood outputs for each
configured dataset label, with \texttt{<variant>} $\in \{\texttt{smoothened},\texttt{PTArcade},\texttt{mock}\}$. \\

\texttt{<detector>\_SNR} & Detector-specific SNR outputs for all enabled
sensitivity curves (e.g.\ \texttt{LISA\_SNR}, \texttt{ET\_SNR}, \texttt{BBO\_SNR}, \texttt{SKA\_20\_yrs\_SNR}). \\

\texttt{f\_peak\_Hz}, \texttt{h2OmegaGW\_peak} & GW peak frequency and amplitude. \\
\texttt{f\_pivot\_Hz}, \texttt{h2OmegaGW\_at\_pivot} & Pivot frequency and GW amplitude at that pivot. \\
\texttt{DNeff\_GW} & GW contribution to $\Delta N_{\text{eff}}$. \\

\texttt{[masses]} & Zero-temperature mass spectrum entries (model-dependent). \\

\bottomrule
\end{tabular}
\caption{Standard columns in \texttt{output\_table.csv}. Model parameters and
mass-spectrum entries are placeholders that expand according to the chosen
model and the configured \texttt{MassSpectrum}.}
\label{tab:csv_columns}
\end{table}

\begin{table}
  \centering
  \footnotesize
  \begin{tabular}{p{0.25\linewidth}p{0.1\linewidth}p{0.55\linewidth}}
    \toprule
    Parameter & Default & Role \\
    \midrule
    \texttt{internal\_scale} & 1000 & Reference scale in internal units;
    usually the vev in internal units. \\
    \texttt{Tmax\_factor} & 2.5 & Sets
    $T_{\text{max}}=\texttt{Tmax\_factor}\times \texttt{internal\_scale}$
    up to which the phases are traced. \\
    \texttt{diftol} & 1 & Merge phases whose minima distance in field space
                          is less than \texttt{diftol}\,$\times$\texttt{internal\_scale}. \\
    \texttt{Z2\_cutof\_factor} & $-5$ & Discard far-negative
    mirror phases for $\mathbb{Z}_2$ symmetries. \\
    \texttt{tracing\_derivative\_order} & 4 & Finite-difference
    order for $\partial \bm{\phi}/\partial T$ and Hessians. Either 2 or 4. \\
    \texttt{tracing\_field\_accuracy} & $10^{-3}$ & Accuracy in the
    fields during phase tracing: Tolerance in the minimisation and
    field space finite difference step \texttt{x\_eps}. \\
    \texttt{tracing\_temp\_accuracy} & $10^{-3}$ & Temperature finite
    difference step size in gradient computation $\partial V / \partial T$ (\texttt{T\_eps}). \\
    \texttt{nucleation\_Ttol} & $10^{-8}$ & Relative root-finding tolerance for $T_\text{nuc}$. \\
    \texttt{approx\_strength\_\allowbreak threshold} & $10^{-4}$ & Treat transitions with approximate strength
    $\Delta V/T^4$ below this value as second order \\
    \texttt{do\_tachyon\_test} & True & Check if there are
    tachyonic instabilities at $T=T_0$ during model initialisation. \\
    \texttt{tracing\_args.dtstart} & $10^{-4}$ & Relative initial
    trace step in temperature ($\Delta T = $ \texttt{dtstart}$\times($\texttt{Tmax} - \texttt{Tmin})). \\
    \texttt{tracing\_args.dtabsMax} & 20.0 & Maximal
    temperature step in the tracing (relative to \texttt{dtstart}). \\
    \texttt{tracing\_args.dtfracMax} & 0.25 & Max fractional
    temperature step ($\Delta T/T$). \\
    \texttt{tracing\_args.dtmin} & $10^{-6}$ & Minimum step
    size before declaring a phase endpoint. \\
    \texttt{tracing\_args.minratio} & $10^{-4}$ & Smallest eigenvalue
    ratio before treating an extremum as a saddle point. \\
    \texttt{tracing\_args.tjump} & $10^{-5}$ & Relative temperature
    jump after phase vanishes to search for new phase,
    scaled by (\texttt{Tmax} - \texttt{Tmin}). \\
    \texttt{tracing\_args.deltaX\_tol} & 1.2 & Tolerance
    factor for field-space step errors. \\
    \bottomrule
  \end{tabular}
  \caption{Default values in \texttt{TracingConf}.}
  \label{tab:tracingconf_defaults}
\end{table}

\begin{table}
  \centering
  \footnotesize
  \begin{tabular}{p{0.4\linewidth}p{0.1\linewidth}p{0.40\linewidth}}
    \toprule
    Parameter & Default & Role \\
    \midrule
    \texttt{tunneling\_findProfile\_params.phitol} & $10^{-10}$ & Accuracy target for the profile solver. \\
    \texttt{tunneling\_findProfile\_params.xtol} & $10^{-10}$ & Root-finding tolerance in field space. \\
    \texttt{tunneling\_init\_params.rscale} & \texttt{None} & Bubble radius scale. If none,
                                                              determined from the oscillation time around the potential barrier.\\
    \texttt{tunneling\_findProfile\_params.rmin} & $10^{-4}$ & Minimum radial coordinate for the
    bounce profile (relative to the bubble radius scale \texttt{rscale}). \\
    \texttt{tunneling\_findProfile\_params.rmax} & $10^{4}$ & Maximum radial coordinate for the
    bounce profile (relative to the bubble radius scale \texttt{rscale}). \\
    \texttt{V\_spline\_samples} & 1000 & Samples used to spline the potential along the path. \\
    \texttt{tunneling\_init\_params.phi\_eps} & $10^{-3}$ & Finite difference step in field space for numerical
                                                            derivatives of the potential in the bounce equation. \\
    \texttt{tunneling\_findProfile\_params.npoints} & 500 & Number of radial samples in the 1D profile. \\
    \texttt{deformation\_deform\_params.fRatioConv} & $2\times 10^{-2}$ & Convergence criterion: Stop when the maximal force on the
                                                                     path divided by the maximal pot.~gradient is
                                                                     smaller than \texttt{fRatioConv}. \\

    \texttt{deformation\_deform\_params.startstep} & $2\times 10^{-3}$ & Initial path-deformation step size. \\
    \texttt{deformation\_deform\_params.converge\_0} & 5.0 & Increase the convergence criterion of the first
                                                             iteration by the factor \texttt{converge\_0}. \\
    \texttt{deformation\_deform\_params.\allowbreak fRatioIncrease} & 5.0 & The maximum fractional amount the normal force ratio
                                                                can increase before raising an error. \\
    \texttt{deformation\_deform\_params.maxiter} & 500 & Maximum number of deformation iterations. \\
    \bottomrule
  \end{tabular}
  \caption{Default values for tunnelling and path-deformation control.}
  \label{tab:tunneling_params}
\end{table}

\begin{table}
  \centering
  \footnotesize
  \begin{tabular}{p{0.34\linewidth}p{0.16\linewidth}p{0.42\linewidth}}
    \toprule
    Parameter & Default & Role \\
    \midrule
    \texttt{wall\_velocity} & \texttt{"LTE"} & Wall velocity prescription
                                               (\texttt{"LTE"} or \texttt{float}
                                               between 0 and 1). \\
    \texttt{bw\_collisions} & \texttt{"off"} & Bubble-collision source model:
                                               If \texttt{"off"} do not include contribution;
                                               if \texttt{"NLO"}, compute from NLO friction;
                                               if \texttt{"full"}, set $\kappa_{\mathrm{col}} = 1$. \\
    \texttt{epsilon\_turbulence} & 0.1 & Fraction of kinetic energy converted to turbulence. If
    0, do not include turbulence as a GW source.\\
    \texttt{sound\_speed} & \texttt{"compute"} & Sound-speed prescription
    (\texttt{"compute"} or value between $ 0 < c_\text{s}\leq 1$). \\
    \texttt{check\_if\_T0\_global\_min} & False & Enforce the $T=0$ global-minimum check. \\
    \bottomrule
  \end{tabular}
  \caption{Default of the \texttt{GWConf} settings. They control the computation of the
    bubble wall velocity, efficiency factors and contributions to the GW spectrum.}
  \label{tab:gwconf_defaults}
\end{table}

\begin{table}
  \centering
  \footnotesize
  \begin{tabular}{p{0.25\linewidth}p{0.17\linewidth}p{0.5\linewidth}}
    \toprule
    Parameter & Default & Role \\
    \midrule
    \texttt{algorithm\_mode} & \texttt{adaptive\_\allowbreak step\_size} & Solver mode:
    \texttt{adaptive\_step\_size} or \texttt{fixed\_step\_size}. \\
    \texttt{integral\_method} & \texttt{ode} & Backend used for the percolation integral
    evaluation. Options: \texttt{ode} or \texttt{double\_integral}. \\
    \texttt{time\_temperature\_mode} & \texttt{sound\_speed} & Choice of time-temperature
    relation, using either \texttt{sounds\_speed} or \texttt{bag\_model}. \\
    \texttt{f\_perc} & 0.28957 & True-vacuum fraction at percolation. \\
    \texttt{f\_final} & 0.99 & Completion threshold for $P(T_\text{f})$. \\
    \texttt{maxit} & 10 & Maximum iterations in steps 2 and 3 of the percolation solver. \\
    \midrule
    \texttt{n\_action} & 30 & Fixed number of action support points used in \texttt{fixed\_step\_size} mode. \\
    \texttt{weight} & $2/3$ & Fraction of points of the temperature grid to lay between
                              $T_\text{nuc}$ and $T_{\perc}^{\text{approx}}$. \\
    \texttt{rel\_increment} & 0.10 & Relative expansion of the temperature interval when
    $P(T)$ does not bracket the target range. \\
    \texttt{max\_boundary\_ratio} & 0.45 & Maximum fraction of grid points with
    $P_\text{t}=0$ or $P_\text{t}=1$ before expanding the interval. \\
    \midrule
    \texttt{f\_start} & $10^{-3}$ & Initial $P_\text{t}$, starting point
    for the ODE solver. \\
    \texttt{n\_action\_min} & 15 & Minimum number of action support points
    constructed before percolation step 2. \\
    \texttt{n\_action\_increment} & 5 & Number of additional action support
    points added per percolation step 2 iteration. \\
    \texttt{n\_action\_max} & 60 & Maximum total number of action support
    points allowed across iterations. \\
    \texttt{max\_action\_temperatures} & 100 & Absolute upper limit on
    distinct temperatures at which the action may be evaluated in one run. \\
    \texttt{large\_delta\_p\_refine\_thr} & 0.1 & If the change in neighbouring
    $P(T)$ values exceeds this threshold, additional refinement points are
    inserted. \\
    \texttt{large\_delta\_p\_success\_thr} & 0.2 & If the change in neighbouring
    $P(T)$ values exceeds this threshold, the solution is not accepted as converged. \\
    \texttt{jitter\_GH4\_threshold} & 1.0 & Tolerance in $\log_{10}(\Gamma/H^4)$
    above which unresolved action jitter is flagged. \\
    \texttt{jitter\_rescue} & \texttt{True} & If enabled, reruns the calculation
    with tighter tunnelling settings when jitter is detected. \\
    \texttt{n\_jitter\_save} & 20 & Maximum number of jitter diagnostic samples
    written out. \\
    \texttt{acc\_tperc} & $10^{-2}$ & Relative accuracy target for $T_{\perc}$. \\
    \texttt{acc\_tfinal} & $10^{-2}$ & Relative accuracy target for $T_\text{f}$. \\
    \texttt{acc\_rh} & $10^{-2}$ & Relative accuracy target for $RH$. \\
    \bottomrule
  \end{tabular}
  \caption{Default values in \texttt{PercolationConf}. The second (third) block only affects the
  \texttt{fixed\_step\_size} (\texttt{adaptive\_step\_size}) mode.}
  \label{tab:percolationconf_defaults}
\end{table}

\newpage

\vspace*{0.5cm}
\printbibliography 

\end{document}